\documentclass[preprint,3p,times]{elsarticle}
\usepackage{amssymb}

\usepackage{amsmath}
\usepackage{subcaption}
\usepackage{caption}
\usepackage{xcolor}
\usepackage{float}
\usepackage{graphicx}
\usepackage{bm}
\usepackage{algorithm}
\usepackage{algpseudocode}
\usepackage{url}

\algrenewcommand{\algorithmiccomment}[1]{\hskip0.5em // \textit{#1} }

\newtheorem{theorem}{Theorem}[section]

  { \end{itemize} }
  { \end{enumerate} }

\newcommand{\HW}[1]{} 

\newcommand{\IDEE}[1]{}
 
\allowdisplaybreaks[3]

\def\ED{\end{document}}

\newenvironment{TC} {\left \{\begin{array}{ll}} {\end{array} \right.}
\newcounter{pcounter} 
 
\newcommand{\HH}{{\mathcal H}}

\newcommand{\LL}{{\mathcal L}}
\newcommand{\MM}{{\mathcal M}}

\newcommand{\YY}{{\mathcal Y}}

\newcommand{\R}{\mathbb{R}}

\renewcommand{\phi}{\varphi}

\newcommand{\Ome}{\Omega}

\DeclareMathOperator*{\cof}{cof}

\def\XXint#1#2#3{{\setbox0=\hbox{$#1{#2#3}{\int}$}
\vcenter{\hbox{$#2#3$}}\kern-.5\wd0}}

\usepackage{color}
\definecolor{verylightblue}{rgb}{0.95, 0.95, 0.95}  
\definecolor{lightblue}{rgb}{0.7, 0.7, 1}

\definecolor{eqyellow}{rgb}{0.9375,0.8984,0.5469}
\definecolor{subeqyellow}{rgb}{1,0.9373,0.8353}

\definecolor{mygreen}{rgb}{0.3, 0.6, 0.3} 
\definecolor{verylightgreen}{rgb}{0.95, 0.95, 0.95} 
\definecolor{verydarkgreen}{rgb}{0, 0.5, 0}
\definecolor{darkgreen}{rgb}{0.85, 0.85, 0.85}  
\definecolor{mydarkgreen}{rgb}{0, 0.5, 0} 

\definecolor{mybrown}{rgb}{0.85, 0.4, 0.3}
\definecolor{verylightbrown}{rgb}{0.98, 0.72, 0.58}
\definecolor{verydarkbrown}{rgb}{0.44, 0.26, 0.26}

\definecolor{orange}{rgb}{1, 0.5, 0}

\definecolor{BurntOrange}{rgb}{1,0.356,0}

\definecolor{mydarkred}{rgb}{1,0.086,0.255}

\definecolor{RoseVYDP}{rgb}{0.84,0.086,0.255}

\definecolor{dgreen}{rgb}{0, 0.8, 0.5}     

\definecolor{CanaryBRT}{rgb}{1,0.76,0.26}

\definecolor{cyan}{rgb}{0, 1, 1}
\definecolor{verylightgray}{rgb}{0.95, 0.95, 0.95}
\definecolor{verylightgray}{rgb}{0.95, 0.95, 0.95}
\definecolor{verylightred}{rgb}{1, 0.8, 0.78}
\definecolor{verylightyellow}{rgb}{0.99, 0.98, 0.5}

\newcommand{\ignore}[1]{{}}

\begin{document}

\newcommand{\beq}{\begin{equation}}
\newcommand{\beql}[1]{\begin{equation}\label{#1}}
\newcommand{\eeq}{\end{equation}}
\newcommand{\bea}{\begin{eqnarray}}
\newcommand{\beal}[1]{\begin{eqnarray}\label{#1}}
\newcommand{\eea}{\end{eqnarray}}
\newcommand{\bean}{\begin{eqnarray*}}
\newcommand{\eean}{\end{eqnarray*}}

\newcommand{\del}{\delta}

\newcommand{\mbf}[1]{\mathbf{#1}}
\newcommand{\sbf}[1]{\boldsymbol{#1}}

\newcommand{\intL}{\int_{\cal L}}

\newcommand{\beps}{\sbf{\varepsilon}}
\newcommand{\bsig}{\sbf{\sigma}}
\newcommand{\blam}{\sbf{\lambda}}
\newcommand{\bmu}{\sbf{\mu}}
\newcommand{\ba}{\mbf{a}}
\newcommand{\bi}{\mbf{i}}
\newcommand{\bB}{\mbf{B}}
\newcommand{\bS}{\mbf{S}}
\newcommand{\bV}{\mbf{V}}
\newcommand{\bA}{\mbf{A}}
\newcommand{\bE}{\mbf{E}}
\newcommand{\bU}{\mbf{U}}
\newcommand{\bR}{\mbf{R}}
\newcommand{\bX}{\mbf{X}}
\newcommand{\bK}{\mbf{K}}
\newcommand{\bd}{\mbf{d}}
\newcommand{\bs}{\mbf{s}}
\newcommand{\be}{\mbf{e}}
\newcommand{\bff}{\mbf{f}}
\newcommand{\bF}{\mbf{F}}
\newcommand{\bg}{\mbf{g}}
\newcommand{\bG}{\mbf{G}}
\newcommand{\bL}{\mbf{L}}
\newcommand{\bM}{\mbf{M}}
\newcommand{\bH}{\mbf{H}}
\newcommand{\bkron}{\mbf{1}}
\newcommand{\sigm}{\sigma_{\rm m}}
\newcommand{\epsm}{\varepsilon_{\rm m}}
\newcommand{\bI}{\mbf{I}}
\newcommand{\bIv}{\mbf{I}_{\rm K}}
\newcommand{\bId}{\mbf{I}_{\rm D}}
\newcommand{\bIs}{\mbf{I}_{\rm S}}
\newcommand{\bCe}{\mbf{C}_{\rm e}}
\newcommand{\bDe}{\mbf{D}_{\rm e}}
\newcommand{\bC}{\mbf{C}}
\newcommand{\bD}{\mbf{D}}
\newcommand{\bu}{\mbf{u}}
\newcommand{\bv}{\mbf{v}}
\newcommand{\bw}{\mbf{w}}
\newcommand{\bz}{\mbf{z}}
\newcommand{\bT}{\mbf{T}}
\newcommand{\bt}{\mbf{t}}
\newcommand{\bn}{\mbf{n}}
\newcommand{\bN}{\mbf{N}}
\newcommand{\bx}{\mbf{x}}
\newcommand{\bq}{\mbf{q}}
\newcommand{\bh}{\mbf{h}}

\newcommand{\cE}{{\cal E}}
\newcommand{\cH}{{\cal H}}
\newcommand{\cA}{{\cal A}}
\newcommand{\cD}{{\cal D}}

\newcommand{\G}{{\rm I\!G}}

\newcommand{\bnab}{\sbf{\nabla}}
\newcommand{\bnabs}{\sbf{\nabla}_{\rm s}}

\newcommand{\third}{\mbox{$\frac 1 3$}}
\newcommand{\half}{\mbox{$\frac 1 2$}}
\newcommand{\dV}{\,\mbox{d}V}
\newcommand{\dx}{\,\mbox{d}x}
\newcommand{\dxi}{\,\mbox{d}\xi}
\newcommand{\deta}{\,{\rm d}\eta}
\newcommand{\dS}{\,\mbox{d}S}
\newcommand{\ds}{\,\mbox{d}s}
\newcommand{\dA}{\,\mbox{d}A}
\newcommand{\dt}{\,\mbox{d}t}
\newcommand{\pard}[2]{\frac{\partial #1}{\partial #2}}
\newcommand{\parder}[2]{\frac{\partial #1}{\partial #2}}

\begin{frontmatter}


\title{Surface-Polyconvex Models for Soft Elastic Solids}

\author[1,3]{Martin Hor\'{a}k}
\ead{martin.horak@cvut.cz}

\author[1]{Michal \v{S}mejkal\corref{cor1}}
\ead{michal.smejkal@fsv.cvut.cz}

\author[3,2]{Martin Kru\v{z}\'{i}k}
\ead{kruzik@utia.cas.cz,kruzik@fsv.cvut.cz}

\cortext[cor1]{Corresponding author}

\address[1]{Czech Technical University in Prague, Faculty of Civil Engineering, Department of Mechanics, Thákurova 2077/7, 166 29 Prague 6, Czechia}
\address[2]{Czech Technical University in Prague, Faculty of Civil Engineering, Department of Physics, Thákurova 2077/7, 166 29 Prague 6, Czechia}
\address[3]{Czech Academy of Sciences, Institute of Information Theory and Automation, Pod vod\'{a}renskou
v\v{e}\v{z}\'{\i}~4, 182~00~Prague~8, Czechia}

\begin{abstract}
Soft solids with surface energy exhibit complex mechanical behavior, necessitating advanced constitutive models to capture the interplay between bulk and surface mechanics. This interplay has profound implications for material design and emerging technologies. In this work, we set up variational models for bulk-surface elasticity and explore a novel class of surface-polyconvex constitutive models that account for surface energy while ensuring the existence of minimizers. 

These models are implemented within a finite element framework and validated through benchmark problems and applications, including, e.g., the liquid bridge problem and the Rayleigh-Plateau instability,
for which the surface energy plays the dominant role.
The results demonstrate the ability of surface-polyconvex models to accurately capture surface-driven phenomena, establishing them as a powerful tool for advancing the mechanics of soft materials in both engineering and biological applications.

\end{abstract}



\end{frontmatter}


\section{Introduction}
The trend toward miniaturization has driven advancements in mechanical, optical, and electronic devices, leading to increasingly smaller structures such as nanowires, nanobeams, liquid crystals, gels, and thin films. However, macroscale theories and technologies do not directly translate to smaller scales due to size effects. It is well established that surface and interface energies are among the main drivers of the size effects \cite{eremeyev2016effective}. The contrast between bulk and surface properties arises primarily because surface atoms have fewer neighboring bonds compared to those within the bulk and are further affected by phenomena like surface oxidation, aging, and coatings. Moreover, the surface effects intensify as materials decrease in size, with surface area scaling quadratically and volume cubically. This phenomenon is especially significant for soft solids, as highlighted in recent studies \cite{style2017elastocapillarity,bico2018elastocapillarity,xu2018surface,shao2018extracting,andreotti2016solid}, where surface effects are evident even at micron scales.

Describing the behavior of small-scale structures often requires first-principle methods such as Molecular Dynamics (MD), where surface effects are intrinsically included. However, MD is inefficient even for nanoscale structures, e.g., nanoelectromechanical systems, since even though the characteristic length scale may be in the order of nanometers, the size of the whole structure may be in micrometers \cite{fish2010multiscale}. In these scenarios, MD is not feasible, as it is limited to simulating molecular systems with no more than $10^6$ atoms over timescales of only tens to hundreds of nanoseconds \cite{roccatano2018short}. To address these limitations, enhancing the underlying continuum theory presents an efficient alternative.
To this end, Gurtin and Murdoch \cite{gurtin1975continuum} developed a unified continuum mechanics theory that incorporates surface effects, assuming that surface energy depends on surface strains. Initially based on small-strain elasticity theory, this approach was successful in characterizing hard nanomaterials \cite{wei2006finite,sharma2003effect,mozaffari2020surface,miller2000size}. Recent studies, e.g., \ \cite{style2017elastocapillarity,bico2018elastocapillarity,xu2018surface,shao2018extracting,andreotti2016solid,wu2018effect}, have demonstrated that surface energy also plays an important role in the small-scale mechanics of soft solids, significantly affecting the behavior of solids, such as gels. 
However, despite a comprehensive understanding of the physical origins behind the surface properties of stiff solids, a systematic exploration of the surface properties of soft solids is still at an early stage of development.
The significance of surface stress can be quantified through the elastocapillary length scale, defined as the ratio of surface energy to the shear elastic modulus \cite{bico2018elastocapillarity}. Surface effects become paramount for structures with characteristic sizes on the order of or smaller than the elastocapillary length scale.

The ability to manipulate and engineer soft materials is dependent on a thorough understanding of their behavior. Insights into the interactions between elastocapillary effects and bulk properties could lead to the development of novel technologies, such as innovative actuation methods in soft robotics. Consequently, the primary goal in this area of research is to formulate appropriate continuum models for surfaces. While numerous studies on soft solids have focused on isotropic and strain-independent fluid-like surface energies (see \cite{mora2015softening,mishra2018effect,xuan2017plateau,wang2020axisymmetric,riccobelli2020surface}), recent experimental evidence has shown that in some soft solids, the surface energy is strain-dependent \cite{xu2017direct,jensen2017strain,hui2020surface,heyden2021contact,heyden2021robust}. This phenomenon, known as the Shuttleworth effect \cite{shuttleworth1950surface}, is often overlooked in soft solid modeling. Notably, the theory of deformable bodies with general deformation-dependent boundary potentials is summarized in \cite{steinmann2008boundary}, where the duality between deformational and configurational mechanics is highlighted, while its extension to thermomechanics is presented in \cite{javili2013thermomechanics}.

It should be noted that in soft gels, the strain-dependent surface stress can far exceed the surface tension at zero stress \cite{style2017elastocapillarity,nicolas2021surface}. However, these empirical findings offer only a rudimentary insight into the fundamental mechanisms governing the impact of strain-dependent surface energies. Consequently, a deeper theoretical understanding of surface constitutive modeling is essential for advancing the field of modeling of soft materials. Additionally, the pronounced Shuttleworth effect introduces unique phenomena not found in liquids, such as the ability to manipulate the overall behavior of the material by adjusting the elastocapillary length, which can be extensively varied by applying different strain conditions. This opens promising avenues for the creation of new materials, for instance the development of adaptable adhesion and wetting systems for use in robotics, sensors, and sophisticated drug delivery systems \cite{wang2021bioinspired}.

A fundamental aspect in describing material behavior involves establishing a constitutive law. Specifically in mechanics, this refers to the relationship between stress and strain. Developing a comprehensive model for the constitutive behavior of soft surfaces presents significant challenges due to the pronounced geometric nonlinearity of soft materials. 
The form of elastic constitutive equations for soft material surfaces, which consider strain-dependent energy in conditions of large deformation, remains a subject of ongoing research \cite{style2017elastocapillarity}. For instance, the consideration of a neo-Hookean-type surface energy is noted in \cite{javili2010finite} or \cite{heyden2021contact}. However, while the theory of large deformations is well-established for bulk materials, its extension to surface materials has not yet been sufficiently investigated.

Only a mathematically well-formulated theory can yield a model that is predictive and physically relevant. Various constitutive constraints designed to ensure material stability have been suggested previously; for a summary, refer to \cite{marsden1994mathematical}. The strong mathematical foundations for large elastic deformations rely on the direct method of the calculus of variations. It is widely recognized that bulk energy densities relevant to mechanics cannot be convex, as this conflicts with physical principles such as the objectivity of the stored energy function. Typically, this issue is addressed by examining the weak lower semi-continuity of the energy functional, which leads to the quasiconvexity of the stored energy density. 
However,  quasi-convexity is a non-local condition that is difficult to verify. Thus, the stored energy density is often designed to fulfill a stronger condition, denoted as polyconvexity.
Introduced in \cite{ball1976convexity}, polyconvexity is a well-established principle for bulk materials and has been further applied to thermo-mechanical \cite{silhavy2013mechanics} and electro-mechanical problems \cite{horak2022polyconvex,gil2016new}. Additionally, polyconvexity is also physically meaningful as it involves the deformation gradient, its cofactor, and its determinant, which map infinitesimal vectors, areas, and volumes. 

Furthermore, polyconvexity leads to the Legendre-Hadamard condition, which ensures the presence of waves with real wave speeds, thereby guaranteeing material stability. It should be emphasized that polyconvexity characterizes most realistic models, including those based on statistical mechanics. In particular, models lacking polyconvexity, such as the Saint Venant-Kirchhoff model, are prone to nonphysical instabilities, as discussed in \cite{horak2020gradient}. However, polyconvex energy densities do not accurately represent certain materials, such as nematic elastomers, shape memory alloys, and magnetostrictive and ferroelectric materials, which are known to exhibit fine structures. The emergence of microstructures in these materials is attributed to the inability to find an exact optimum, necessitating increasingly detailed oscillations in optimizing sequences. To address this, the concept of gradient polyconvexity has been introduced recently in \cite{benesova2018note} and further explored in \cite{horak2022gradient}.

The concept of surface\footnote{Note that the condition is called interface polyconvexity in \cite{silhavy2011equilibrium}.} polyconvexity was introduced much later by Šilhavý in \cite{silhavy2011equilibrium, Silhavy-2010b} in the context of shape memory alloys, yet it has remained largely unnoticed within the continuum mechanics community. Remarkably, according to the condition outlined in \cite{silhavy2011equilibrium}, none of the strain-dependent constitutive models for surfaces documented in the literature are surface-polyconvex. 

Therefore, in this contribution, we explore surface-polyconvex constitutive models in the context of soft solids with surface energies guaranteeing mathematical well-posedness and the existence of minimizers. 

\section{Nonlinear surface elasticity}

\subsection{Kinematics}\label{sec:kin}
We consider a bounded domain in the reference, i.e., undeformed configuration $\Omega \subset \mathbb{R}^3$ with sufficiently regular boundary $\Gamma:=\partial \Omega$. Moreover, the material is assumed to possess surface energy on the part of the boundary $S \subset \Gamma$ with smooth boundary $\partial S$. 
The reference domain ${\Omega}$ is deformed by a deformation mapping $\bm{\varphi}:{\Omega} \to {\Omega}^{\bm{\varphi}}$ into the deformed configuration ${\Omega}^{\bm{\varphi}}\subset \mathbb{R}^3$. The deformation $\bm{\varphi}$ is assumed to be sufficiently smooth, injective, and orientation-preserving. {Moreover, we assume that the deformation mapping $\bm{\varphi}$ can be continuously extended to the closure of~$\Omega$ and we will not distinguish between $\bm{\varphi}$ and its extension.}
The material and spatial points are denoted by $\bm{X}$ and $\bm{x} = \bm{\varphi}(\bm{X})$ respectively. The surface deformation map $\hat{\bm{\varphi}}:\Gamma \to \bm{\varphi}(\Gamma)$ mapping the material points on the surface is defined by restricting ${\bm{\varphi}}$ to the boundary, i.e.,\
\bea
\hat{\bm{\varphi}}(\bm{X}) = {\bm{\varphi}}(\bm{X}), \quad \text{for }  \bm{X} \in \Gamma
\eea
sometimes denoted by $\hat{\bm{\varphi}} = \bm{\varphi}\rvert_\Gamma$. 

\subsubsection{Bulk deformation}
The deformation gradient 
$\bm{F}: \mathbb{R}^3 \to \mathbb{R}^3$ maps the infinitesimal vectors from the undeformed to the deformed configuration and is defined by 
\bea
\bm{F} = \nabla {\bm{\varphi}}
\eea
where $\nabla$ denotes the gradient with respect to the reference configuration. 
The Jacobian $J$, i.e., determinant of deformation gradient, transforms infinitesimal volume elements from the reference configuration {$\mathrm{d}V$} to infinitesimal volume elements  in the current configuration $\mathrm{d}v$
\beq
\mathrm{d}v = \det\bm{F}\mathrm{d}V
\eeq
To ensure orientation-preservation, 
$J :=\det\bm{F}>0$
\footnote{Rigorously, the Jacobian is defined as the limit:  
\begin{equation}
J(\bm{X}) = \lim_{r\to 0} \frac{\mathcal{L}^3(\bm{\varphi}(B(\bm{X},r)))}{\mathcal{L}^3(B(\bm{X},r))}
\end{equation}
where \( B(\bm{X},r) \) is an open ball centered at \( \bm{X} \) with radius \( r>0 \), and \( \mathcal{L}^3 \) denotes the three-dimensional Lebesgue measure, representing volume. This definition characterizes the local volumetric change under the deformation \( \bm{\varphi} \), with \( J(\bm{X}) \) measuring the ratio of the deformed volume to the original volume in an infinitesimally small neighborhood around \( \bm{X} \).}


Having established the mapping of infinitesimal vectors and volumes, it remains to state the relation between the oriented infinitesimal area elements {$\bm{n} \mathrm{d}a$} in the actual configuration and infinitesimal area elements $\bm{N}\mathrm{d}A$ in the reference configuration. Such a relation is described by Nanson's formula
\bea
\bm{n} \mathrm{d}a = \text{cof} \bm{F}\bm{N} \mathrm{d}A \label{eq25}
\eea
where $\bm{N}$ and $\bm{n}$ are the unit outward normals to an infinitesimal surface in the reference and deformed configuration and 
\bea
\text{cof} \bm{F} := J \bm{F}^{-T} 
\eea
denotes the cofactor tensor of $\bm{F}$. Alternatively, the cofactor can be expressed by
\bea \label{eq:cofTensorCrossProduct}
\text{cof} \bm{F} = \frac{1}{2} \bm{F} \bm{\times} \bm{F}\label{eq42}
\eea
where the tensor cross product  $\bm{\times}$ is defined using the Levi-Civita tensor $\bm{\epsilon}$ by 
\bea\label{eq:tensorCrossProduct}
 \left[ \bm{F} \bm{\times}\bm{F} \right]_{ij} = [\bm{\epsilon}]_{imn} [\bm{\epsilon}]_{jop}[\bm{F}]_{mo}[\bm{F}]_{np} \label{eq45}
\eea
as introduced in \cite{de1982vektor} and later reintroduced, e.g., in \cite{bonet2016tensor}.
Here, the indices denote tensor components with respect to some arbitrary basis in $\mathbb{R}^3$. The advantage of the latter definition of the cofactor lies in the fact that it can also be applied to rank-deficient tensors.  

\subsubsection{Surface deformation}\label{surfDef}
To define deformation measures relevant for the boundary surface, we first need to establish how tensor fields on the boundary are differentiated. Assume that a tensor field $\hat{\bm{Q}}: \Gamma \to W$ maps material points $\bm{X}$ on the boundary surface $\Gamma$ to a point in a finite-dimensional vector space $W$. For our needs $W=\mathbb{R}$, $W=\mathbb{R}^3$, or $W=\mathbb{R}^3 \times \mathbb{R}^3$, meaning $\hat{\bm{Q}}$ is a scalar, first order, or second order tensor field. The surface gradient $\hat{\nabla}$ of surface tensor field $\hat{\bm{Q}}$ at $\bm{X}$, i.e.,  $\hat{\nabla} \hat{\bm{Q}}(\bm{X}): \mathbb{R}^3\to W$  is defined by
\bea
\lim_{\substack{\bm{Y}\to \bm{X} \\ \bm{Y}\in \Gamma, \bm{Y} \neq \bm{X}}}\dfrac{ \left|\left| \hat{\bm{Q}}(\bm{Y}) -\hat{\bm{Q}}(\bm{X}) - \hat{\nabla} \hat{\bm{Q}}(\bm{X})\left( \bm{X} - \bm{Y} \right)\right|\right|}{\left|\left|\bm{X} - \bm{Y}\right|\right|} = 0
\eea
and is required to satisfy 
\bea
\hat{\nabla} \hat{\bm{Q}}(\bm{X}) \hat{\bm{I}} = \hat{\nabla} \hat{\bm{Q}}(\bm{X}) \label{eq26}
\eea
where $\hat{\bm{I}} = \bm{I} - \bm{N}\otimes \bm{N}$ is the orthogonal projection tensor onto the tangent space $T_{\bm{X}}\Gamma$ of the boundary surface $\Gamma$ at $\bm{X}$ with $\bm{I}$ being the identity tensor in $\mathbb{R}^3$. Note that the definition follows the convention from \cite{silhavy2013mechanics, silhavy2013direct}, which differs from the formulation by other authors, such as Gurtin and Murdoch in \cite{gurtin1975continuum}. Therefore, the surface deformation gradient is defined as a linear transformation from $\mathbb{R}^3$, and not only from the tangent space $T_{\bm{X}}\Gamma$, which possesses certain advantages as will become evident in the subsequent sections.

Let the mapping $\hat{\bm{Q}}(\bm{X})$ be a restriction of smooth mapping ${\bm{Q}}(\bm{X})$, with ${\bm{Q}}:\Omega\to W$, defined in the neighborhood of $\bm{X}\in \mathbb{R}^3$, the surface gradient is then expressed as projection of the full gradient on the surface tangent space  
\bea
\hat{\nabla} \hat{\bm{Q}} = {\nabla} {\bm{Q}} \hat{\bm{I}}
\eea

Applying the previous definitions to the surface deformation mapping $\hat{\bm{\varphi}}$ (setting $\hat{\bm{Q}}=\hat{\bm{\varphi}}$ and $W = \mathbb{R}^3$), we can define the surface deformation gradient $\hat{\bm{F}}:\mathbb{R}^3 \to \mathbb{R}^3$ as
\bea
\hat{\bm{F}}=\hat{\nabla} \hat{\bm{\varphi}}
\eea
which transforms surface projections of vectors from the reference configuration to the tangential vectors in the deformed configuration. Note that the surface deformation gradient is rank-deficient, but a pseudo-inverse $\hat{\bm{F}}^{-1} :\mathbb{R}^3 \to \mathbb{R}^3$ exists and fulfills
\bea
\hat{\bm{F}}^{-1} \hat{\bm{F}} = \hat{\bm{I}}, \ \ \ \hat{\bm{F}} \hat{\bm{F}}^{-1}  = \hat{\bm{i}} \label{eq39}
\eea
with $\hat{\bm{i}} = \bm{I}-\bm{n}\otimes \bm{n}$ being the projection tensor onto the deformed tangent space  $T_{\bm{x}}\bm{\varphi}(\Gamma)$. In addition, from \eqref{eq39}, we can easily derive the following identities
\bea
\hat{\bm{F}} \hat{\bm{I}} = \hat{\bm{F}}, \ \ \ \hat{\bm{i}}\hat{\bm{F}}  = \hat{\bm{F}}, \ \ \  \hat{\bm{F}}^{-1} \hat{\bm{i}} = \hat{\bm{F}}^{-1}, \ \ \  \hat{\bm{I}}\hat{\bm{F}}^{-1}  = \hat{\bm{F}}^{-1}\label{eq40}
\eea

If the mapping $\hat{\bm{\varphi}}$ can be extended by $\bm{\varphi}$ also in the neighborhood of the boundary surface, the surface deformation gradient and its pseudo-inverse can be expressed in terms of the bulk deformation gradient as
\bea
\hat{\bm{F}} = \bm{F}\hat{\bm{I}}, \ \ \ \hat{\bm{F}}^{-1} = \bm{F}^{-1}\hat{\bm{i}} \label{eq41}
\eea

As a consequence of condition \eqref{eq26}, the equation
\begin{equation}
\hat{\bm{F}} \bm{N} = \bm{0}
\end{equation}
holds, indicating that the surface deformation gradient maps the normal vector of the reference surface to the zero vector. 
As already discussed, in the case of bulk deformation, the Jacobian $J=\det \bm{F}$ is a fundamental quantity characterizing the change of infinitesimal volumes. Analogously, for a boundary surface, a quantity characterizing the change of infinitesimal surface area, i.e., surface Jacobian, denoted by  $\hat{J}$, is highly important in the modeling of the behavior of elastic surfaces. 
As shown in \ref{App:SurfaceJacobian}, alternative formulas for evaluation of $\hat{J}$ can be derived:
\bea
\hat{J} = \left|\left| \text{cof} \bm{F}\bm{N} \right|\right| = \left|\left| \text{cof} \hat{\bm{F}} \bm{N} \right|\right| = \left|\left| \text{cof} \hat{\bm{F}}  \right|\right| = \sqrt{\text{tr} \left(\text{cof}\hat{\bm{C}}\right)  } = \sqrt{\underbrace{\frac{1}{2}\text{tr}\left(\hat{\bm{C}}\right)^2 -\frac{1}{2}\text{tr}\left(\hat{\bm{C}}^T\hat{\bm{C}}\right)}_{I_2(\hat{\bm{C}})}}\label{eq43} 
\eea
 It is important to note that the surface Jacobian can be expressed as the square root of the surface right Cauchy-Green deformation tensor, i.e.,
\beq
\hat{J} = \sqrt{I_2(\hat{\bm{C}})}
\eeq
where $\hat{\bm{C}} = \hat{\bm{F}}^T\hat{\bm{F}}$.

\subsection{Governing equations}
In the present paper we consider both the bulk as well as the surface to be made out of hyperelastic materials, i.e.,\ materials for which the stored energy density functions exist. For the bulk material, the stored energy density is described by the constitutive relation
\bea
\Psi_b(\bm{X}) = \tilde\Psi_b\left(\bm{F}(\bm{X})\right), \ \ \  \bm{X} \in \Omega
\eea
where $\tilde\Psi_b:\text{Lin}(\mathbb{R}^3,\mathbb{R}^3)_+ \to \mathbb{R}$ is the bulk response function. $\text{Lin}(A,B)$ denotes a set of all linear transformations between vector spaces $A$ and $B$; in the particular case $\text{Lin}(\mathbb{R}^3,\mathbb{R}^3)$ is the set of all second-order tensors and the subscript $+$ means a restriction to second-order tensors with positive determinant.
On the part $S$ of the boundary surface, the surface stored energy density is governed by constitutive law  
\bea
\Psi_s(\bm{X}) = \tilde\Psi_s\left(\hat{\bm{F}}(\bm{X}),\bm{N}(\bm{X})\right), \ \ \  \bm{X} \in S
\eea
where $\tilde\Psi_s: G \to \mathbb{R}$ is the surface energy response function 
with 
\bea
G = \left\{   \left(\hat{\bm{F}}, \bm{N}\right) \in \text{Lin}(\mathbb{R}^3,\mathbb{R}^3)\times \mathbb{R}^3:   \hat{\bm{F}} \bm{N}=\bm{0}  \right\}\label{eq34}
\eea
Details related to the restrictions of the particular form of the response functions are discussed later in Section \ref{Sec:poly}. 

The total energy associated with the deformed body 
\bea \label{Eq:E_total}
E(\bm{\varphi}) = E_b(\bm{\varphi}) + 
E_s(\bm{\varphi}) + E_{ext}(\bm{\varphi}) \label{eq27}
\eea
is composed of the stored energy in the bulk material
\bea
E_b(\bm{\varphi}) = \int_{\Omega}  \Psi_b \mathrm{d}\mathcal{V}
\eea
stored energy in the material of the surface
\bea
E_s(\bm{\varphi})=E_s(\hat{\bm{\varphi}})=\int_{S}  \Psi_s \mathrm{d}\mathcal{A} \label{eq57}
\eea
and of the potential of external forces 
\bea
E_{ext}(\bm{\varphi}) = - \int_{\Omega}  \bm{\varphi}\cdot \bm{b} \mathrm{d}\mathcal{V}-\int_{ \Gamma}  \bm{\varphi}\cdot \bm{s}  \mathrm{d}\mathcal{A} 
\eea
where $\bm{b}:\Omega \to \mathbb{R}^3$ stands for the body forces and $\bm{s}:\Gamma \to \mathbb{R}^3$ denotes the applied surface tractions.

Relying on the principle of stationary potential energy, the solution of the problem is sought as the stationary point of the energy functional \eqref{eq27}. In the stationary point, the first variation of the energy functional
\bea
\delta E(\bm{\varphi},\delta\bm{\varphi}) = \dfrac{\mathrm{d}\left( E(\bm{\varphi} + t \delta\bm{\varphi})  \right)   }{\mathrm{d}t} \bigg \rvert_{t=0}, \ \ \ t \in \mathbb{R}
\eea
must vanish for arbitrary variation $\delta\bm{\varphi}: \Omega \to \mathbb{R}^3$ satisfying $\delta\bm{\varphi} = \bm{0}$ on the Dirichlet boundary $\Gamma_d$.

The variation $\delta E_b(\bm{\varphi},\delta\bm{\varphi})$ of the energy contribution from the bulk and the variation of the potential of external forces {$\delta E_{ext}(\bm{\varphi},\delta\bm{\varphi})$} are obtained by standard procedure and read
\bea
\delta E_b(\bm{\varphi},\delta\bm{\varphi}) &=&  \int_{\Omega}  {\bm{P}} : \left( \nabla \delta{\bm{\varphi}}\right) \mathrm{d}\mathcal{V}  =  - \int_{\Omega}  \delta\bm{\varphi}\cdot \text{Div}\bm{P} \mathrm{d}\mathcal{V}+\int_{\Gamma}  \delta\bm{\varphi}\cdot \bm{P}\bm{N} \mathrm{d}\mathcal{A} \label{eq31}\\ 
\delta E_{ext}(\bm{\varphi},\delta\bm{\varphi}) &=&  - \int_{\Omega}  \delta\bm{\varphi}\cdot \bm{b} \mathrm{d}\mathcal{V}-\int_{ \Gamma}  \delta\bm{\varphi}\cdot \bm{s}  \mathrm{d}\mathcal{A}   \label{eq32}
\eea
where $\bm{P}$ is the First Piola-Kirchhoff stress governed by the constitutive relation
\bea
\bm{P}(\bm{X}) = \tilde{\bm{P}}(\bm{F}(\bm{X})), \ \ \ \bm{X} \in \Omega
\eea
The response function for $\tilde{\bm{P}}$ is connected to the stored energy density response function by
\bea
\tilde{\bm{P}} = \mathrm{D} \tilde \Psi_b
\eea

where $\mathrm{D}$ denotes the derivative of a function with respect to its arguments; see \ref{Appendix:derivatives} for its definition. Specifically, $\mathrm{D} \tilde \Psi_b$ represents the derivative of $\tilde \Psi_b$ with respect to $\bm{F}$, while the derivative of $\tilde \Psi_s$ consists of two components: the partial derivatives with respect to $\hat{\bm{F}}$ and  $\bm{N}$, expressed as $\mathrm{D}  \tilde \Psi_s = ( \partial_{\hat{\bm{F}}} \tilde \Psi_s,\partial_{{\bm{N}}}  \tilde \Psi_s )$.

Evaluating the variation of the surface part of the energy we obtain
\bea
\delta E_s(\bm{\varphi},\delta\bm{\varphi}) = \delta E_s(\hat{\bm{\varphi}},\delta\hat{\bm{\varphi}}) = \int_{S}  \partial_{\hat{\bm{F}}} \tilde \Psi_s :\left( \hat\nabla \delta\hat{\bm{\varphi}}\right) \mathrm{d}\mathcal{A}
 = \int_{S}  \partial_{\hat{\bm{F}}} \tilde \Psi_s :\left( \hat\nabla \delta\hat{\bm{\varphi}}\hat{\bm{I}}\right)  \mathrm{d}\mathcal{A}
 = \int_{S}  \hat{\bm{P}} :\left( \hat\nabla \delta\hat{\bm{\varphi}}\right) \mathrm{d}\mathcal{A} \label{eq28}
\eea
where the surface first Piola-Kirchhoff stress $\hat{\bm{P}}$ is given by constitutive response function 
\bea
\hat{\bm{P}}(\bm{X}) = \tilde{\hat{\bm{P}}}\left(\hat{\bm{F}}(\bm{X})\right), \ \ \ \bm{X} \in S
\eea
which is expressed in terms of the surface stored energy function as
\bea
\tilde{\hat{\bm{P}}} = \partial_{\hat{\bm{F}}} \tilde \Psi_s \hat{\bm{I}} \label{eq51}
\eea

Since $\hat{\bm{P}}$ is superficial, i.e.,\ $\hat{\bm{P}} = \hat{\bm{P}}\hat{\bm{I}}$, the surface divergence theorem \eqref{eq29} can be applied and the variation \eqref{eq28} is transformed to
\bea
\delta E_s(\hat{\bm{\varphi}},\delta\hat{\bm{\varphi}}) =\int_{S}  \hat{\bm{P}} :\left( \hat\nabla \delta\hat{\bm{\varphi}}\right) \mathrm{d}\mathcal{A}= - \int_{S}  \delta\hat{\bm{\varphi}}\cdot \widehat{\text{Div}}\hat{\bm{P}} \mathrm{d}\mathcal{A}+\int_{\partial S}  \delta\hat{\bm{\varphi}}\cdot \hat{\bm{P}} \hat{\bm{N}} \mathrm{d}\mathcal{L} 
\eea
Collecting the individual contributions of the variation of the stored energy functional leads to
\bea
\delta E(\bm{\varphi},\delta\bm{\varphi}) = - \int_{\Omega}  \delta\bm{\varphi}\cdot \left( \text{Div}\bm{P} + \bm{b}\right)\mathrm{d}\mathcal{V}+\int_{\Gamma \setminus S}  \delta\bm{\varphi}\cdot\left( \bm{P}\bm{N}-\bm{s} \right) \mathrm{d}\mathcal{A}  
-\int_{S}  \delta\bm{\varphi}\cdot\left(\widehat{\text{Div}}\hat{\bm{P}}+\bm{s}  -\bm{P}\bm{N}\right) \mathrm{d}\mathcal{A}  
+
\int_{\partial S}  \delta\hat{\bm{\varphi}}\cdot \hat{\bm{P}} \hat{\bm{N}} \mathrm{d}\mathcal{L} \label{eq30}
\eea
Subsequently, requiring the energy functional to be stationary, i.e.,\ $\delta E(\bm{\varphi},\delta\bm{\varphi}) = 0$ for an arbitrary variation $\delta\bm{\varphi}$, the strong form of the equilibrium equations together with the boundary conditions are acquired
\bea
\text{Div}\bm{P} + \bm{b} = \bm{0} \ &&\text{in } \Omega \label{eq:equil1} \\
 \bm{P}\bm{N}=\bm{s} \ &&\text{on } \partial\Omega \setminus S \\ 
 \widehat{\text{Div}}\hat{\bm{P}}+\bm{s}  -\bm{P}\bm{N} = \bm{0} \ &&\text{on }  S \label{eq:equil3} \\
\hat{\bm{P}} \hat{\bm{N}} =  \bm{0}\ &&\text{on } \partial S  \label{eq:equil2}
\eea
To obtain the individual equations, the variations $\delta\bm{\varphi}$ are sequentially chosen such that all the integrals in \eqref{eq30} except one vanish and by arbitrariness of $\delta\bm{\varphi}$ the previous equations must hold. 

The weak form of equilibrium is obtained by collecting the individual variation terms before applying the divergence and surface divergence theorems. Particularly, adding the first part of Equation \eqref{eq31} to Equations \eqref{eq32}, and \eqref{eq28} and expressing $\delta E(\bm{\varphi},\delta\bm{\varphi}) = 0$ yields
\bea
\int_{\Omega}  {\bm{P}} : \left( \nabla \delta{\bm{\varphi}}\right) \mathrm{d}\mathcal{V}   +
 \int_{S}  \hat{\bm{P}} : \left( \hat\nabla \delta\hat{\bm{\varphi}}\right) \mathrm{d}\mathcal{A} =   \int_{\Omega}  \delta\bm{\varphi}\cdot \bm{b} \mathrm{d}\mathcal{V}+\int_{ \Gamma}  \delta\bm{\varphi}\cdot \bm{s}  \mathrm{d}\mathcal{A} \label{eq33}
\eea
which will be the starting point for the numerical formulation based on the finite element method introduced later in Section \ref{Sec:FEM}.

\subsection{Constitutive equations {for} surfaces} \label{sec:const}
In the previous section, the surface energy response function was considered of a general form $\tilde\Psi_s\left(\hat{\bm{F}},\bm{N}\right)$.
However, such a broad description can be further restricted by considering invariance with respect to two classes of transformation, which are the change of observer and the change of reference configuration. The first one is closely connected to the principle of material frame indifference, while the latter one is related to the notion of material symmetry. 

\subsubsection{Material frame indifference}
The principle of material frame indifference states that the energy response function must remain invariant under a change of observer.
Denoting the quantities associated with the changed deformed configuration by $(\cdot)^*$, we require
\bea
\tilde\Psi_s\left(\hat{\bm{F}},\bm{N}\right) = \tilde\Psi_s\left(\hat{\bm{F}}^*,\bm{N}^*\right)
\eea
to hold. If the change of observer transformation is described by a rotation $\bm{Q}\in $ {SO}(3), where {SO}(3) denotes the special orthogonal group, the relation between the original and modified quantities is
$\bm{N}^* = \bm{N}$ and $ \hat{\bm{F}}^* =\bm{Q} \hat{\bm{F}}$
and the previous equation is rephrased to
\bea \label{eq:Psi(F)=Psi(QF)}
\tilde\Psi_s\left(\hat{\bm{F}},\bm{N}\right) = \tilde\Psi_s\left(\bm{Q}\hat{\bm{F}},\bm{N}\right)
\eea
To further analyze the implications of frame indifference, we consider the polar decomposition of $\hat{\bm{F}}$.
Since $\hat{\bm{F}}$ is rank-deficient, its right polar decomposition, i.e., decomposition in product of a rotation tensor $\hat{\bm{R}}$ and a stretch tensor $\hat{\bm{U}}$ such as $\hat{\bm{F}} =\hat{\bm{R}}\hat{\bm{U}} $ can be computed by considering the singular value decomposition (SVD) as 
\beq\label{eq:SVD}
\hat{\bm{F}} = \tilde{\bm{U}} \bm{\Sigma} \tilde{\bm{V}}^T.
\eeq
The rotation tensor can be then recovered as $\hat{\bm{R}}=\tilde{\bm{U}}\tilde{\bm{V}}^T$ and the stretch tensor is given by $\hat{\bm{U}}=\tilde{\bm{V}}\bm{\Sigma}\tilde{\bm{V}}^T$, see \cite{poya2023variational}.

Having established the polar decomposition, the following arguments closely follow the classical arguments used for bulk.  First, we can rewrite  \eqref{eq:Psi(F)=Psi(QF)} using the polar decomposition as
\bea \label{eq:Psi(F)=Psi(QRU)}
\tilde\Psi_s\left(\hat{\bm{F}},\bm{N}\right) = \tilde\Psi_s\left(\bm{Q}\hat{\bm{R}}\hat{\bm{U}},\bm{N}\right)
\eea
Since the above relation holds for all $\bm{Q}\in \mathrm{SO}(3)$, it must also hold for the special choice $\bm{Q} = \hat{\bm{R}}^T$.  Substituting for this relation for $\bm{Q}$ back to \eqref{eq:Psi(F)=Psi(QRU)}, we get
\bea \label{eq:Psi(F)=Psi(U)}
\tilde\Psi_s\left(\hat{\bm{F}},\bm{N}\right) = \tilde\Psi_s\left(\hat{\bm{U}},\bm{N}\right)
\eea
Thus, to satisfy the principle of material frame indifference, the response function depends on $\hat{\bm{F}}$ only through $\hat{\bm{U}}$. Note that the right stretch tensor can also be computed as $\hat{\bm{U}} = \sqrt{\hat{\bm{F}}^T\hat{\bm{F}}}$.
\subsubsection{Change of reference configuration}
To proceed further, we assume the surface material possesses certain symmetry, meaning the response of the material remains invariant with respect to homogeneous change of the reference configuration $\bm{\Lambda}: \Omega \to \Omega^{\#}$ defined by
\bea
\bm{X}^{\#} = \bm{\Lambda}(\bm{X}) =\bm{H}^{-1}\bm{X} + \bm{X}^\#_0
\eea
where $\bm{X}^{\#}$ denotes the material point in the changed reference configuration, $\bm{H} \in \text{Lin}(\mathbb{R}^3,\mathbb{R}^3)_+$ and $\bm{X}^\#_0$ is a constant vector. The deformation mapping $\bm{\varphi}^{\#}:\Omega^{\#} \to \Omega^{\bm{\varphi}}$ is connected to the original deformation by
\bea
\bm{\varphi}^{\#}(\bm{X}^{\#}) = \bm{\varphi}\left(\bm{H}\left(\bm{X}^{\#}-\bm{X}^\#_0\right)\right)
\eea
The geometrical and kinematic quantities in the changed reference configuration are connected to the original ones by
\begin{gather}
\bm{N}^\# = \frac{\bm{H}^T\bm{N}}{|| \bm{H}^T\bm{N}||}, \ \ \ \ \  \  \ \ \ \ \Hat{\bm{I}}^\# = \bm{I} - \frac{(\bm{H}^T\bm{N}) \otimes (\bm{H}^T\bm{N}) }{|| \bm{H}^T\bm{N}||^2}\\
{\bm{F}}^\# ={\bm{F}} \bm{H}, \  \ \ \ \ \  \ \ \ \
\Hat{\bm{F}}^\# = {\bm{F}}^\# \Hat{\bm{I}}^\# = {\bm{F}} \bm{H} \Hat{\bm{I}}^\#={\bm{F}} \Hat{\bm{I}}\bm{H} \Hat{\bm{I}}^\# =\Hat{\bm{F}} \bm{H} \Hat{\bm{I}}^\#, \  \ \ \ \ \  \ \ \ \
\Hat{\bm{C}}^\# =  \Hat{\bm{I}}^\# {\bm{H}}^T \Hat{\bm{C}}{\bm{H}} \Hat{\bm{I}}^\# \label{eq56}
\end{gather}
where in $\eqref{eq56}_2$ we used that $\bm{N} \cdot \bm{H} \Hat{\bm{I}}^\# = \bm{0}$ and therefore $\bm{H} \Hat{\bm{I}}^\# = \Hat{\bm{I}}\bm{H} \Hat{\bm{I}}^\#$.

Using the change of variables formula, the energy contained in some surface region $\mathcal{R}$ can be expressed in terms of energy response function $\tilde\Psi_s^\#\left(\hat{\bm{F}}^\#,\bm{N}^\#\right)$ by
\bea
\int_{\mathcal{R}}  \tilde\Psi_s\left(\hat{\bm{F}}(\bm{X}),\bm{N}(\bm{X})\right) \mathrm{d}\mathcal{A}= \int_{\mathcal{R}^\#}  \tilde\Psi_s^\#\left(\hat{\bm{F}}^\#(\bm{X}^\#),\bm{N}^\#(\bm{X}^\#)\right) \mathrm{d}\mathcal{A}^\# = \int_{\mathcal{R}}  \tilde\Psi_s^\#\left(\Hat{\bm{F}}(\bm{X}) \bm{H} \Hat{\bm{I}}^\#,\frac{\bm{H}^T\bm{N}(\bm{X})}{|| \bm{H}^T\bm{N}(\bm{X})||}\right) ||\text{cof}~\bm{H}^{-1} \bm{N} ||\mathrm{d}\mathcal{A} 
\eea
where $\mathcal{R}^\# = \bm{\Lambda}(\mathcal{R})$. Since the integration region is arbitrary, we obtain
\bea
\tilde\Psi_s\left(\hat{\bm{F}},\bm{N}\right) =  \tilde\Psi_s^\#\left(\Hat{\bm{F}} \bm{H} \Hat{\bm{I}}^\#,\frac{\bm{H}^T\bm{N}}{|| \bm{H}^T\bm{N}||}\right) ||\text{cof}~\bm{H}^{-1} \bm{N} || \label{eq58}
\eea
that must always hold. However, if the material obeys certain symmetry characterized by a symmetry group $\mathcal{G}$, the response function defined in the configuration $\Omega^\#$ can be replaced by the original response function in $\Omega$. Consequently, we define the symmetry group $\mathcal{G}$ as the set of all $\bm{H}$ for which the following condition holds:
\bea  \label{eq:ChangeOfReference}
\tilde\Psi_s\left(\hat{\bm{F}},\bm{N}\right) =  \tilde\Psi_s\left(\Hat{\bm{F}} \bm{H} \Hat{\bm{I}}^\#,\frac{\bm{H}^T\bm{N}}{|| \bm{H}^T\bm{N}||}\right) ||\text{cof}~\bm{H}^{-1} \bm{N} || \label{eq59}
\eea

In the following, we investigate three particular cases of the symmetry group $\mathcal{G}$, firstly the so-called Special Linear Group  $\mathrm{SL}(3)$ describing fluid-like materials, secondly the group associated with isotropic solids, and finally, the symmetry group related to an anisotropic material with one family of fibers.
\subsubsection{Fluid-like material}
In the case of a fluid-like constitutive law for the surface, the group is $\mathrm{SL}(3) = \{\bm{H} \in{\text{Lin}(\mathbb{R}^3,\mathbb{R}^3)}, \mathrm{det}\;\bm{H} = 1\}$
The material symmetry condition for the fluid-like surface can be written as
\bea \label{eq:ChangeOfReferenceFluid}
\tilde\Psi_s\left(\hat{\bm{U}},\bm{N}\right) =  \tilde\Psi_s\left(\Hat{\bm{U}} \bm{H} \Hat{\bm{I}}^\#,\frac{\bm{H}^T\bm{N}}{|| \bm{H}^T\bm{N}||}\right) ||\mathrm{cof}~\bm{H}^{-1} \bm{N} || \label{eq59b}
\eea
where we combined reduced constitutive relation \eqref{eq:Psi(F)=Psi(U)} obtained from the material frame indifference with the general formula for the change of reference configuration \eqref{eq:ChangeOfReference}.
Since the equation above has to be valid for all $\bm{H} \in \mathrm{SL}(3)$, it must hold also for the following particular choice of $\tilde{\bm{H}}$
\beq
\tilde{\bm{H}} = \hat{\bm{U}}^{-1} + \hat{J}\bm{N}\otimes\bm{N}
\eeq

To show that the determinant of this particular choice is equal to one, we make use of an identity for the determinant of the sum of two second-order tensors, see \cite{podio1991surface}:
\bea
\mathrm{det}\tilde{\bm{H}} &=& \mathrm{det}(\hat{\bm{U}}^{-1} + \hat{J}\bm{N}\otimes\bm{N}) =  \mathrm{det}(\hat{\bm{U}}^{-1}) + \mathrm{cof}(\hat{\bm{U}}^{-1}):(\hat{J}\bm{N}\otimes\bm{N}) + \hat{\bm{U}}^{-1}:\mathrm{cof}(\hat{J}\bm{N}\otimes\bm{N}) + \mathrm{det}(\hat{J}\bm{N}\otimes\bm{N}) 
\\
&=&\mathrm{cof}(\hat{\bm{U}}^{-1}):(\hat{J}\bm{N}\otimes\bm{N}) = \hat{J}\underbrace{\bm{N}\cdot\mathrm{cof}(\hat{\bm{U}}^{-1})\bm{N}}_{1/\hat{J}} = \hat{J}\frac{1}{\hat{J}} = 1
\eea
Here, the first and last terms vanish because they are determinants of rank-deficient tensors, while the third term is also zero due to the cofactor of a rank-one tensor being zero.

Before substituting our specific choice of $\bm{H}$ into \eqref{eq:ChangeOfReferenceFluid}, we first state the following useful identities:
\bea
\frac{\tilde{\bm{H}}^T\bm{N}}{||\tilde{\bm{H}}^T\bm{N}||} &=& \frac{\left(\hat{\bm{U}}^{-1} + \hat{J}\bm{N}\otimes\bm{N}\right)^T\bm{N}}{||\left(\hat{\bm{U}}^{-1} + \hat{J}\bm{N}\otimes\bm{N}\right)^T\bm{N}||} = \frac{\hat{J}\bm{N}}{||\hat{J}\bm{N}||} = \bm{N}
\\
\Hat{\bm{I}}^\# &=& \bm{I} - \frac{(\tilde{\bm{H}}^T\bm{N}) \otimes (\tilde{\bm{H}}^T\bm{N}) }{|| \tilde{\bm{H}}^T\bm{N}||^2} = \Hat{\bm{I}}
\\
\Hat{\bm{U}} \tilde{\bm{H}} \Hat{\bm{I}}^\# &=& \Hat{\bm{U}} \tilde{\bm{H}} \Hat{\bm{I}} = \Hat{\bm{U}} \hat{\tilde{\bm{H}}} = \Hat{\bm{U}} \hat{\bm{U}}^{-1} = \Hat{\bm{I}}
\\
\tilde{\bm{H}}^{-1} &=& \hat{\bm{U}} + \frac{1}{\hat{J}}\bm{N}\otimes\bm{N}
\\
||\mathrm{cof}~\tilde{\bm{H}}^{-1} \bm{N} || &=&   ||\mathrm{cof}(\hat{\bm{U}} + \frac{1}{\hat{J}}\bm{N}\otimes\bm{N}) \bm{N} || = ||\mathrm{cof}\hat{\bm{U}} \bm{N} || = \hat{J}
\eea
The first identity shows that the transformed normal vector $\tilde{\bm{H}}^T \bm{N}$, when normalized, remains equal to $\bm{N}$. This follows from the specific structure of $\tilde{\bm{H}} = \hat{\bm{U}}^{-1} + \hat{J} \bm{N} \otimes \bm{N}$ and the fact that $\hat{\bm{U}}^{-1} \bm{N} = 0$, which implies $\tilde{\bm{H}}^T \bm{N} = \hat{J} \bm{N}$. Normalizing gives $\bm{N}$, since $\hat{J} > 0$. The second identity confirms that the projection operator $\Hat{\bm{I}}^\#$, constructed from $\tilde{\bm{H}}^T \bm{N}$, reduces to the standard projection $\Hat{\bm{I}} = \bm{I} - \bm{N} \otimes \bm{N}$. This is because the outer product term simplifies due to $\tilde{\bm{H}}^T \bm{N} = \hat{J} \bm{N}$. The third identity demonstrates that applying the composition of $\Hat{\bm{U}}$ with $\tilde{\bm{H}}$ results in the identity on the tangent plane. Specifically, we use that $\hat{\tilde{\bm{H}}} = \hat{\bm{U}}^{-1}$, so that $\Hat{\bm{U}} \hat{\tilde{\bm{H}}} = \Hat{\bm{U}} \hat{\bm{U}}^{-1} = \Hat{\bm{I}}$. The fourth identity gives the inverse of $\tilde{\bm{H}}$, showing that it takes the form $\hat{\bm{U}} + \frac{1}{\hat{J}} \bm{N} \otimes \bm{N}$. This is consistent with standard results for rank-one updates of singular matrices where the added term restores the full rank. The fifth identity states that the cofactor of $\tilde{\bm{H}}^{-1}$ applied to $\bm{N}$ has norm equal to $\hat{J}$. Using the fact that the cofactor of a surface tensor is invariant under the rank-one augmentation in the normal direction and that $\bm{N}$ lies in the null space of $\hat{\bm{U}}$, the expression simplifies to $\|\mathrm{cof} \hat{\bm{U}} \bm{N}\| = \hat{J}$.

Finally, substituting $\tilde{\bm{H}}$ into \eqref{eq:ChangeOfReferenceFluid} and using the expressions above we get
\bea \label{eq:ChangeOfReferenceFluidFinal}
\tilde\Psi_s\left(\hat{\bm{U}},\bm{N}\right) =  \tilde\Psi_s\left(\Hat{\bm{U}} \tilde{\bm{H}} \Hat{\bm{I}}^\#,\frac{\tilde{\bm{H}}^T\bm{N}}{|| \tilde{\bm{H}}^T\bm{N}||}\right) ||\mathrm{cof}~\tilde{\bm{H}}^{-1} \bm{N} ||  = \tilde\Psi_s\left(\Hat{\bm{I}},\bm{N}\right)\Hat{J} = \gamma \Hat{J}
\eea
Therefore, the fluid-like material corresponds to the well-known model, usually referred to as surface tension.
\subsubsection{Isotropic material}
In the isotropic case, the group of symmetry is given by $\mathcal{G}=$ SO(3) and thus $\bm{H} = \bm{Q}$. In that case, Equation \eqref{eq59} simplifies to
\bea
\tilde\Psi_s\left(\hat{\bm{F}},\bm{N}\right) =  {\tilde{\Psi}}_s\left(\Hat{\bm{F}} \bm{Q} ,{\bm{Q}^T\bm{N}}\right) 
\eea
since $\Hat{\bm{I}} \bm{Q} \Hat{\bm{I}}^\# = \Hat{\bm{I}} \bm{Q}$. The equality in the second argument—the unit normal vector—holds only if the function depends solely on its norm. As the norm is constant, the function must be independent of the vector $\bm{N}$. Substituting the left polar decomposition, i.e., $\hat{\bm{F}}= \hat{\bm{V}}\hat{\bm{R}}$, where $\hat{\bm{R}}=\tilde{\bm{U}}\tilde{\bm{V}}^T$ and  $\hat{\bm{V}} = \tilde{\bm{U}}\bm{\Sigma}\tilde{\bm{U}}^T$, and choosing $\bm{Q} = \bm{R}^T$, we get
\bea
\tilde\Psi_s\left(\hat{\bm{F}},\bm{N}\right) =  \tilde{\tilde{\Psi}}_s\left(\Hat{\bm{V}}\right) 
\eea
Therefore, for the isotropic material, the response function depends on the surface deformation gradient only through the left stretch tensor $\hat{\bm{V}}$. 

Alternatively, we can construct a response function that depends on the surface deformation gradient only through the surface right Cauchy-Green tensor $\Hat{\bm{C}} = \Hat{\bm{U}}^2$. Consequently, by  combining material frame indifference with isotropy we obtain
\bea
\tilde\Psi_s\left(\hat{\bm{F}},\bm{N}\right) = \bar{\bar{\Psi}}_s\left(\hat{\bm{C}}\right) =  \bar{\bar{\Psi}}_s\left(\bm{Q}^T\Hat{\bm{C}} \bm{Q} \right) , \ \ \ \ \ \forall \bm{Q} \in \text{SO(3)}
\eea

Since the function $\bar{\bar{\Psi}}_s$ is isotropic, based on the representation theorem it can be represented as function of principal invariants of the argument $\hat{\bm{C}}$ such that
\bea
\bar{\bar{\Psi}}_s\left(\hat{\bm{C}}\right) = \hat{\hat{\Psi}}_s\left( I_1(\hat{\bm{C}}),I_2(\hat{\bm{C}}),I_3(\hat{\bm{C}}) \right)
\eea
where 
\bea
I_1(\hat{\bm{C}}) = \text{tr}\hat{\bm{C}} = || \hat{\bm{F}} ||^2, \ \ \ I_2(\hat{\bm{C}}) =\frac{1}{2}\left( \left(\text{tr}\hat{\bm{C}}\right)^2-\text{tr}\left(\hat{\bm{C}}^2\right)\right) = || \text{cof}\hat{\bm{F}} ||^2, \ \ \  I_3(\hat{\bm{C}}) = \text{det}\hat{\bm{C}} = 0 \label{eq48}
\eea
where in ${\eqref{eq48}}_2$ we employed the equality \eqref{eq43}.
Since $I_3$ is always zero, the surface energy response function simplifies to 
\bea
\bar{\bar{\Psi}}_s\left(\hat{\bm{C}}\right) = \tilde{\tilde{\Psi}}_s\left( || \hat{\bm{F}} ||^2,|| \text{cof}\hat{\bm{F}} ||^2\right) \label{eq49}
\eea
To summarize, in the case of isotropic material, the energy response function is governed by a scalar function of two scalar arguments which are the first two principal invariants of the tensor $\hat{\bm{C}}$.

\subsubsection{Anisotropic material} \label{aniso}
Let us now consider a basic anisotropic model for surfaces. We assume that the material surface is made of fibres whose direction is described by a continuous sufficiently well behaved tangential vector field $\bm{a}:S \to \mathbb{R}^3$ which satisfies 
\bea
||\bm{a}(\bm{X})|| = 1, \ \ \ \ \ \ \ \ \ \bm{a}(\bm{X}) \cdot \bm{N}(\bm{X}) = 0 \ \ \ \ \ \ \ \ \ \forall \bm{X} \in S
\eea

The symmetry group $\mathcal{G}^A(\bm{X}) \subset $ SO(3), which generally varies for different material points $\bm{X}$, is then defined by the set of all orthogonal matrices which leave the vector $\bm{a}(\bm{X})$ unchanged (up to a sign), meaning that  
\bea
\mathcal{G}^A(\bm{X}) = \left\{ \bm{Q}'\in \text{SO(3)} : \bm{Q}' \bm{a}(\bm{X}) =  \pm \bm{a}(\bm{X})   \right\} = \left\{ \bm{Q}'\in \text{SO(3)} : {\bm{Q}'}^T \bm{S}(\bm{X}) {\bm{Q}'} =  \bm{S}(\bm{X})   \right\} \label{eq61}
\eea
where $\bm{S} = \bm{a}\otimes \bm{a}$ is the so-called structural tensor field. 

Consequently, Equation \eqref{eq59} is reformulated to 
\bea
\tilde\Psi_s\left(\hat{\bm{F}},\bm{N}\right) =  {\tilde{\Psi}}_s\left(\Hat{\bm{F}}\Hat{\bm{I}} \bm{Q}' ,{\bm{Q}'^T\bm{N}}\right) 
\eea
which must hold for arbitrary $\bm{Q}' \in \mathcal{G}^A$. To stress that application of $\hat{\bm{F}}$ contains projection on the tangent plane, we explicitly wrote $\hat{\bm{F}} \hat{\bm{I}}$ instead of $\hat{\bm{F}}$ in the first argument of the second term. 
The meaning of the equality in the first argument can be explained as follows. The energy stored in the surface material point deformed by $\hat{\bm{F}}$ must be the same as in the point which is firstly rotated around tangential vector $\bm{a}$ by $\bm{Q}'$, then projected back to the original surface and then deformed by $\hat{\bm{F}}$. This equality is fulfilled if the direction of the fibres is aligned with vector $\bm{a}$. 

Similarly, as in the isotropic case, the equality in the second argument is only satisfied if the response function is independent of it, leading to
\bea
\tilde\Psi_s\left(\hat{\bm{F}},\bm{N}\right) = \bar{\Psi}_s\left(\hat{\bm{C}}\right)
\eea

The scalar function $\bar{\Psi}_s(\hat{\bm{C}})$ of symmetric second-order tensorial argument is invariant with respect symmetry group $\mathcal{G}^A$ defined in \eqref{eq61} using the structural tensor ${\bm{S}}$, only if there exists isotropic function $\tilde{\bar{\Psi}}_s(\hat{\bm{C}},{\bm{S}})$, i.e.,
\bea
\tilde{\bar{\Psi}}_s(\hat{\bm{C}},{\bm{S}}) = \tilde{\bar{\Psi}}_s(\bm{Q}\hat{\bm{C}}\bm{Q}^T,\bm{Q}{\bm{S}}\bm{Q}^T)
\eea

Subsequently, following \cite{spencer1971theory}, the function $\tilde{\bar{\Psi}}_s$ can be expressed as function $\tilde{\tilde{\bar{\Psi}}}_s(I_1,I_2,I_4,I_5)$, where $I_i, \ i = \{1,2\}$ are the nonzero principal invariants of $\hat{\bm{C}}$ and 
\bea
I_4 &=& \text{tr}(\hat{\bm{C}} {\bm{S}}) = \bm{a}\cdot \hat{\bm{C}} \bm{a} = || \hat{\bm{F}} \bm{a} ||^2 \label{eq:i4}\\
I_5 &=& \text{tr}(\hat{\bm{C}}^2 {\bm{S}}) = \bm{a}\cdot \hat{\bm{C}}^2 \bm{a} = || \hat{\bm{F}}^T\hat{\bm{F}} \bm{a} ||^2 \label{eq:i5}
\eea
are the generalized invariants.

To summarize, in the case the surface is assumed to be anisotropic, where $\bm{a}$ defines locally the direction of fibres, the surface energy response function is given by
\bea \label{eq62}
\tilde\Psi_s\left(\hat{\bm{F}},\bm{N}\right) = \tilde{\tilde{\bar{\Psi}}}_s\left(|| \hat{\bm{F}} ||^2,|| \text{cof}\hat{\bm{F}} ||^2,|| \hat{\bm{F}} \bm{a} ||^2,|| \hat{\bm{F}}^T\hat{\bm{F}} \bm{a} ||^2 \right)
\eea


\section{Surface-polyconvex  models and existence of minimizers}
\label{Sec:poly}

To guarantee the existence of solutions, we require that the energy functional \eqref{Eq:E_total} is lower semicontinuous in a suitable topology. From the applied analysis perspective, a key question is which stored surface and bulk strain energies ensure that the functional $E$ in \eqref{Eq:E_total} admits minimizers. Thus,  we focus on the polyconvexity condition for both bulk and surface strain energies and highlight its key implication, i.e., the existence of minimizers. To support the subsequent analysis, we reintroduce certain fundamental quantities, such as the deformation $\bm{\varphi}$, within a more mathematically rigorous framework.

\subsection{Existence results}

The spaces $W^{1,p}$, $1 \leq p < \infty$, denote the standard Sobolev space of $L^p$-functions with
weak derivative in $L^p$. Furthermore, $BV$ stands for the space of integrable maps with bounded variations, see, e.g.,~\cite{AmbrosioFuscoPallara-Book,EvansGariepy-Book} for references. For a (measurable) set $\mathcal{E} \subset \R^3$, we
denote its three-dimensional Lebesgue measure by $\LL^3(\mathcal{E})$ and its two-dimensional Hausdorff measure by $\HH^2(\mathcal{E})$. The
space of vector-valued Radon measures on $\Ome$ with values in $Y$ is denoted by $\MM(\Ome,Y)$.

\medskip

To describe the state of the elastic material, we also need to introduce the deformation function $\bm{\varphi} \in
W^{1,p}(\Ome,\R^3)$, $p > 3$, which describes the deformation of the elastic body with respect to the reference
configuration $\Ome$. Assume that $\bm{\varphi} _0\in W^{1,p}(\Ome,\R^3)$ is given and that $\Gamma_D\subset\partial\Ome$ is nonempty and measurable.  We  consider deformations $\bm{\varphi}  \in \YY$, where
\begin{align}  %
  \YY = \Big\{ \bm{y} \in W^{1,p}(\Ome,\R^3) \ :\ \bm{y}=\bm{\varphi} _0 \text{ on } \Gamma_D, \ \det {\nabla} \bm{y} >0 \text{ a.e.}\ , \int_\Ome \det{\nabla} \bm{y}({\bm{X}})\,\mathrm{d} {\mathcal{V} }\leq \mathcal{L}^3(\bm{y}(\Omega)) \Big\} 
\end{align}
where we will always use the assumption $p > 3$. The integral inequality together with the orientation-preservation is
the so-called Ciarlet-Ne\v{c}as condition which ensures invertibility of 
{$\bm{\varphi}$} almost everywhere in $\Ome$
\cite{Ciarlet-Book,CiarletNecas-1987}.

We assume that the specimen in its reference configuration is represented by a bounded
Lipschitz domain $\Ome\subset\R^3$.

Le us recall that the total bulk energy of the specimen has the form
\begin{align} \label{stat-bulk} %
  E_{ b}(\bm\varphi):=\int_\Ome \tilde{\Psi}_{b}({\nabla} \bm{\varphi})\,\mathrm{d}\mathcal{V}
\end{align}
where we consider the specific bulk energy response function $\tilde{\Psi}_b : \text{Lin}(\mathbb{R}^3,\mathbb{R}^3)\to \R\cup\{+\infty\}$ of the specimen.
 Thus, we assume the framework of hyperelasticity, where the first Piola-Kirchhoff stress tensor has a polyconvex potential
\begin{align} \label{W-ass1} %
  \tilde{\Psi}_b(\bm{F}):=
  \begin{cases}
    h(\bm{F},\cof \bm{F},\det \bm{F}) & \mbox{ if } \det \bm{F}>0 \\
    +\infty \mbox{ otherwise}
  \end{cases}
\end{align}
for some convex function $h:\R^{19} \to \R$. We use the following additional standard assumptions on the specific bulk
energy $\tilde{\Psi}_b$. First, we assume the growth condition of the bulk energy, i.e., for some constants $C>0$ and $p>3$
\begin{align}
  &\tilde{\Psi}_b(\bm{F})\ge C(-1+|\bm{F}|^p) \quad\forall \bm{F} \in \text{Lin}(\mathbb{R}^3,\mathbb{R}^3)\  \label{W-ass2} 
\end{align}
Moreover, assume that
\begin{align}
&\lim_{\det \bm{F}\to 0_+} \tilde{\Psi}_b(\bm{F})=+\infty\  \label{W-ass4}
\end{align}
ensuring that the material cannot collapse into zero volume without requiring infinite energy. It prevents unphysical deformations where the body would locally fold or compress into an infinitely small volume, thereby enforcing a form of material integrity.
Finally, we assume the material frame indifference of the bulk energy:
\begin{align}
  &\tilde{\Psi}_b(\bm{R}\bm{F})= \tilde{\Psi}_b(\bm{F}) \quad \forall \bm{R}\in{\rm SO}(3), \bm{F} \in \text{Lin}(\mathbb{R}^3,\mathbb{R}^3)\  \label{W-ass3}
\end{align}

Let us now turn our attention to the surface energy response function  $\tilde\Psi_s: G \to \mathbb{R}$ with 
\bea
G = \left\{   \left(\hat{\bm{F}}, \bm{N}\right) \in \text{Lin}(\mathbb{R}^3,\mathbb{R}^3)\times \mathbb{R}^3:   \hat{\bm{F}} \bm{N}=\bm{0}  \right\} \label{eq34b}
\eea
\v{S}ilhav\'{y} defined the notion of interface polyconvex functions in \cite{silhavy2011equilibrium,Silhavy-2010}.   A function is  interface polyconvex if there exists a \textbf{positively one-homogeneous convex}  function $\Phi$ such that
\bea
\tilde\Psi_s(\hat{\bm{F}},\bm{N}) = \Phi(\bm N, \hat{\bm{F}}\times \bm{N}, \text{cof}\hat{\bm{F}} \bm{N}), \ \ \ \forall (\hat{\bm{F}},\bm{N}) \in G \label{eq50}
\eea
He applied this property to phase transitions in multiphase materials where $\bm{N}$ denotes the unit normal to an unknown interface between phases. Note that we call this condition surface polyconvexity. Moreover, in our setting, $\bm{N}$ is the outer unit normal to $\Ome$, i.e., it is not a variable in our problem. 
Hence, we can restrict to 
\bea
\hat\Psi_s(\hat{\bm{F}}) = \hat\Phi( \hat{\bm{F}}\times \bm{N}, \text{cof}\hat{\bm{F}} \bm{N}), \ \ \ \forall (\hat{\bm{F}},\bm{N}) \in G \label{eq50a}
\eea
Furthermore, there exists a measure $\bm{M}:=(\bm{K},\bm{L})\in \MM(\mathbb{R}^3;\text{Lin}(\mathbb{R}^3,\mathbb{R}^3))\times\MM(\mathbb{R}^3;\mathbb{R}^3)$ with
  \begin{align}\label{measures1}
    \bm{K}:= (\hat{{\nabla}}  \bm{\varphi}\times \bm{N})\HH^2|_{\partial\Ome}  \quad \text{and} \quad 
    \bm{L} := (\cof\hat{{\nabla}}  \bm{\varphi} \bm{N})\HH^2\vert_{\partial\Ome}\ 
  \end{align}

  The surface energy can then be defined using the measure $\bm{M}$ as
  \begin{align} \label{stat-interface} E_s(\bm{\varphi}):=
    \begin{TC}
      \displaystyle \int_\Ome \hat\Phi\left(\frac{\mathrm{d} \bm{M}}{\mathrm{d} |\bm{M}|}\right)\ \mathrm{d}|\bm{M}|\ %
      &\text{for $\bm{\varphi} \in \YY$} \\ \\
      \infty &\text{else}
    \end{TC}
  \end{align}
  were $|\bm{M}|$ denotes the total variation of measure $\bm{M}$. The following result is a subtle generalization of{ \cite[Thm.~6.5]{silhavy2011equilibrium}}. 

  \begin{theorem}\label{Thm:existence}
  Assume that (\ref{stat-bulk}--\ref{stat-interface}) hold. Let  $\Ome\subset\R^3$ be a bounded Lipschitz domain, $p>3$, and $\hat{\Phi}(\bm{K},\bm{L})\ge~c|(\bm{K}, \bm{L}))|$ for every $\bm{K}\in \mathrm{Lin}(\mathbb{R}^3,\mathbb{R}^3)$ and $\bm{L}\in \mathbb{R}^3$ and some $c>0$ independent of $\bm{K}$ and $\bm{L}$.  If $\YY\ne\emptyset$ and $\inf_{\YY}E_b+E_s<+\infty$ then there is a minimizer of $E_b+E_s$ on $\YY$.
      
  \end{theorem}

An important assumption in Theorem~\ref{Thm:existence} is that both the bulk and surface energy densities are polyconvex. It is well known that polyconvexity—and its surface counterpart, surface polyconvexity—are sufficient conditions for ensuring lower semicontinuity of the energy functional with respect to the weak topology. A more general concept, known as quasiconvexity \cite{Morrey1952} is typically used in the bulk; however, this condition is not compatible with the standard growth conditions in elasticity, see \cite{ball2002some}. The concept of surface quasiconvexity has been thoroughly discussed in \cite{silhavy2011equilibrium}; see also \cite{Fonseca-1989, Parry-1987}. Nonetheless, it remains unclear whether quasiconvexity in both the bulk and on the surface is both necessary and sufficient for lower semicontinuity of the total energy functional.

Note that the rank-one convexity of bulk energy is frequently used in the engineering community due to its clear physical interpretation, particularly in the context of wave propagation. However, for the bulk energy, it represents only a necessary—not sufficient—condition for the lower semicontinuity of the energy functional. Moreover, an analogous notion of rank-one convexity for surface energies has not yet been established. Therefore, the polyconvexity for both the bulk and surface energy is crucial to show the existence of minimizers. It is worth noting that a deeper investigation of the mutual interaction between bulk and surface energies may lead to a relaxation of the currently imposed surface-polyconvexity condition for the existence of minimizers.
\subsection{Polyconvex Invariants}
{
In Section \ref{sec:const}, we applied the principle of material frame indifference and concepts of material symmetry to reduce the possible forms of the surface energy only to functions of the principle and generalized invariants. Moreover, in the previous section we discussed the condition of surface polyconvexity, due to which the surface energy response function must be given by positively one-homogeneous convex function $\hat\Phi( \hat{\bm{F}}\times \bm{N}, \text{cof}\hat{\bm{F}} \bm{N})$, see Equation \eqref{eq50a}.  
Next, we combine all of these restrictions to formulate surface polyconvex free energy respecting material symmetry as well as material frame indifference.}
\subsubsection{Fluid-like surface-polyconvex model}
For a fluid-like surface constitutive law, we showed that the energy response function is governed by surface tension law \eqref{eq:ChangeOfReferenceFluidFinal}, i.e.,\ 
\bea
\tilde{{\Psi}}_s\left(\hat{\bm{F}},\bm{N}\right) = \gamma \hat{J} = \gamma || \text{cof}\hat{\bm{F}} \bm{N} ||
\eea
The Frobenius norm is a positively one-homogeneous and convex function; therefore, such a form of energy is surface-polyconvex.  
\subsubsection{Isotropic surface-polyconvex model}
In the isotropic case, we justified that the surface energy must be of form \eqref{eq49}, meaning 
\bea
\tilde{{\Psi}}_s\left(\hat{\bm{F}},\bm{N}\right) = \tilde{\tilde{\Psi}}_s\left( || \hat{\bm{F}} ||^2,|| \text{cof}\hat{\bm{F}} ||^2\right) = \tilde{\tilde{\Psi}}_s\left( || \hat{\bm{F}} \times \bm{N} ||^2,|| \text{cof}\hat{\bm{F}} \bm{N} ||^2\right) 
\eea
where the second equality is obtained after realizing that $|| \hat{\bm{F}} || = || \hat{\bm{F}} \times \bm{N} ||$ and $||  \text{cof}\hat{\bm{F}}|| = ||  \text{cof}\hat{\bm{F}} \bm{N} ||$, which we explained in Equation \eqref{eq43}.

Combining the previous formulation with the requirement of one-homogeneity of function $\hat\Phi( \hat{\bm{F}}\times \bm{N}, \text{cof}\hat{\bm{F}} \bm{N})$ from Equation \eqref{eq50a}, with an assumption of an additive split of the energy into $\Hat{J}$-dependent and $\Hat{\bm{F}}$-dependent terms, we obtain
\bea \label{eq:PolyconvexSurfaceEnergy}
\tilde{{\Psi}}_s\left(\hat{\bm{F}},\bm{N}\right) = \tilde{{\Psi}}^I_s\left(\hat{\bm{F}},\bm{N}\right) 
= \alpha \left|\left|\hat{\bm{F}} \times \bm{N}\right|\right| + \gamma || \text{cof}\hat{\bm{F}}  \bm{N} ||
= \alpha \left|\left|\hat{\bm{F}} \right|\right| + \gamma \hat{J} = \alpha \sqrt{I_1(\hat{\bm{C}})} + \gamma \sqrt{I_2(\hat{\bm{C}})}
\eea
where $\alpha$ and $\gamma$ are non-negative constants characterizing the contribution of each term.
The part $\gamma\hat{J}$ corresponds to the well-known fluid-like surface tension model. In the case only this term is present, the stored energy density per undeformed area is proportional to the area of the deformed surface, since $\hat{J}=\mathrm{d}a/\mathrm{d}A$. Consequently, in the absence of any bulk energy, the material tends to deform such that its deformed surface is minimized, independently of the level of deformation. 
On the other hand, the term  $\alpha \left|\left|\hat{\bm{F}} \right|\right|$ represents the deformation-dependent part of the surface-polyconvex energy. 

The invariants of $\hat{\bm{C}}$ themselves are not surface-polyconvex due to the requirement of one-homogeneity; however, their square roots satisfy the condition of surface polyconvexity. Alternatively, the surface energy can be expressed in terms of the singular values of the surface deformation gradient $\hat{\lambda}_1$ and $\hat{\lambda}_2$. Moreover, the surface energy proposed in~\eqref{eq:PolyconvexSurfaceEnergy} can be generalized by considering a sum of arbitrary isotropic norms, which yield convex functions that are positively one-homogeneous in $||\hat{\bm{F}}||$. Based on this observation, we propose an Ogden-type surface-polyconvex model of the form
\begin{equation} \label{eq:PolyconvexSurfaceOgden}
\tilde{\Psi}_s\left(\hat{\bm{F}}, \bm{N} \right) = {\tilde{\Psi}}^O_s\left(\hat{\lambda}_1, \hat{\lambda}_2 \right) = \sum_{I=1}^N \alpha_I \left( \hat{\lambda}_1^{\beta_I} + \hat{\lambda}_2^{\beta_I} \right)^{1/\beta_I} + \gamma \hat{\lambda}_1 \hat{\lambda}_2,
\end{equation}
where $\alpha_I$, $\beta_I$, and $\gamma$ are model parameters. The model reduces to~\eqref{eq:PolyconvexSurfaceEnergy} for $N = 1$, $\alpha_1 = \alpha$, and $\beta_1 = 2$, which will be used in the following sections.

The response function of the first Piola-Kirchhoff surface stress is obtained from Equation \eqref{eq51} as
\bea
\tilde{\hat{\bm{P}}} =\tilde{\hat{\bm{P}}}^I = \partial_{\hat{\bm{F}}} \tilde \Psi^I_s \hat{\bm{I}} = \alpha \dfrac{\hat{\bm{F}} }{\left|\left|\hat{\bm{F}} \right|\right| }  + \gamma \underbrace{\frac{1}{\hat{J}} \hat{\bm{F}} \left( \text{tr}\hat{\bm{C}}\hat{\bm{I}} - \hat{\bm{C}} \right)}_{\hat{J}\hat{\bm{F}}^{-T}}
\eea
where the relations for the derivatives $\partial_{\hat{\bm{F}}}|| \hat{\bm{F}}||$ and $\partial_{\hat{\bm{F}}} \hat{J}$ were substituted from Equations \eqref{eq52} and \eqref{eq46}. 

Alternatively, the expression for $\partial_{\hat{\bm{F}}} \hat{J}$ can be taken directly from Equation~\eqref{eq53}. By applying the Piola transformation~\eqref{eq:piola}, we obtain the surface Cauchy stress response function:
\begin{equation}
\tilde{\hat{\bm{\sigma}}} = \frac{1}{\hat{J}} \tilde{\hat{\bm{P}}} \hat{\bm{F}}^T = 
 \frac{\alpha}{\hat{J} \left\| \hat{\bm{F}} \right\|} \hat{\bm{F}} \hat{\bm{F}}^T + \gamma \hat{\bm{i}}.
\end{equation}
The second term in the surface Cauchy stress, associated with the fluid-like contribution scaled by $\gamma$, is independent of deformation. In contrast, the first term depends explicitly on the deformation through $\hat{\bm{F}}$.

It is evident from this expression that the surface Cauchy stress tensor is positive semi-definite, as both $\hat{\bm{F}} \hat{\bm{F}}^T$ and the identity tensor $\hat{\bm{i}}$ are positive semi-definite, and the coefficients $\alpha, \gamma \geq 0$. Evaluating the stress in the reference configuration, where $\hat{\bm{F}} = \hat{\bm{I}}$ and thus $\hat{\bm{F}} \hat{\bm{F}}^T = \hat{\bm{I}} = \hat{\bm{i}}$, and $\hat{J} = 1$ we obtain:
\beq
\tilde{\hat{\bm{\sigma}}}(\hat{\bm{F}} = \hat{\bm{I}}) = \frac{\alpha}{ \left\| \hat{\bm{I}} \right\|} \hat{\bm{i}} + \gamma \hat{\bm{i}} = \left( \frac{\alpha}{\sqrt{2}} + \gamma \right) \hat{\bm{i}}.
\eeq
This shows that the stress in the reference configuration is nonzero unless both $\alpha$ and $\gamma$ vanish, which implies that the model does not admit a stress-free reference state. This reflects an intrinsic pre-stress associated with the surface behavior and is closely related to surface polyconvexity, which inherently precludes compressive surface stresses, see also \cite{Silhavy-2010b}.

The first surface elasticity tensor is obtained by differentiating the first Piola-Kirchhoff surface stress response function with respect to the surface deformation gradient
\bea \label{eq:FirstElasticityTensor}
\tilde{\hat{\bm{A}}} = \tilde{\hat{\bm{A}}}^I = \partial_{\hat{\bm{F}}} \tilde{\hat{\bm{P}}}=\left( \partial_{\hat{\bm{F}}} \otimes \partial_{\hat{\bm{F}}} \right) \tilde \Psi_s \hat{\bm{I}}  &=&
\dfrac{\alpha}{\left|\left|\hat{\bm{F}} \right|\right| } \left( \hat{\bm{I}} ~\overline{\otimes} ~\hat{\bm{I}} - \dfrac{1}{\left|\left|\hat{\bm{F}} \right|\right|^2 } \hat{\bm{F}} {\otimes} \hat{\bm{F}} \right)
+\nonumber \\ 
 &+&   \frac{\gamma }{\hat{J}} \left( -\partial_{\hat{\bm{F}}}  \hat{J} \otimes \partial_{\hat{\bm{F}}}  \hat{J} + \bm{I}~ \overline{\otimes} 
 \left(  \text{tr}\hat{\bm{C}}\hat{\bm{I}} - \hat{\bm{C}}\right) + 2 \hat{\bm{F}} \otimes \hat{\bm{F}} - \hat{\bm{F}} \underline{\otimes}    \hat{\bm{F}}  - \left(\hat{\bm{F}}\hat{\bm{F}}^T \right)\overline{\otimes} \hat{\bm{I}} \right) 
\eea

\subsubsection{Anisotropic polyconvex model}

In Section \ref{aniso}, we reduced the form of the surface energy response function for the anisotropic surface material to expression \eqref{eq62}, which reads
\bea 
\tilde\Psi_s\left(\hat{\bm{F}},\bm{N}\right) = \tilde{\tilde{\bar{\Psi}}}_s\left(|| \hat{\bm{F}} ||^2,|| \text{cof}\hat{\bm{F}} ||^2,|| \hat{\bm{F}} \bm{a} ||^2,|| \hat{\bm{F}}^T\hat{\bm{F}} \bm{a} ||^2 \right)
\eea
This formulation can be further specialized by the constraint of one-homogeneity in  $\hat{\bm{F}}\times \bm{N}$, and $\text{cof}\hat{\bm{F}} \bm{N}$. Since $\hat{\bm{F}} = - (\hat{\bm{F}}\times\bm{N})\times \bm{N}$, $\hat{\bm{F}}$ is linear in $\hat{\bm{F}}\times\bm{N}$ and consequently if we assume additive decomposition into individual terms, the surface energy response function must be given by
\bea \label{eq_anisoGen}
\tilde{{\Psi}}_s\left(\hat{\bm{F}},\bm{N}\right) 
= \tilde{{\Psi}}_s^I\left(\hat{\bm{F}},\bm{N}\right) + \tilde{{\Psi}}_s^{AI}\left(\hat{\bm{F}},\bm{N}\right) =\alpha \left|\left|\hat{\bm{F}} \right|\right| + \gamma \hat{J} +
\eta \left|\left| \hat{\bm{F}} \bm{a} \right|\right| + \beta  \sqrt{\left|\left| \hat{\bm{F}}^T \hat{\bm{F}} \bm{a} \right|\right|} = \alpha \sqrt{I_1} + \gamma \sqrt{I_2} +\eta  \sqrt{I_4} + \beta \sqrt[4]{I_5}
\eea
where $\tilde{{\Psi}}_s^{AI} = \eta || \hat{\bm{F}} \bm{a} || + \beta  \sqrt{|| \hat{\bm{F}}^T \hat{\bm{F}} \bm{a} ||}$ denotes the anisotropic part of the surface energy response function. Analogously to the isotropic case, the generalized invariants $I_4$ and $I_5$ defined in Equations \eqref{eq:i4} and \eqref{eq:i5} do not fulfill the one-homogeneity property. Consequently, the surface energy density must depend on the surface-polyconvex invariants $\sqrt{I_4}$ and $\sqrt[4]{I_5}$. 

The surface first Piola-Kirchhoff stress tensor is then expressed using Equations  \eqref{eq51}, \eqref{eq63}, and \eqref{eq64} 
\bea
\tilde{\hat{\bm{P}}} = \partial_{\hat{\bm{F}}} \tilde \Psi_s \hat{\bm{I}} = \tilde{\hat{\bm{P}}}^I + \eta\dfrac{\hat{\bm{F}} \hat{\bm{a}} \otimes \hat{\bm{a}} }{\left|\left|\hat{\bm{F}}\hat{\bm{a}}  \right|\right| }+ 
 \frac{\beta}{2\left|\left|\hat{\bm{C}}\hat{\bm{a}}  \right|\right| \sqrt{\left|\left|\hat{\bm{C}}\hat{\bm{a}}  \right|\right| }} \hat{\bm{F}}\left( \hat{\bm{C}}\hat{\bm{a}} \otimes  \hat{\bm{a}}+ \hat{\bm{a}}\otimes\hat{\bm{C}}\hat{\bm{a}}  \right)
\eea

The first elasticity tensor response function reads as 
\bea \label{eq:FirstElasticityTensorAniso}
\tilde{\hat{\bm{A}}}  =\left( \partial_{\hat{\bm{F}}} \otimes \partial_{\hat{\bm{F}}} \right) \tilde \Psi_s \hat{\bm{I}} = \tilde{\hat{\bm{A}}}^I + \eta(\partial_{\hat{\bm{F}}} \otimes \partial_{\hat{\bm{F}}} ) \sqrt{\left|\left|\hat{\bm{C}} \hat{\bm{a}} \right|\right|} + 
\beta(\partial_{\hat{\bm{F}}} \otimes \partial_{\hat{\bm{F}}} ) ||\hat{\bm{F}}\hat{\bm{a}} ||
\eea
where the expressions for the second derivatives are skipped for brevity and can be found in Equations \eqref{eq65} and \eqref{eq66}.

\section{Numerical examples}

\subsection{Finite element implementation}\label{Sec:FEM}
To test the proposed surface material models, the surface elasticity framework was implemented into OOFEM open-source finite element software \cite{patzak2012oofem}. The derivation of the discretized weak form of equilibrium equations for the continua with boundary energies can be found, e.g., in \cite{javili2010finite}. The main extensions arise in the evaluation of nodal internal forces and of the element stiffness matrix, where additional contributions of the surface energy appear. To account for these terms, we introduced additional surface elements, which are assigned to the faces of the elements located at the surface.
Here we just briefly revise the key points. 

The weak form of equilibrium for bodies with surface energy is given in Equation \eqref{eq33}, and serves as the starting point for the discretization procedure.
The volume $\Omega$ and the surface $S$ are divided into 
$N_{b,el}$ volume and $N_{s,el}$ surface non-overlapping elements
\bea
\Omega = \bigcup_{\beta = 1}^{N_{b,el}} \Omega^\beta, \
S = \bigcup_{\gamma = 1}^{N_s,el} S^\gamma \label{eq18}
\eea
Using the Bubnov-Galerkin approximation, both the geometry and the test fields are projected onto a finite-dimensional subspace. For each bulk element with nodal reference coordinates $\bm{X}^{i}$, nodal deformed coordinates $\bm{x}^{i}$, and nodal test field values $\delta{\bm{\varphi}}^{i}$ with local numbering $i = 1,\dots, N_{b,no}$, where $N_{b,no}$ is the number of nodes per element, the positions and test field are expressed by
\bea
\bm{X}(\bm{\xi}) \approx \sum_{i=1}^{N_{b,no}} N^i(\bm{\xi})\bm{X}^{i}, \ 
\bm{x}(\bm{\xi}) \approx \sum_{i=1}^{N_{b,no}} N^i(\bm{\xi})\bm{x}^{i}, \ 
\delta{\bm{\varphi}}(\bm{\xi}) \approx \sum_{i=1}^{N_{b,no}} N^i(\bm{\xi})\delta{\bm{\varphi}}^{i} \label{eq20}
\eea
where $N^i$ is the $i$-th shape function and $\bm{\xi}$ stands for the column matrix of parameter coordinates associated with the bulk element. Similarly, the positions and test field on the surface element with nodal coordinates $\bm{X}^{j}$ and $\bm{x}^{j}$ and nodal test field values $\delta{\bm{\varphi}}^{j}$ with local numbering $j = 1,\dots, N_{s,no}$, where $N_{s,no}$ denotes the number of nodes per surface element, are approximated by
\bea
\bm{X}(\bm{\xi_s}) \approx \sum_{j=1}^{N_{s,no}} N_s^j(\bm{\xi_s})\bm{X}^{j}, \ 
\bm{x}(\bm{\xi_s}) \approx \sum_{j=1}^{N_{s,no}} N_s^j(\bm{\xi_s})\bm{x}^{j}, \ 
\delta{\bm{\varphi}}(\bm{\xi_s}) \approx \sum_{j=1}^{N_{s,no}} N_s^j(\bm{\xi_s})\delta{\bm{\varphi}}^{j} \label{eq21}
\eea
where $N_s^j$ denotes the $j$-th shape function on the surface and the components of $\bm{\xi_s}$ are the parameter coordinates of the surface elements. Note that $i$ and $j$ denote the local nodal numbering referring to individual elements.

Using decomposition in \eqref{eq18}, the integrals in \eqref{eq33} can be split into sum over elements, shape function approximations from Equations \eqref{eq20} and \eqref{eq21} are then substituted and using the arbitrariness of the test function coefficients, the approximated discrete residual form of equilibrium is obtained
\bea
\bm{R}(\bm{x})= \bm{0} \label{eq22}
\eea
The residual vector $\bm{R} = \left[\bm{R}^1, \dots,\bm{R}^J,\dots, \bm{R}^{N_{no}}  \right]^T$ is composed of the nodal residuals $\bm{R}^J$, which are assembled in a standard way from the nodal residual contributions of each adjacent element. $N_{no}$ stands for the overall number of nodes.
If node $J$ is located at the surface, there are two types of contribution to the nodal residual $\bm{R}^J$, one from the bulk elements
\bea
\bm{R}_\beta^J = \int_{\Omega^\beta} \bm{P} \cdot \nabla N^i~ \mathrm{d}\mathcal{V} -  \int_{\Omega^\beta} \bm{b} N^i~ \mathrm{d}\mathcal{V} 
\eea
and the other from the surface elements
\bea
\bm{R}_\gamma^J = \int_{S^\gamma} \hat{\bm{P}} \cdot \hat{\nabla} N_s^j~ \mathrm{d}\mathcal{A} \label{eq23}
\eea
Here $i$ and $j$ denote the local indexing of the global nodal index $J$. If node $J$ belongs only to a bulk element, only the $\bm{R}_\beta^J$ contribution is present. 
The global vector of the unknown deformed positions has the form $\bm{x}= \left[\bm{x}^1, \dots,\bm{x}^{N_{no}}   \right]^T$.

For the solution of Equation \eqref{eq22}, the Newton-Raphson method is applied, for which the tangent stiffness matrix
\bea
\bm{K}(\bm{x}) = \dfrac{\partial \bm{R}(\bm{x})}{\partial \bm{x}}
\eea
needs to be expressed. The submatrix $\bm{K}^{IJ}={\partial \bm{R}^I}/{\partial \bm{x}^J}$ associated to global nodes $I$ and $J$ is assembled from the nodal contributions of each adjacent element. If both nodes belong to a single surface element, the Cartesian components $\left[\bm{K}_{\gamma }^{IJ}\right]_{ac}$ of the surface element contribution read
\bea
\left[\bm{K}_{\gamma }^{IJ}\right]_{ac}= \int_{S^\gamma} \left[\hat{\nabla} N_s^i\right]_{b} \left[\tilde{\hat{\bm{A}}}\right]_{abcd}  \left[\hat{\nabla} N_s^j\right]_{d}~ \mathrm{d}{\mathcal{A}} \label{eq24}
\eea
and the Cartesian components of the bulk element contribution are
\bea
\left[\bm{K}_{\beta }^{IJ}\right]_{ac}= \int_{\Omega^\beta} \left[\nabla N^i\right]_{b} \left[\tilde{\bm{A}}\right]_{abcd}  \left[\nabla N^j\right]_{d}~ \mathrm{d}\mathcal{V} 
\eea
$i$ and $j$ denote the local indices corresponding to global nodal indices $I$ and $J$. The fourth order surface material tensor $\tilde{\hat{\bm{A}}} = \partial \hat{\bm{P}} / \partial \Hat{\bm{F}}$ was already defined for isotropic and anisotropic models in \eqref{eq:FirstElasticityTensor} and \eqref{eq:FirstElasticityTensorAniso}, respectively. $\tilde{\bm{A}}$ is the classical bulk first elasticity tensor.  If at least one of the nodes $I$ or $J$ belongs purely to a bulk element, only the bulk contribution $\bm{K}_{\beta }^{IJ}$ applies. 

The evaluation of the terms $\hat{\nabla} N_s^i$ is particularly simple if the surface elements are planar since the reference unit normal $\bm{N}$ is constant over the element. The surface gradient is then obtained by projecting the full gradient $\hat{\nabla} N_s^i = \nabla N_s^i (\bm{I}-\bm{N}\otimes\bm{N})$ by a constant projection tensor $\bm{I}-\bm{N}\otimes\bm{N}$. 

In our implementation, the individual surface element contributions \eqref{eq23} and \eqref{eq24} are firstly evaluated in the Cartesian system aligned with the surface element and then transformed to the global Cartesian system during the assembly process.  

\subsection{Bifurcation tracking}\label{secBif}
To study the effect of surface energy on the presence of instabilities in soft materials, path following and branch switching methods were implemented into the above-mentioned finite element OOFEM software. In this section, we only briefly discuss the applied procedures, since this area is not the main goal of this publication. The concepts were inspired by \cite{Crisfield1997-hm}, \cite{de2012nonlinear}, and \cite{farrell2015deflation}.

Following \cite{Crisfield1997-hm}, the solution $\Bar{\bm{x}}$ of the discretized weak form of equilibrium equations $\bm{R}(\bm{x})=\bm{0}$ is stable if the tangent stiffness matrix $\bm{K}(\bar{\bm{x}}) = \partial \bm{R} (\bm{x})/ \partial \bm{x} \vert_{\Bar{\bm{x}}}$ is positive definite. To check the positive definiteness, the lowest eigenvalue $\lambda_{min}$ of the tangent stiffness matrix is evaluated, and its sign determines if the solution is stable (positive sign) or unstable (negative sign). Alternatively, when solving the linear system using the Cholesky LDLT factorization $\bm{K} = \bm{L}\bm{D}\bm{L}^T$, the signs of the diagonal entries of $\bm{D}$ provide some information about the spectrum of $\bm{K}$. 
{Specifically, the number of positive, zero, and negative elements of the diagonal\footnote{Note that $\bm{D}$ may be block diagonal in general.} matrix $\bm{D}$ corresponds to the number of positive, zero, and negative eigenvalues of the stiffness matrix. }
Since matrix $\bm{D}$ is computed as part of the factorization process, inspecting the sign of its smallest diagonal entry offers a more computationally efficient way to assess stability than calculating the eigenvalues, see \cite{cheng1998modified, faltus2025towards}. Both approaches were implemented and made available as options to assess the stability of the solution. The Eigen library \cite{eigenweb} was used to solve the system of linear equations and to access the diagonal entries of matrix $\bm{D}$ when using the Cholesky factorization, while the Spectra library \cite{spectraweb} was employed for computing the eigenvalues and eigenvectors of the tangent stiffness matrix.
To find a singular point, the parameter controlling the simulation (e.g., surface material parameter $\gamma$ or $\alpha$ as used in Example \ref{Sec:Stability analysis and bifurcation}) is increased, and in each step, the positive definiteness of the tangent stiffness matrix evaluated at the converged state is checked by one of the methods described above. Once the matrix loses positive definiteness, the solution becomes unstable, which indicates that a singular point was passed. To find the singular point accurately, the increments of the control parameter must be sufficiently fine. 

After the singular point is reached, the behavior conceptually differs if the singular point is classified as a limit point or a bifurcation point. We are mainly interested in the latter case, for which multiple solutions exist and "meet" in the bifurcation point. A branch switching technique needs to be applied to find a different solution than the already determined unstable one, denoted by $\bm{x}_0$. The standard methods are based on modification of the unstable solution vector in the direction of the eigenvector $\bm{v}_{min}$ associated with the lowest eigenvalue $\lambda_{min}$. A different approach, referred to as the deflation method \cite{farrell2015deflation}, is based on modifying the residual vector during the iterations such that its norm tends to infinity if the solution vector equals the unstable one. In its basic form, the residual vector is modified to 
\bea
\bm{R}^m(\bm{x}) =\bm{R}(\bm{x}) \left( \frac{1}{||\bm{x}-\bm{x}_0||^2} +1 \right)
\eea
and the equation $\bm{R}^m(\bm{x}) = \bm{0}$ is then solved by the standard Newton-Raphson method. Such modification indeed ensures that $||\bm{R}^m|| \to \infty$ if $\bm{x} \to \bm{x}_0$ and $\bm{R}^m \to \bm{R}$ if $\bm{x}$ is far from the unstable solution $\bm{x}_0$. 
Unfortunately, none of these methods worked robustly enough in our case. Most of the time, the issue came from the fact that during the Newton-Raphson iterations, some elements became inverted, which terminated the simulation. We proposed a procedure that worked robustly in our case. The procedure consisted of a combination of the standard Newton-Raphson method, an exact line search, and the deflation method. For us, the exact line search starting from point $\bm{x}_s$ in direction $\bm{d}$ means solving exactly a projected residual equation $\bm{d}^T \bm{R}(\bm{x}_s + \beta \bm{d}) = 0$ 
for the unknown multiplier $\beta$. To ensure the algorithm does not end up in the unstable solution $\bm{x}_0$, the exact line search is applied on the modified residual equation, leading to a solution of $\bm{d}^T \bm{R}^m(\bm{x}_s +\beta \bm{d}) ={0}$.

For clarity, we now describe the complete bifurcation tracking routine. 
Assume the first unstable solution $\bm{x}_0$ was just found. Afterwards, the eigenvector $\bm{v}_{min}$ corresponding to eigenvalue $\lambda_{min}$ of the tangent stiffness matrix  $\bm{K}(\bm{x}_0)$ evaluated at the unstable solution is computed. If the bifurcation point is sufficiently close to the found unstable solution, the direction of the eigenvector points towards a different desired solution $\bm{x}_d$. To get closer to it, we perform an exact line search with the deflation method in this direction, meaning we solve equation $\bm{v}_{min}^T \bm{R}^m(\bm{x}_0 + \beta \bm{v}_{min}) ={0}$ for the unknown multiplier $\beta$ by the standard Newton method. Note that by this procedure, we found a point 
located at a line crossing the original unstable solution and pointing in the direction of the eigenvector for which the residual vector is orthogonal to the direction of the eigenvector. This does not mean the residual vector is zero, only its projection on the eigenvector direction vanishes. To proceed further, the Newton-Raphson method is applied. In each iteration step $j$, the direction of the increment $\Delta \bm{x}_{j+1}$ is obtained by solving the linearized system of equations $\bm{K}(\bm{x}_{j+1})\Delta \bm{x}_{j+1} = -\bm{R}(\bm{x}_{j+1})$ and the length of the step is determined by the previously described exact line search combined with the deflation method. This process is repeated until a convergence is reached. If that is the case, it means a different solution $\bm{x}_d$ has been found and its stability is checked by the methods described above. If the solution is unstable, the procedure is repeated with the difference that when applying the deflation method during the line search, all previously identified unstable solutions are excluded.
If the solution is stable, the algorithm can proceed with the next loading step. If the increments of the control parameter are small enough, no special treatment is necessary for the subsequent steps. On the other hand, if the increment is too large, the Newton-Raphson method may converge to the unstable solution again, or the simulation may fail completely due to inverted elements. A potential remedy to inverted elements is the so-called discretisation-aware load stepping proposed in \cite{poya2025generalised}. The main steps of the algorithm are summarized in Algorithm \ref{pseudoCode}.     
\begin{algorithm}
\caption{Pseudo-code bifurcation tracking}\label{pseudoCode}
\begin{algorithmic}[1]
\For{$i=1:N$} \Comment {Loop over control parameter increments}
\State $\gamma = \gamma_i$
\State Find solution $\Bar{\bm{x}}$ of $\bm{R}(\bm{x})=\bm{0}$ using standard Newton-Raphson method
\State Evaluate $\bm{K}(\Bar{\bm{x}})$ and check positive definiteness
    \If{is positive definite}
        \State continue
    \Else \Comment{Bifurcation point passed}
        \State $\bm{x}_0=\Bar{\bm{x}}$ is unstable solution
        \State Compute lowest eigenvector $\bm{v}_{min}$ of $\bm{K}(\bm{x}_0)$ 
        \State $j = 0$,\ $\Delta \bm{x}_{j+1} = \bm{v}_{min},\ \bm{x}_{j+1} =\bm{x}_0$
        \While {not converged}
            \State $j\gets j+1$
            \State \textbf{Do exact line search} in direction $\Delta \bm{x}_j$ to find length of the step $\beta$
                \State \hskip1.5em Define the deflated residuum $\bm{R}^m(\bm{x}) =\bm{R}(\bm{x})\left( 1+\frac{1}{||\bm{x}-\bm{x}_0||^2}\right) $ 
                \State \hskip1.5em Solve projected scalar equation $\Delta \bm{x}_{j}^T \bm{R}^m(\bm{x}_j+\beta \Delta\bm{x}_{j}) = 0$ for the multiplier $\beta$ by Newton's method 
                \State \hskip1.5em Update the state $\bm{x}_{j+1} = \bm{x}_j +\beta\Delta \bm{x}_j$
            \State Find next direction $\Delta\bm{x}_{j+1}$ by solving linearized equation $\bm{K}(\bm{x}_{j+1})\Delta \bm{x}_{j+1} = -\bm{R}(\bm{x}_{j+1})$
        \EndWhile
    \EndIf
\EndFor
\end{algorithmic}
\end{algorithm}

\subsection{Numerical examples}
In this section, we test the newly proposed isotropic and anisotropic surface-polyconvex models \eqref{eq:PolyconvexSurfaceEnergy} and \eqref{eq_anisoGen} on a series of examples. To gain better insight into the role of the deformation-dependent contribution, we separate the model into its deformation-dependent and deformation-independent parts and study their individual contributions. Specifically, for the surface energy, we consider either the fluid-like surface tension model $\tilde\Psi_s = \gamma \hat{J}$, or the newly proposed model $\tilde\Psi_s = \alpha \|\hat{\bm{F}}\|$. In the example focused on anisotropy, we further include anisotropic contributions, so that the surface strain energy takes the general form given in Equation~\eqref{eq_anisoGen}. This setup enables a detailed comparison of the mechanical response across different deformation scenarios, highlighting the distinct effects of each model.

We begin by examining the well-known problem of a cylindrical surface spanning between two circular boundaries, with no supporting bulk material. When modeled using a standard fluid-like surface tension formulation, this configuration is commonly referred to as a liquid bridge. Due to its symmetry and simplicity, the problem admits an analytical or semi-analytical solution, or can be reduced to a form suitable for basic numerical techniques. These solutions serve as reference results against which we compare the outcomes obtained from finite element (FE) analysis.

In the second example, we simulate the deformation of an incompressible cube whose surface is endowed with surface energy, modeled using either of the two formulations under investigation. In the case of the surface tension model, the body is expected to minimize its surface area, leading to a deformation toward a spherical shape. 

In addition to offering valuable insight into the distinct behavior of deformation-independent and deformation-dependent surface stress models, the first two examples also serve to validate our finite element implementation, as they allow for analytical or semi-analytical solutions against which the numerical results can be compared.

In the third example, we examine the effect of two anisotropic surface material models on the deformation of an incompressible sphere.

Finally, we investigate how both surface material models influence the onset and evolution of instabilities in soft materials. Our focus is on a long, compressible cylinder subjected to axial stretching—a variation of the classical Plateau–Rayleigh instability, originally formulated for an infinite incompressible liquid cylinder with surface tension acting on its boundary. The onset of instability can be determined semi-analytically by solving the incremental equations of elasticity with a deformation-dependent boundary potential, following the approach developed in \cite{yu2024incremental}. In addition to comparing the bifurcation points obtained via finite element (FE) analysis with the semi-analytical predictions, we also explore the post-bifurcation behavior of the system.

For all the cases, we consider the bulk energy to be a variant of the compressible neo-Hookean material 
\bea
\tilde\Psi_b(\bm{F}) = \frac{1}{2} \mu \left(\bm{F}:\bm{F}-3 -2 \ln J  \right)+\frac{1}{2} \kappa \left( \frac{1}{2}\left(J^2-1 \right)-\ln J \right)
\eea
where $\kappa$ and $\mu$ are the Lamé parameters. In the small-strain limit, these are connected to the Poisson ratio by ${\kappa}/{\mu} = 2 \nu/(1-2\nu)$.

\subsubsection{Liquid bridge without bulk energy}\label{BridgeExam}
We start with a toy problem of a cylinder with length $L$ and radius $R$ fixed at the ends. The structure is depicted in Figure \ref{fig1}. 
With the use of rotational symmetry, only one section can be modeled, resulting in the scheme depicted in Figure \ref{fig2}. 
This first example provides valuable insight into the behavior of the two surface models, as it allows for the derivation of analytical and semi-analytical reference solutions. In parallel, we solve the problem numerically using OOFEM finite element code, enabling a direct comparison that supports both validation and a deeper understanding of the different mechanical responses associated with the deformation-dependent and deformation-independent formulations. 
\begin{figure}[ht]
\centering%
\begin{subfigure}[b]{0.4\textwidth}
\includegraphics[width=\textwidth, keepaspectratio=true]{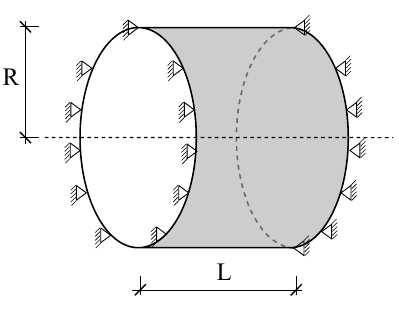}
\subcaption{Full model}
\label{fig1}
\end{subfigure}
\begin{subfigure}[b]{0.4\textwidth}
\includegraphics[width=\textwidth, keepaspectratio=true]{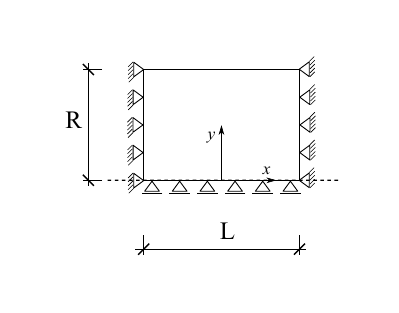}
\subcaption{Axisymmetric view}
\label{fig2}
\end{subfigure}
\caption{Scheme of cylinder.}\label{Scheme1}     
\end{figure} 
As there is no bulk material, such an example corresponds to a thin film spanned between the supports.

In the case of the classical surface tension model with energy $\tilde\Psi_s = \gamma \hat{J}$, the analytical solution of the deformed cylinder can be derived and is identical to the solution of minimal surface, i.e.,\ to the catenoid. In the reduced axisymmetric formulation (see Figure\ \ref{fig2}), the deformed shape is described by 
\beq
y = C\cosh \frac{x}{C}
\eeq
where the constant $C$ is obtained by solving the equation
\beq
C\cosh \frac{L/2 }{C} = R 
\eeq
which needs to be done numerically. For $L/R = 3/2.5$ the value of $C/R$ is 0.7451. A detailed derivation can be found in \ref{apA1}. The shape is displayed in Figure \ref{fig3} by the green solid line. 
\begin{figure}[ht]
\centering%
\includegraphics[width=0.5\textwidth, keepaspectratio=true]{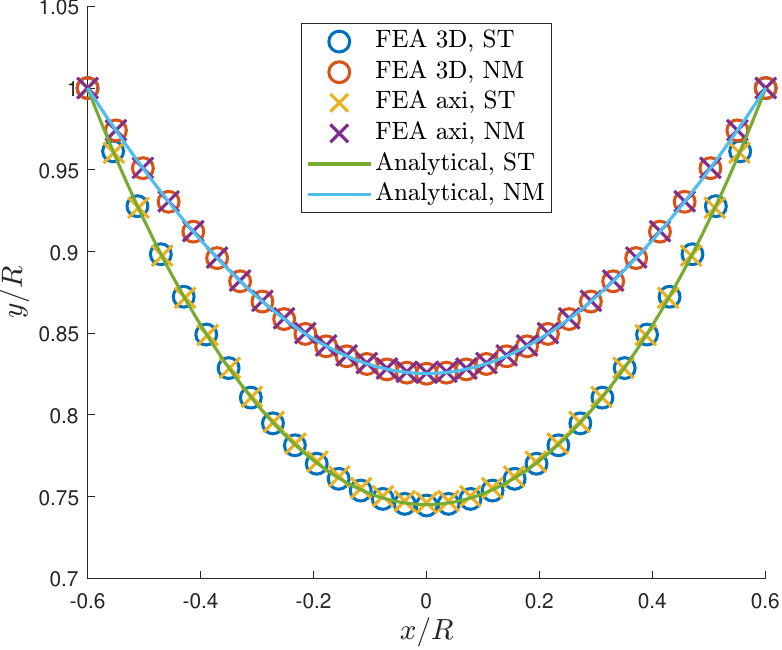}
\caption{Deformed shapes of liquid bridge for the cases $\tilde\Psi_s = \gamma\hat{J}$ and $\tilde\Psi_s = \alpha||\hat{\bm{F}}||$ and no bulk material.}     
\label{fig3}
\end{figure} 

Alternatively in the case of the deformation-dependent model with energy $\tilde\Psi_s = \alpha||\hat{\bm{F}}||$, the problem of finding the deformed shape can be posed as the solution of a surface with minimal stretches, since $||\hat{\bm{F}}|| = \sqrt{\hat{\lambda}_1^2+\hat{\lambda}_2^2}$, with $\hat{\lambda}_1$ and $\hat{\lambda}_2$ being the principal surface stretches. Such a problem can be simplified to the solution of two ordinary differential equations of the unknown horizontal and vertical displacements $u(x)$ and $w(x)$, which can then easily be solved numerically. The details are provided in \ref{apA2}. For the ratio of the parameter $L/R = 3/2.5$, the deformed shape is visualized in Figure \ref{fig3} by the blue solid line. 

Having established the reference solutions for both considered models, we can proceed by comparing them to the results of the FE simulation. Both 3D as well as 2D axisymmetric simulations were conducted. Even though the test case is supposed to have no bulk material, for numerical reasons it is necessary to initially also consider bulk material with nonzero shear modulus $\mu$ and gradually increase the ratio of the surface material parameter ($\gamma$ for the surface tension model or $\alpha$ for the novel model) to the shear modulus. Therefore, the simulation is controlled by the surface material parameter, which is increased until the contribution of the bulk energy becomes negligible compared to the energy of the surface. 

The first Lamé parameter $\kappa$ is set to zero.
The initial magnitudes of the dimensionless surface parameters are $\tilde\gamma = \gamma/(\mu R)=\tilde\alpha = \alpha/(\mu R) =0.4$, which are then quadratically increased in 30 steps. Consequently, at the end of the gradual increase, the parameters reach the value of $\tilde\gamma=\tilde\alpha=360$. Since such a value is much larger than 1, one can conclude that surface energy plays the dominant role and governs the deformation. 

For the full 3D simulation, only a quarter of the cylinder was modeled using 4410 linear brick elements for the bulk (147 elements per side and 30 elements per length) and 420 linear quadrilateral elements for the surface with an overall 5239 nodes. For the axisymmetric simulation, we used 600 2D linear axisymmetric elements for the bulk (20 elements per side and 30 elements per length) and 30 linear axisymmetric line elements for the surface; altogether, the mesh consisted of 651 nodes. The deformed positions of one line of nodes on the surface are marked in Figure \ref{fig3}; circles are used for the results of 3D simulation and crosses are used for the results obtained with axisymmetric elements. We can conclude that the match between the simulation and the analytical solution is almost perfect for both considered surface material models. The deformed shapes obtained from the 3D simulation are then visualized in Figure \ref{fig6}. 
\begin{figure}[ht]
\centering%
\begin{subfigure}[b]{0.49\textwidth}
\includegraphics[width=\textwidth, keepaspectratio=true]{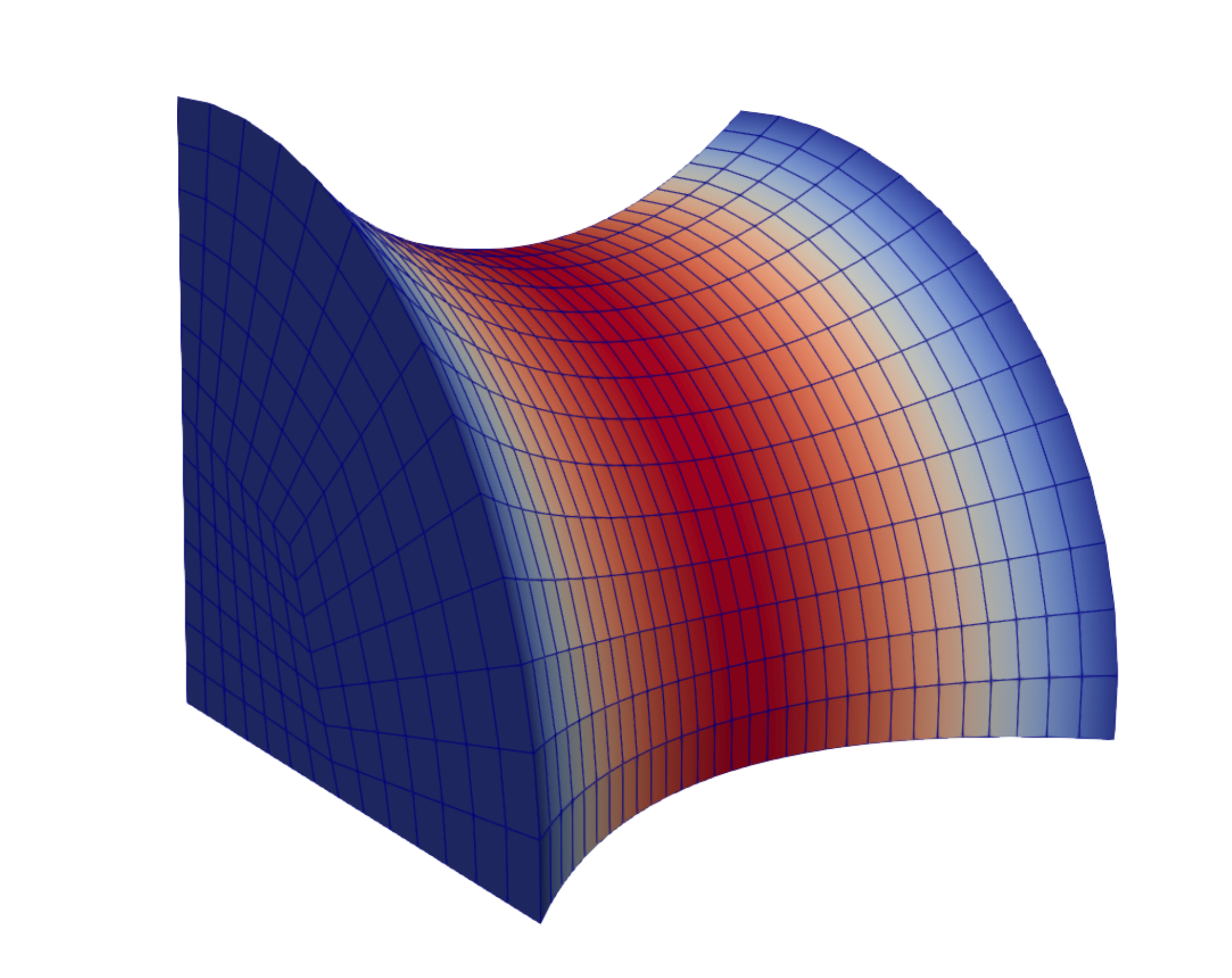}
\subcaption{Surface tension model with energy $\tilde\Psi_s = \gamma\hat{J}$}
\label{fig4}
\end{subfigure}
\begin{subfigure}[b]{0.49\textwidth}
\includegraphics[width=\textwidth, keepaspectratio=true]{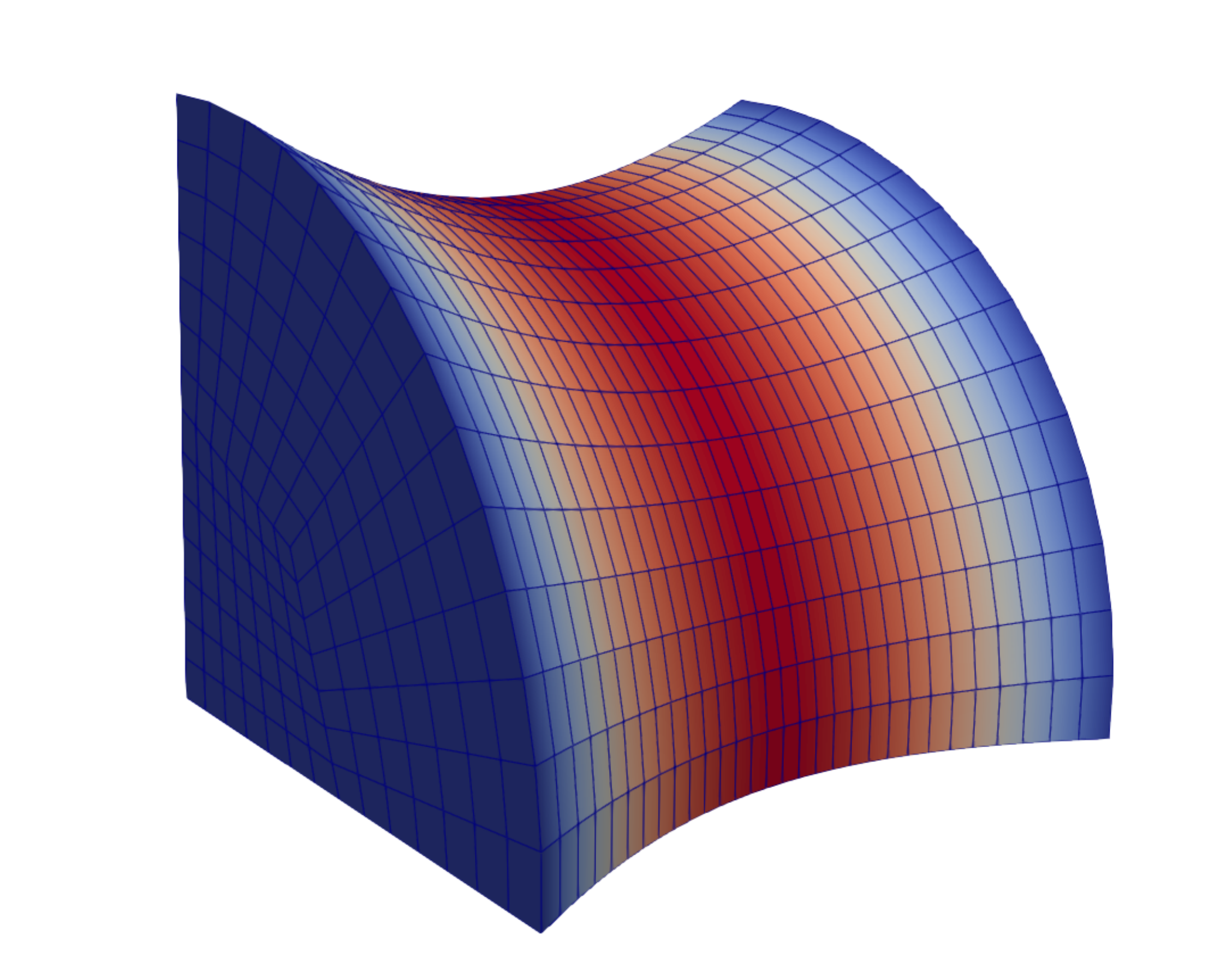}
\subcaption{Novel model with energy $\tilde\Psi_s =\alpha ||\hat{\bm{F}}||$}
\label{fig5}
\end{subfigure}
\caption{Deformed shapes of liquid bridge for the cases $\tilde\Psi_s = \gamma\hat{J}$ and $\tilde\Psi_s = \alpha||\hat{\bm{F}}||$ and no bulk material obtained using 3D FE simulation.}   
\label{fig6}
\end{figure} 
\subsubsection{Liquid bridge without bulk energy with stretching}
While the two models exhibit similar quantitative behavior when the deformation is driven solely by surface energy, their responses differ significantly in the presence of pre-existing deformation. To demonstrate the deformation dependence of the proposed model, we consider a cylinder similar to that in the previous example, but prestretched longitudinally to a prescribed initial stretch $\lambda_{1,0}$ prior to the application of the surface material parameters. For modeling purposes, this effect can be achieved by initially considering a cylinder of length $L/\lambda_{1,0}$ to which the stretch $\lambda_{1,0}$ is applied to deform it to length $L$. Subsequently, we increase the ratio of the surface material parameters to the shear modulus as in the previous test case and compare the resulting deformed shapes. The individual steps are schematically illustrated in Figure \ref{fig7}.    
\begin{figure}[ht]
\centering%
\includegraphics[width=\textwidth, keepaspectratio=true]{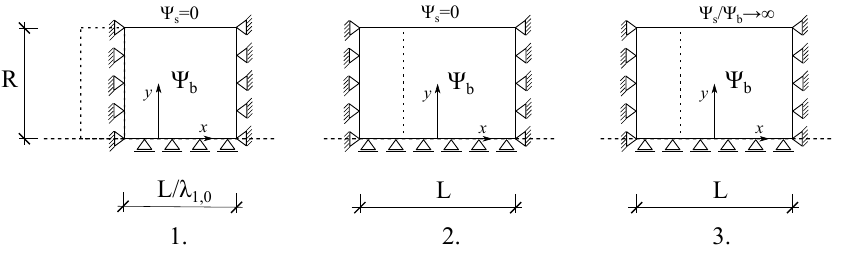}
\caption{Illustration of the individual steps done to simulate the deformation of a prestretched cylinder.}     
\label{fig7}
\end{figure} 
For this test case, the finite element simulations were performed only using the axisymmetric elements.  
In Figure \ref{fig8} the results obtained for prestretch $\lambda_{1,0}=1.5$ are visualized and compared to the previously shown analytical solutions.  
\begin{figure}[ht]
\centering%
\includegraphics[width=0.5\textwidth, keepaspectratio=true]{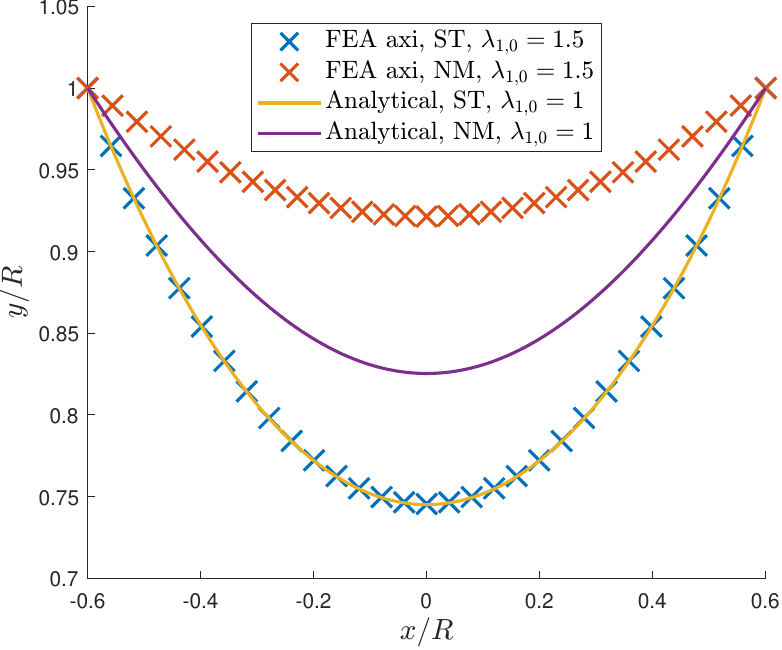}
\caption{Deformed shapes of liquid bridge with prestretch $\lambda_{1,0}=1.5$ evaluated using FE analysis for the cases $\tilde\Psi_s = \gamma\hat{J}$ and $\tilde\Psi_s = \alpha||\hat{\bm{F}}||$ and no bulk material compared to the analytically evaluated deformed shapes without prestretch.}     
\label{fig8}
\end{figure} 

As anticipated, for the surface tension model, the deformed shape of the prestretched specimen matches the analytical solution of the unstretched liquid bridge, which verifies that the surface tension model is deformation-independent. On the other hand, for the proposed model, the prestretch increases the model stiffness, causing a reduction of the vertical deformation of the prestretched liquid bridge, confirming deformation dependence of the proposed material model. 

\subsubsection{Incompressible cube with isotropic surface energy}
\label{sec:cube}
Here we investigate how an incompressible cube of side $a$ deforms if its surface possesses surface energy. Thus, in comparison with the previous example, we investigate the bulk-surface interaction. Note that we examine the same energies that were considered in the previous example. The aim is to investigate how an incompressible cube deforms when the contribution of the surface energy is increased compared to the deviatoric part of the bulk energy. Such an effect can be quantified by the ratio of the surface material parameter to the bulk parameter $\mu$, which we use as the "loading" parameter governing the deformation. 

We now describe the "loading" procedure for the case with nonzero surface parameter $\gamma$, but the same holds also for the other case with nonzero $\alpha$. 
The dimensionless "loading" parameter $\tilde\gamma = \gamma/(\mu r)$ is increased, starting at $\tilde\gamma =  0.01$ and ending at $\tilde\gamma = 0.01\cdot 11^{3} = 13.31$. In between the initial and final steps, the loading parameter is increased in $11$ steps governed by time function $t^{3}$, where $t$ denotes the discrete time.  
The length factor $r=a/2$ is used for normalization. Note that we use the same symbol $\tilde\gamma$ for the dimensionless version of $\gamma$ as in the previous example, even though the meaning is slightly different due to the different geometry.  

In the second test case, the simulations were controlled by increasing the parameter $\tilde\alpha = \alpha/(\mu r)$. The exact values of the parameters were selected as identical to those used for the surface tension model and thus can be obtained by replacing $\gamma$ by $\alpha$ in the above expressions. 

The mesh consists of 4598 nonuniform 10-node quadratic tetrahedron elements and 1572 6-node quadratic triangular surface elements. To prevent volumetric locking, the $u-p$ mixed formulation was applied with a linear approximation of the nodal pressures, see, e.g., \cite{kadapa2022linearized}. 

The results of the simulations for both load cases are visualized in Figure \ref{fig_cube_all}. In the first row, that is, in Figures~\ref{fig:cubeSTa}-\ref{fig:cubeSTd}, the deformed shapes of the cube with the surface tension material model are visualized for various values of the dimensionless surface tension parameter $\tilde\gamma$. 
In the second row, i.e.,\ in Figures \ref{fig:cubeFa}-\ref{fig:cubeFd}, the results corresponding to the novel deformation-dependent model are displayed for increasing parameter $\tilde\alpha$. Clearly, in both cases, the cube tends to deform into a sphere. In the case of the surface tension model, this is expected and well-known behavior; the body tends to deform in such a way that the volume is preserved and the energy of the surface, which is proportional to the deformed area, is minimized. The energy minimum is indeed achieved by deforming the cube into an object with the minimal surface with preserved volume, that is, a sphere of radius $R=(3/4/\pi a^3)^{1/3}$. 
However, it is not obvious that, in the case of the deformation-dependent model, the body tends to deform into a sphere.
Moreover, the fact that the outline of the final shape in both cases is a sphere does not necessarily mean the deformations are identical. By comparing the deformed nodal positions for both cases, one may conclude that they do not match.   

To get an idea of how the deformed shapes differ from a sphere, the outline of a diagonal section of the cube, highlighted in Figure \ref{fig:scheme_cube}, is visualized for various parameters $\tilde \gamma$ or $\tilde \alpha$; see Figures \ref{fig:cube_comp_a} and \ref{fig:cube_comp_b}. Clearly, in the initial stage, the shape of the outline is similar to a rectangle of side lengths $2\sqrt{a}$ and $a$, and with increasing "loading" parameters, it gets closer to the shape of a circle with radius $R=(3/4/\pi a^3)^{1/3}$.   
\begin{figure}[ht]
\centering%
\begin{subfigure}[b]{0.49\textwidth}
\centering%
\includegraphics[width=0.8\textwidth, keepaspectratio=true]{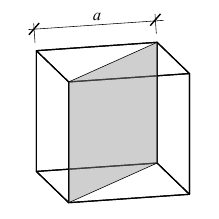}
\subcaption{Cube analyzed in example \ref{sec:cube}. A diagonal section highlighted in gray }
\label{fig:scheme_cube}
\end{subfigure}
\begin{subfigure}[b]{0.49\textwidth}
\centering%
\includegraphics[width=0.8\textwidth, keepaspectratio=true]{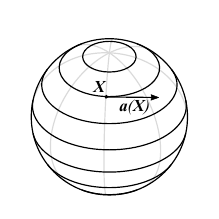}
\subcaption{Schematic view of the vector field $\bm{a}$ defining the direction of anisotropy of the surface strain dependent anisotropic material model}
\label{fig:scheme_sphere}
\end{subfigure}
\caption{Schemes of cube and sphere in examples \ref{sec:cube} and \ref{sec:sphere}.}  
\label{fig:schemes}
\end{figure}

\begin{figure}[ht]
\centering%
\begin{subfigure}[b]{0.195\textwidth}
\includegraphics[width=\textwidth, keepaspectratio=true]{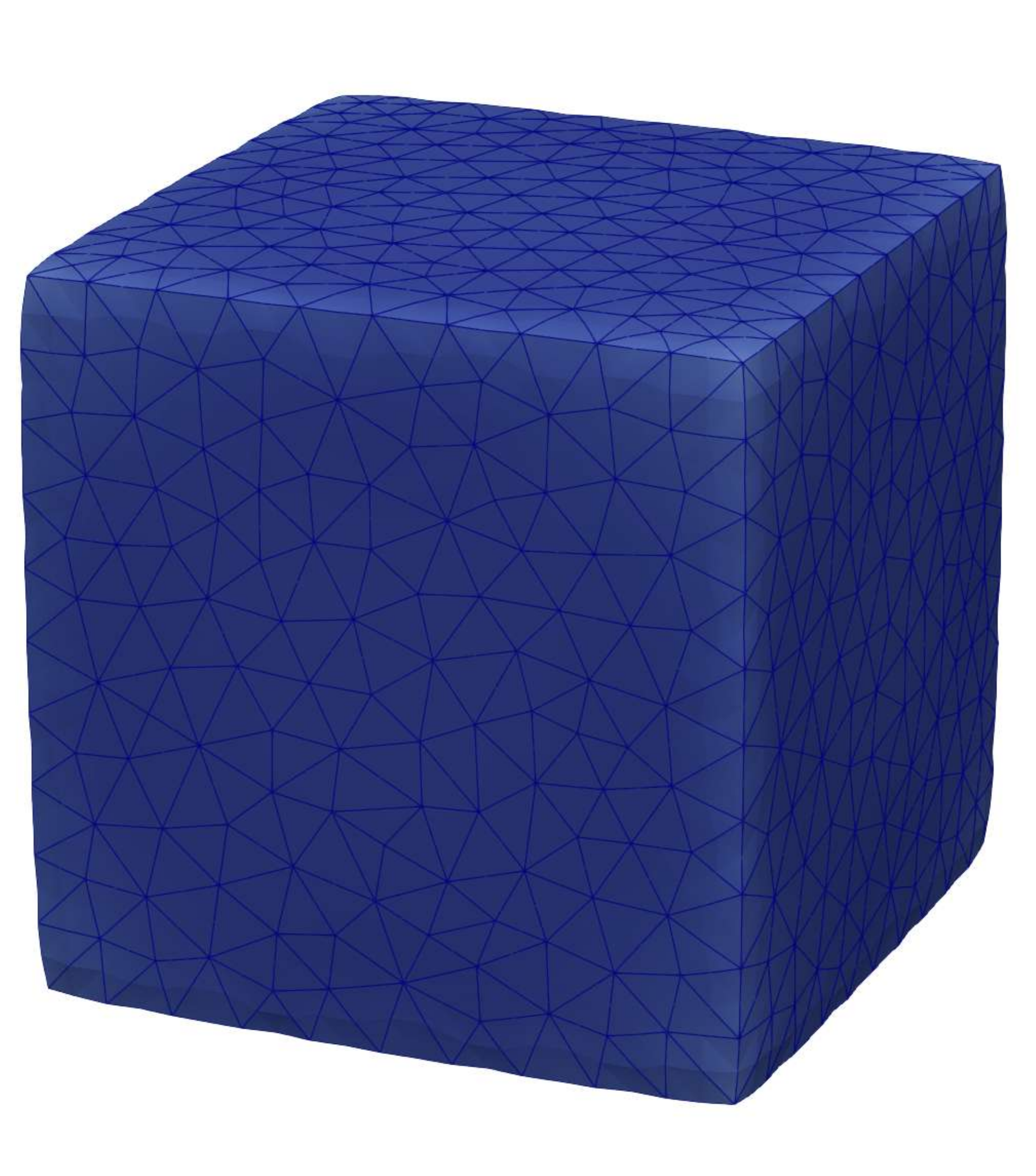}
\subcaption{$\tilde\gamma=0.08$, $\tilde\alpha=0$}
\label{fig:cubeSTa}
\end{subfigure}
\begin{subfigure}[b]{0.195\textwidth}
\includegraphics[width=\textwidth, keepaspectratio=true]{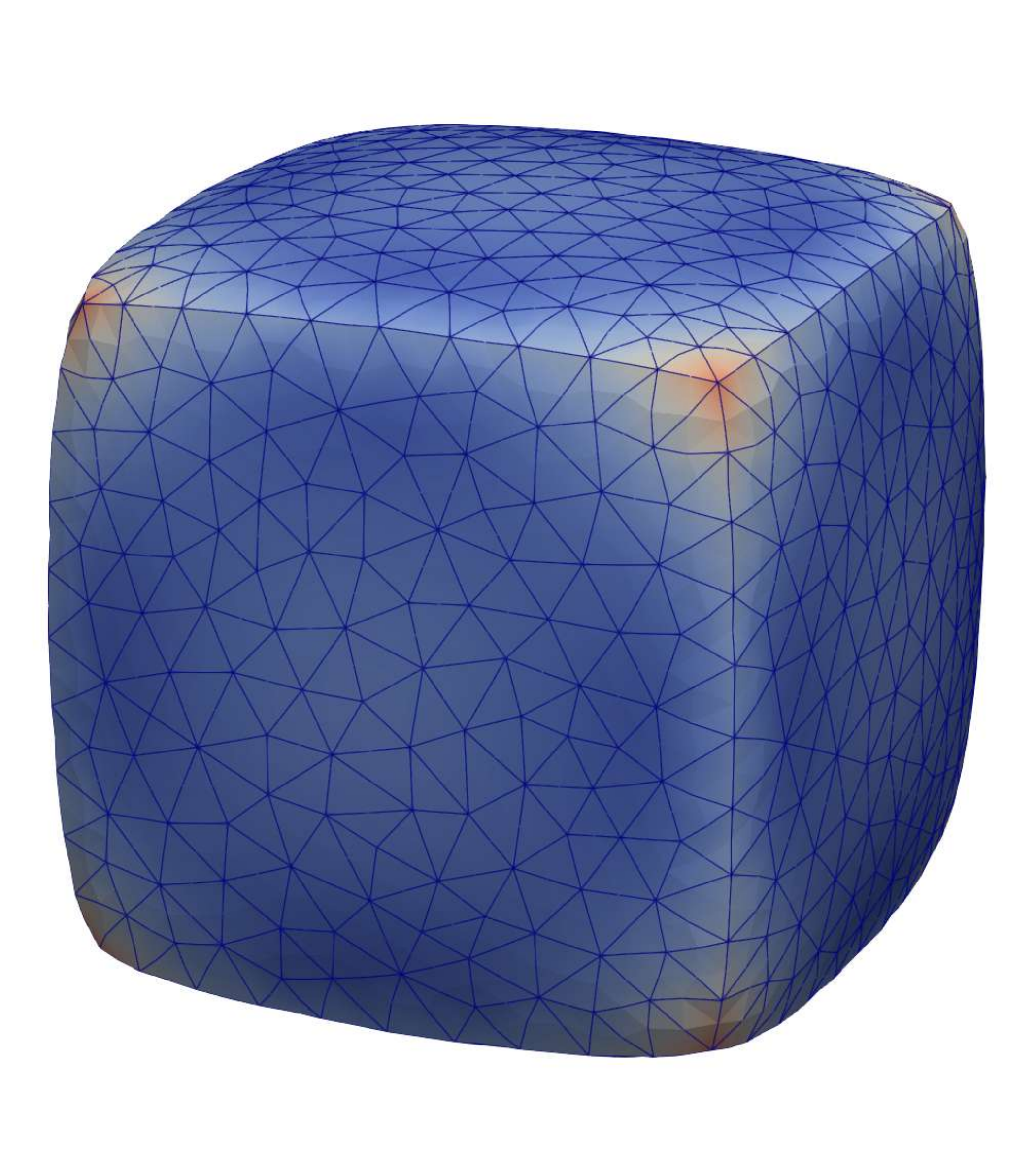}
\subcaption{$\tilde\gamma=0.64$, $\tilde\alpha=0$}
\end{subfigure}
\begin{subfigure}[b]{0.195\textwidth}
\includegraphics[width=\textwidth, keepaspectratio=true]{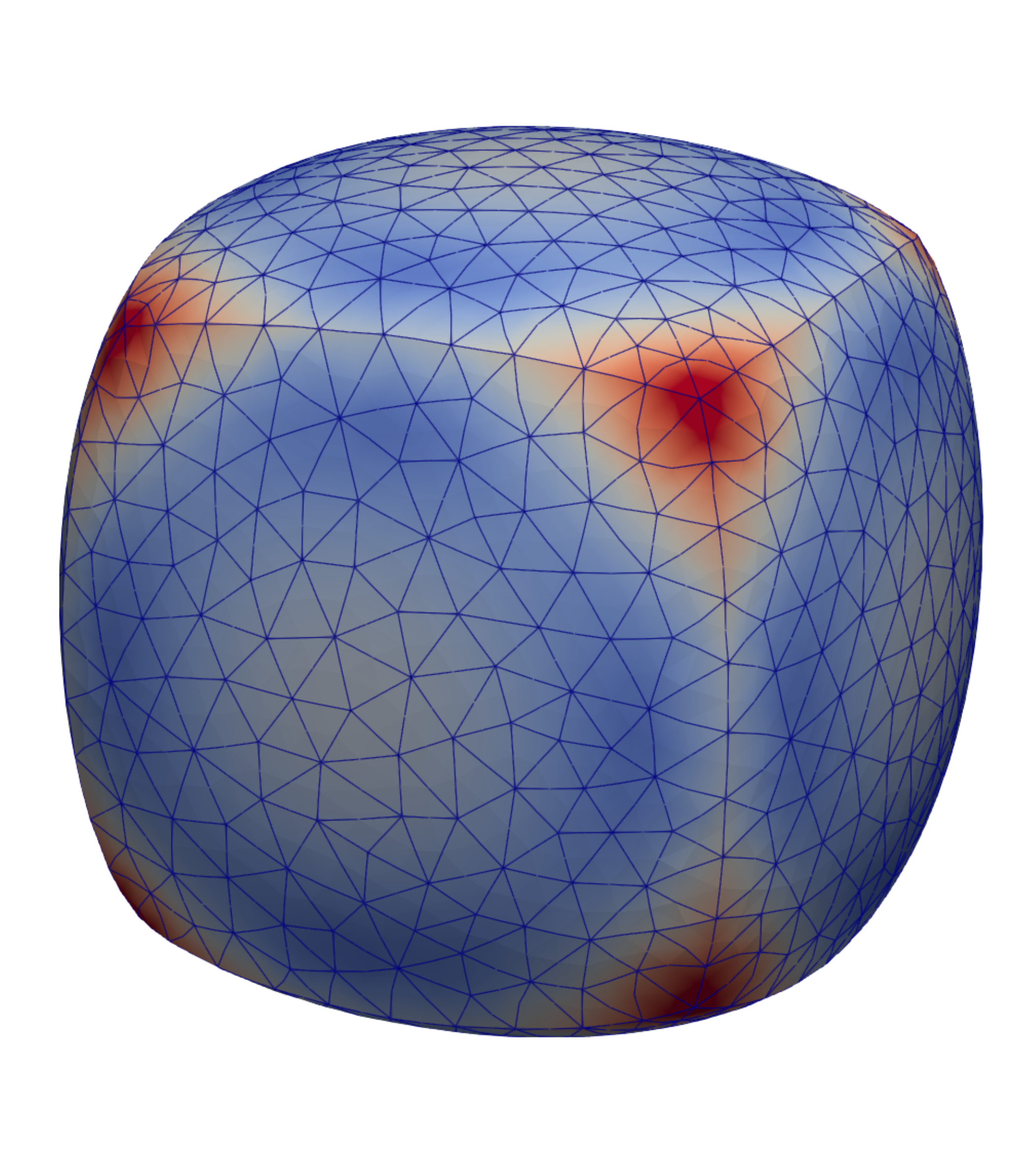}
\subcaption{$\tilde\gamma=2.16$, $\tilde\alpha=0$}
\end{subfigure}
\begin{subfigure}[b]{0.195\textwidth}
\includegraphics[width=\textwidth, keepaspectratio=true]{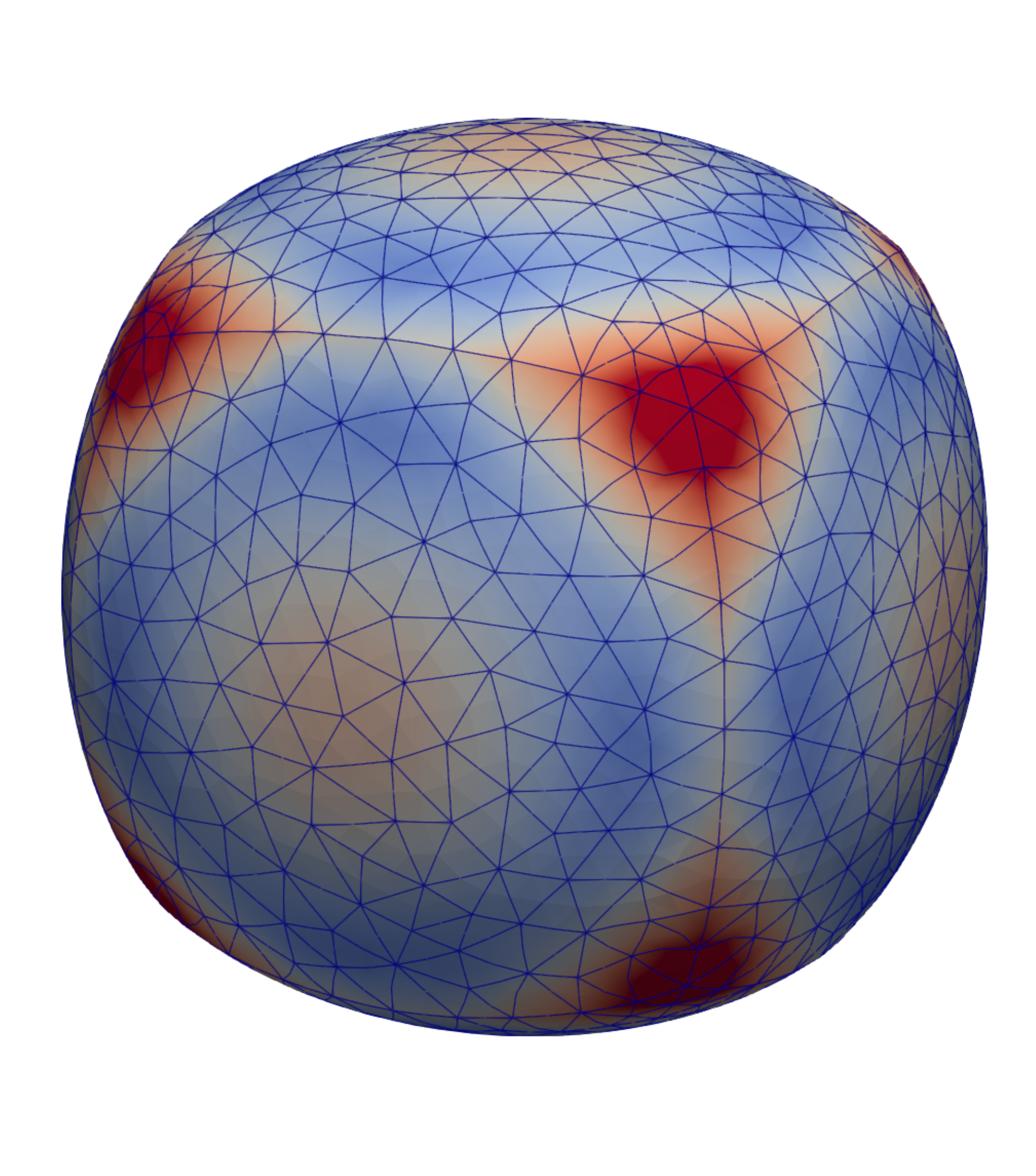}
\subcaption{$\tilde\gamma=5.12$, $\tilde\alpha=0$}
\end{subfigure} 
\begin{subfigure}[b]{0.195\textwidth}
\includegraphics[width=\textwidth, keepaspectratio=true]{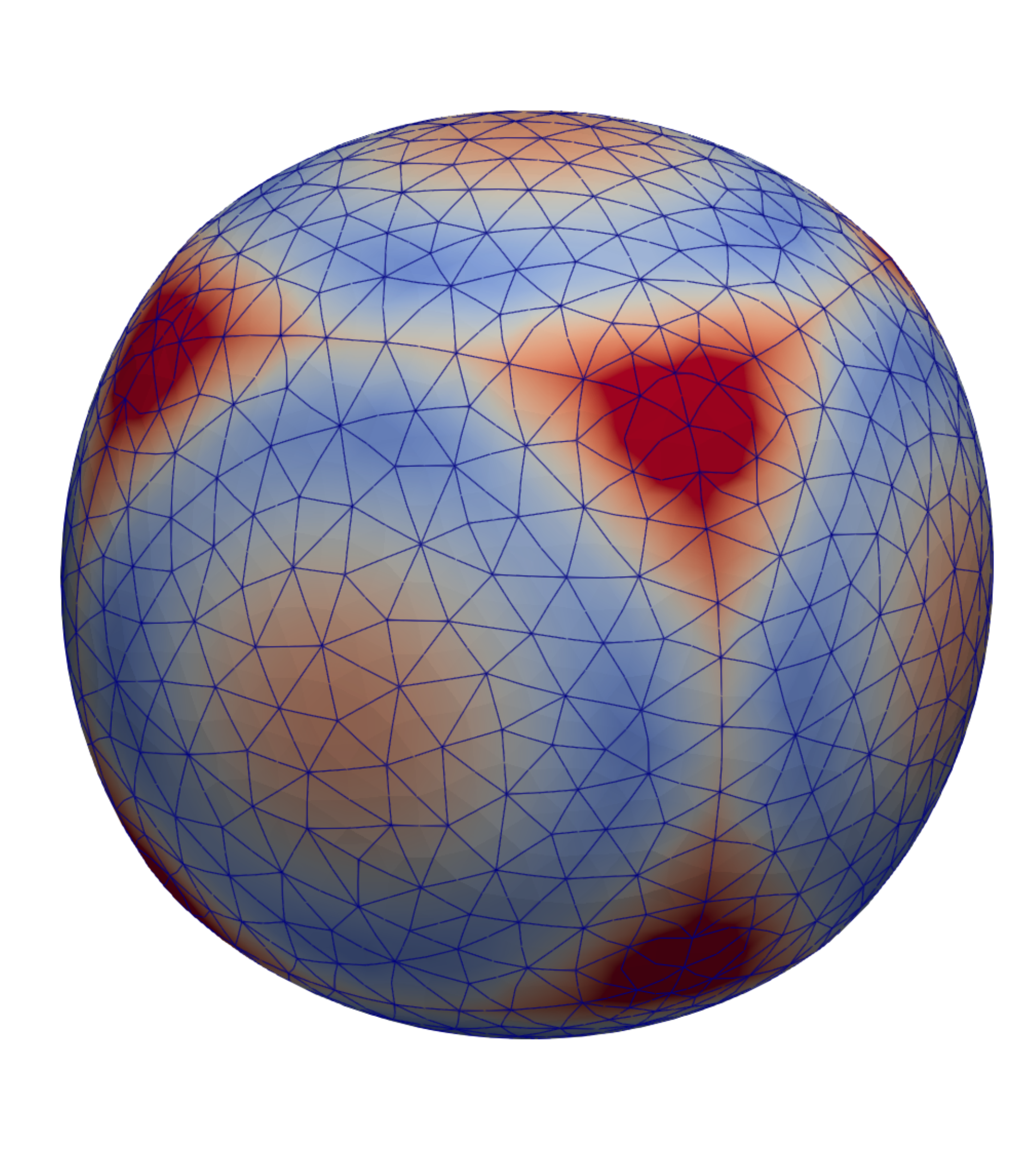}
\subcaption{$\tilde\gamma=13.31$, $\tilde\alpha=0$}
\label{fig:cubeSTd}
\end{subfigure} 
\\ [0.5cm] 
\begin{subfigure}[b]{0.195\textwidth}
\includegraphics[width=\textwidth, keepaspectratio=true]{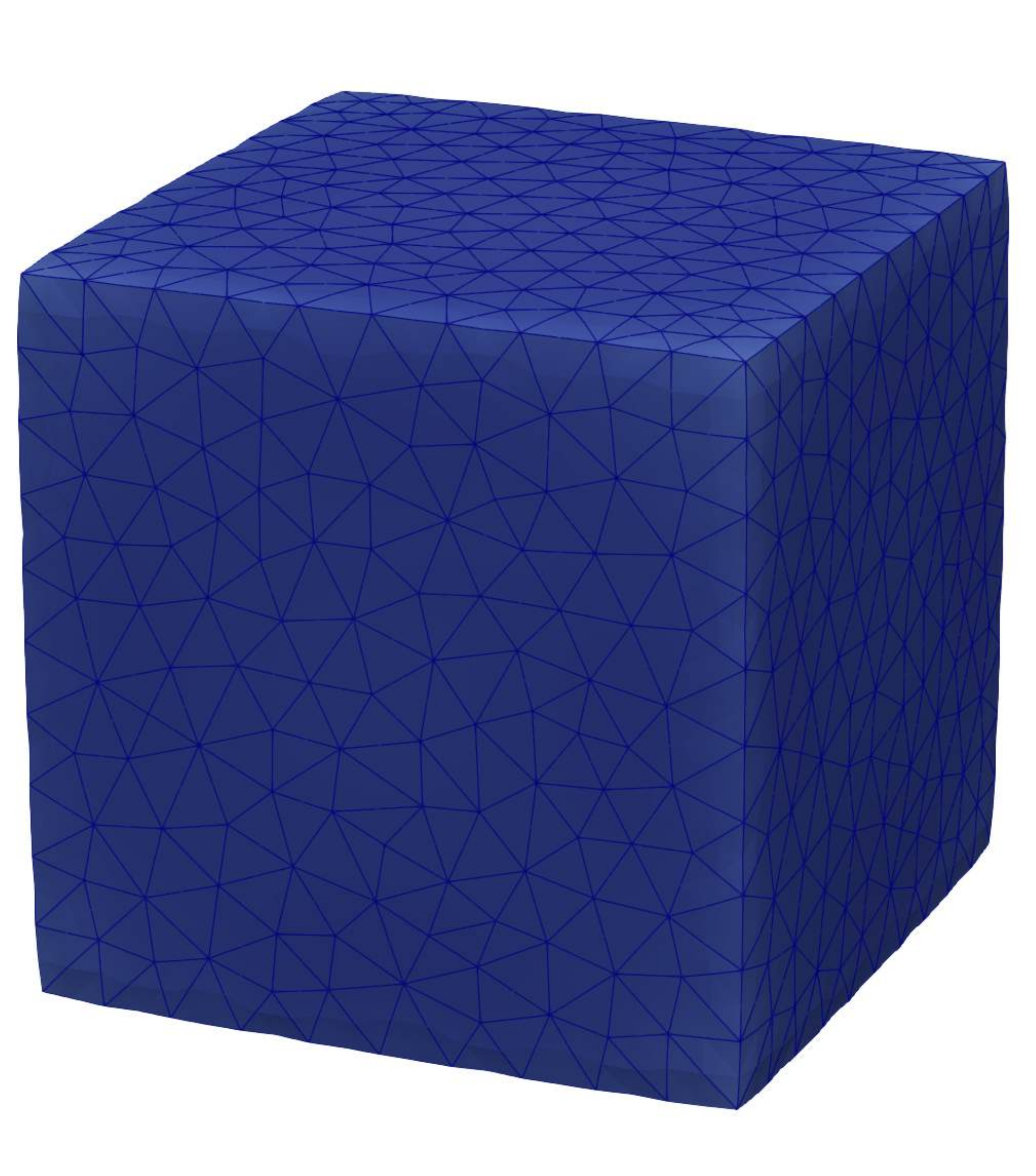}
\subcaption{$\tilde\alpha=0.08$, $\tilde\gamma=0$}
\label{fig:cubeFa}
\end{subfigure}
\begin{subfigure}[b]{0.195\textwidth}
\includegraphics[width=\textwidth, keepaspectratio=true]{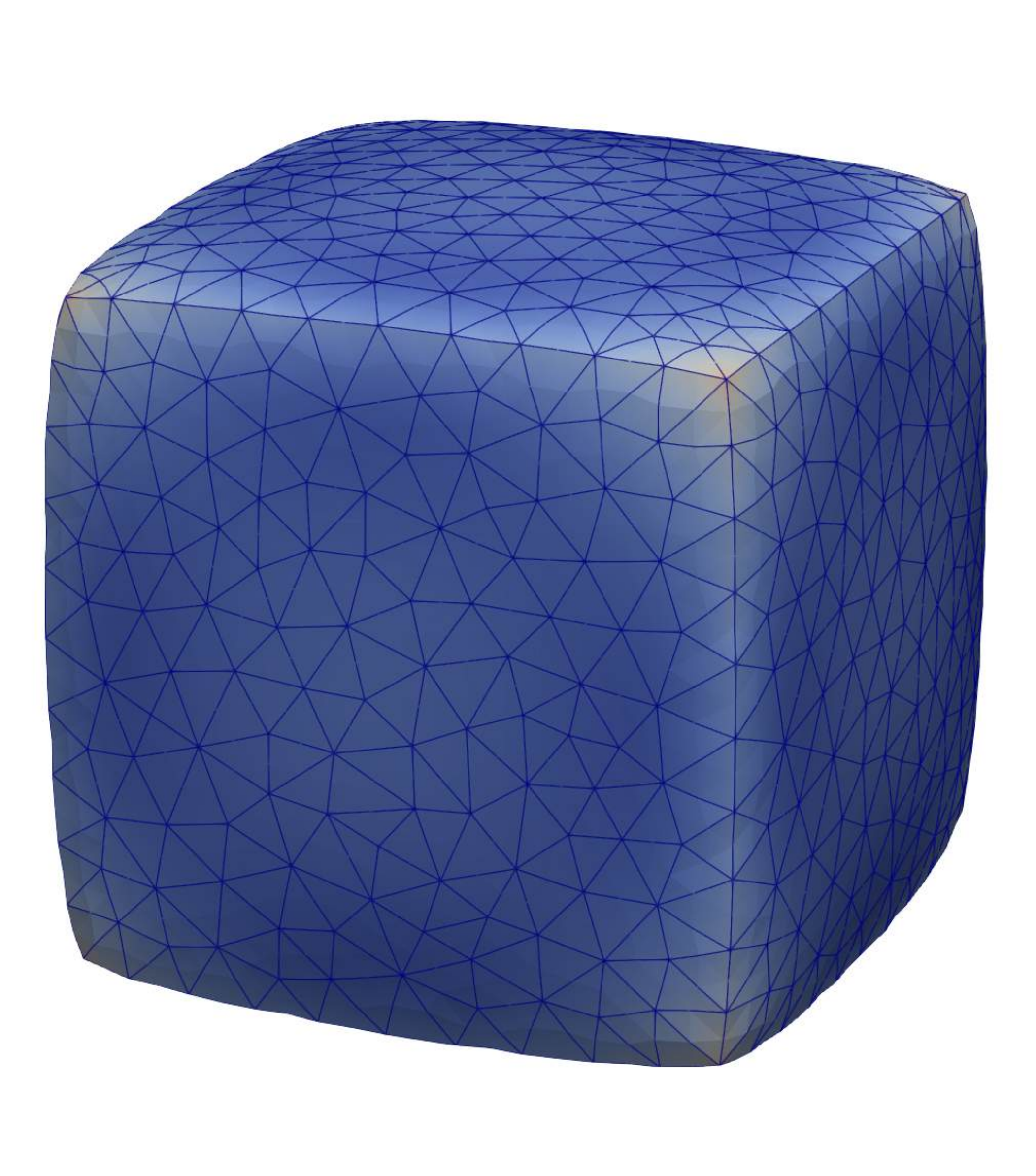}
\subcaption{$\tilde\alpha=0.64$, $\tilde\gamma=0$}
\end{subfigure}
\begin{subfigure}[b]{0.195\textwidth}
\includegraphics[width=\textwidth, keepaspectratio=true]{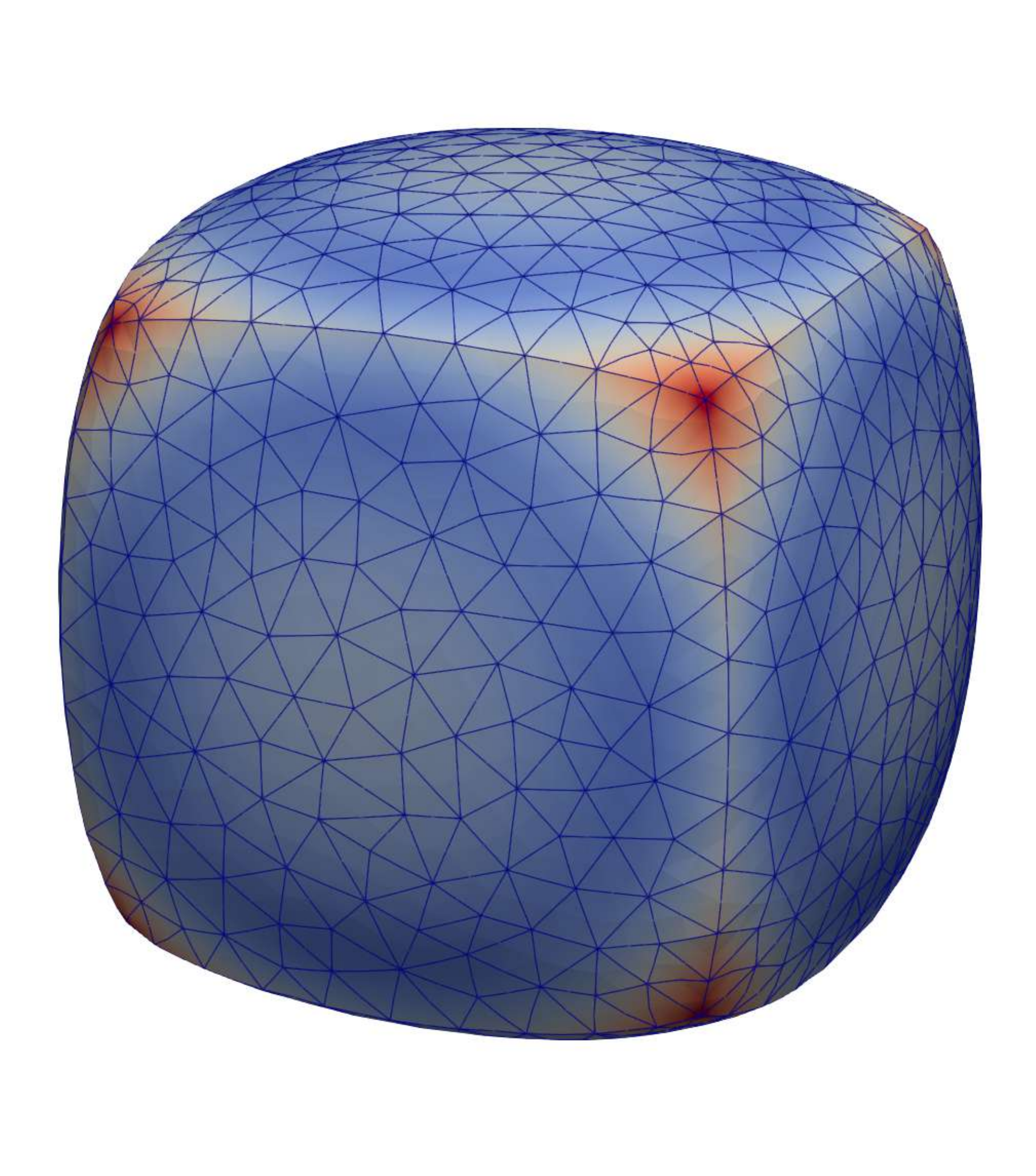}
\subcaption{$\tilde\alpha=2.16$, $\tilde\gamma=0$}
\end{subfigure}
\begin{subfigure}[b]{0.195\textwidth}
\includegraphics[width=\textwidth, keepaspectratio=true]{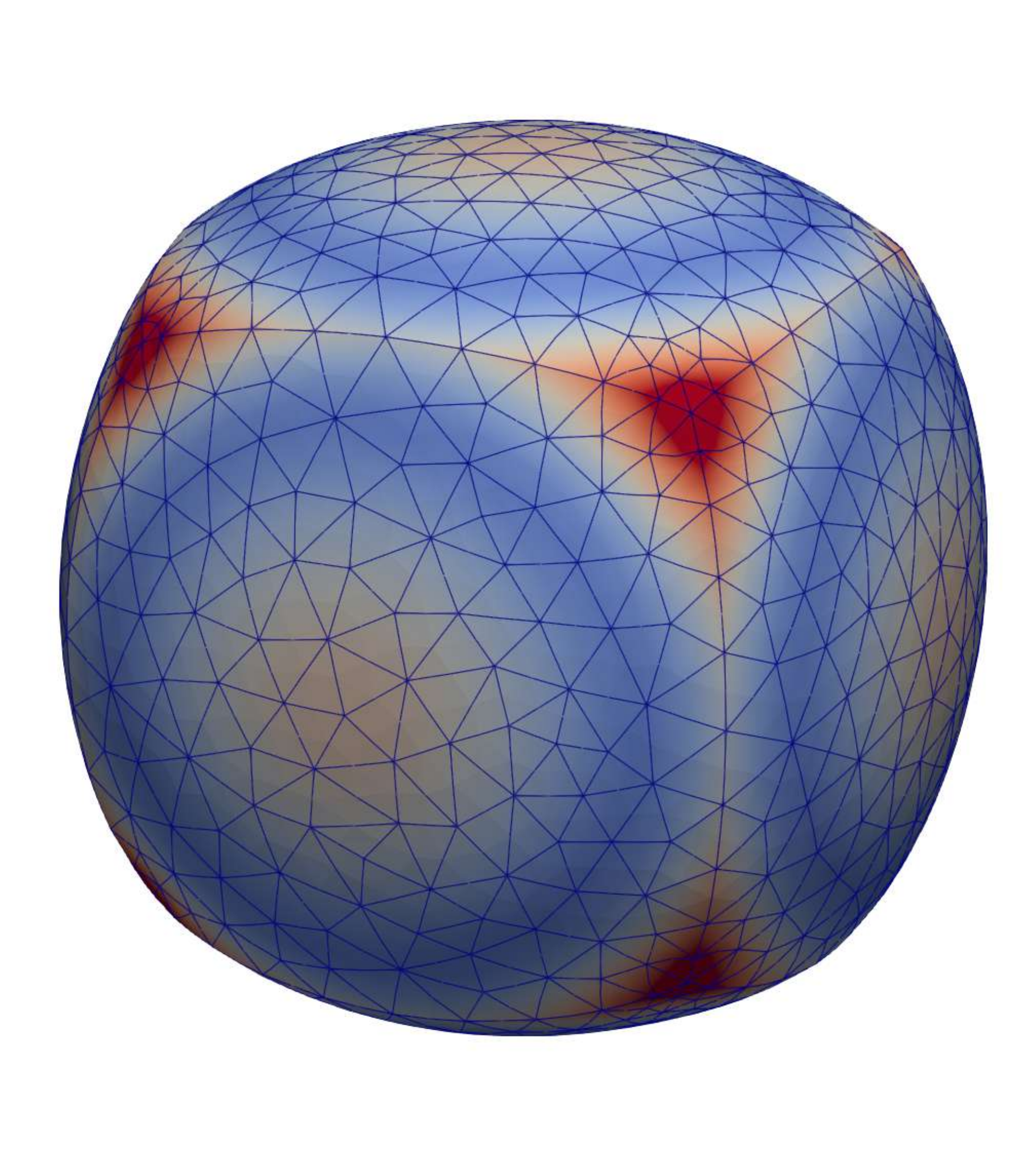}
\subcaption{$\tilde\alpha=5.12$, $\tilde\gamma=0$}
\end{subfigure} 
\begin{subfigure}[b]{0.195\textwidth}
\includegraphics[width=\textwidth, keepaspectratio=true]{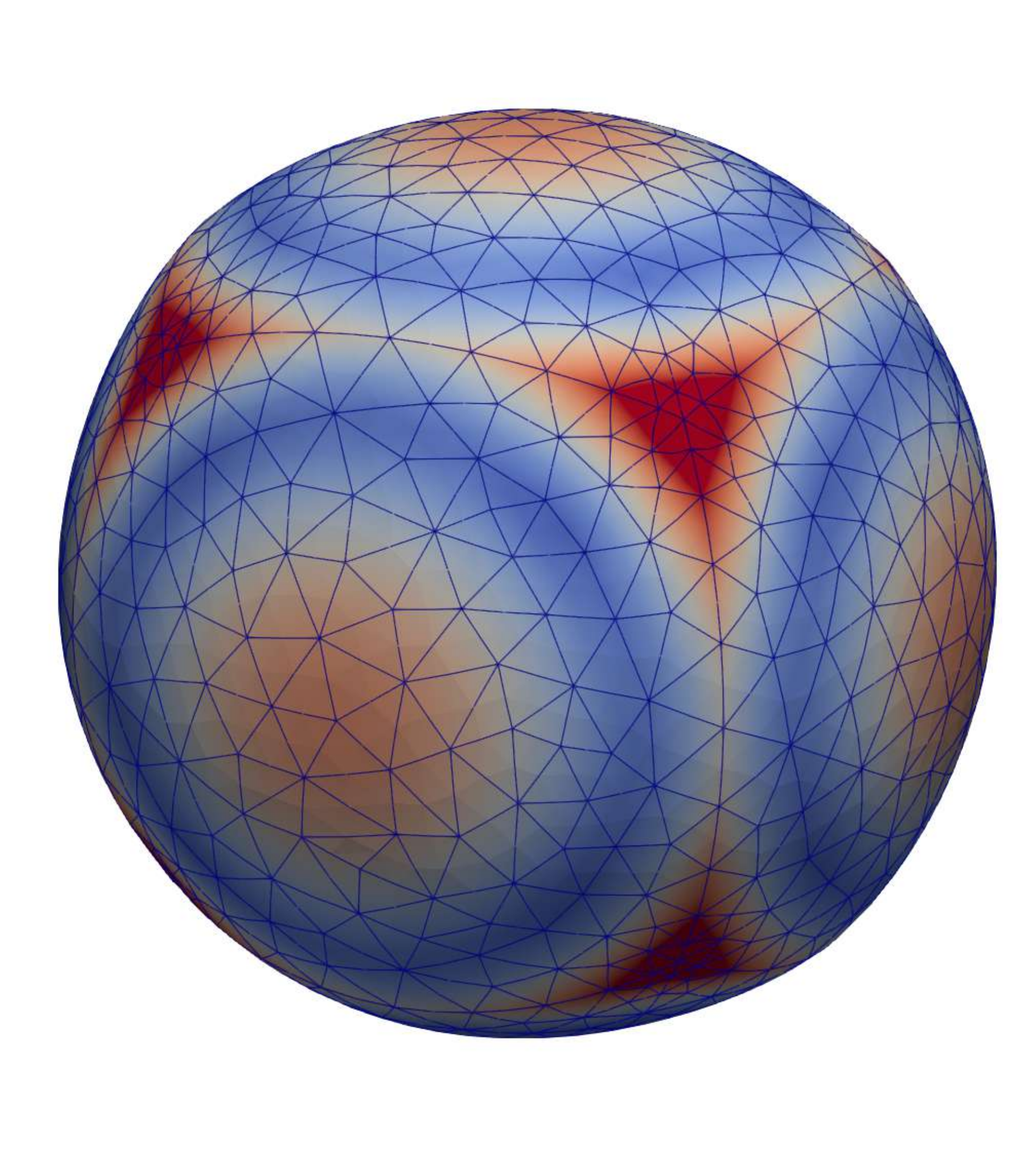}
\subcaption{$\tilde\alpha=13.31$, $\tilde\gamma=0$}
\label{fig:cubeFd}
\end{subfigure} 
\\ 
 [0.5cm] 
\begin{subfigure}[b]{0.49\textwidth}
\includegraphics[width=\textwidth, keepaspectratio=true]{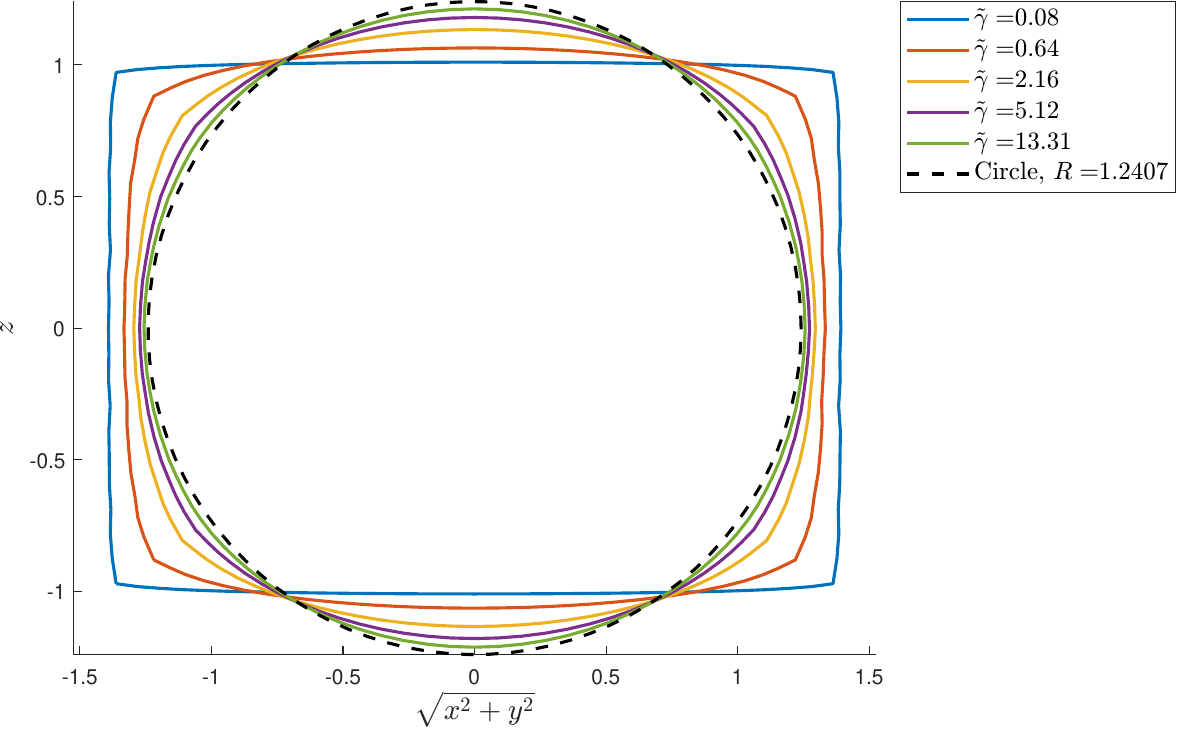}
\subcaption{Outline of the diagonal section of the cube with the surface tension model applied on the surface.}
\label{fig:cube_comp_a}
\end{subfigure}
\begin{subfigure}[b]{0.49\textwidth}
\includegraphics[width=\textwidth, keepaspectratio=true]{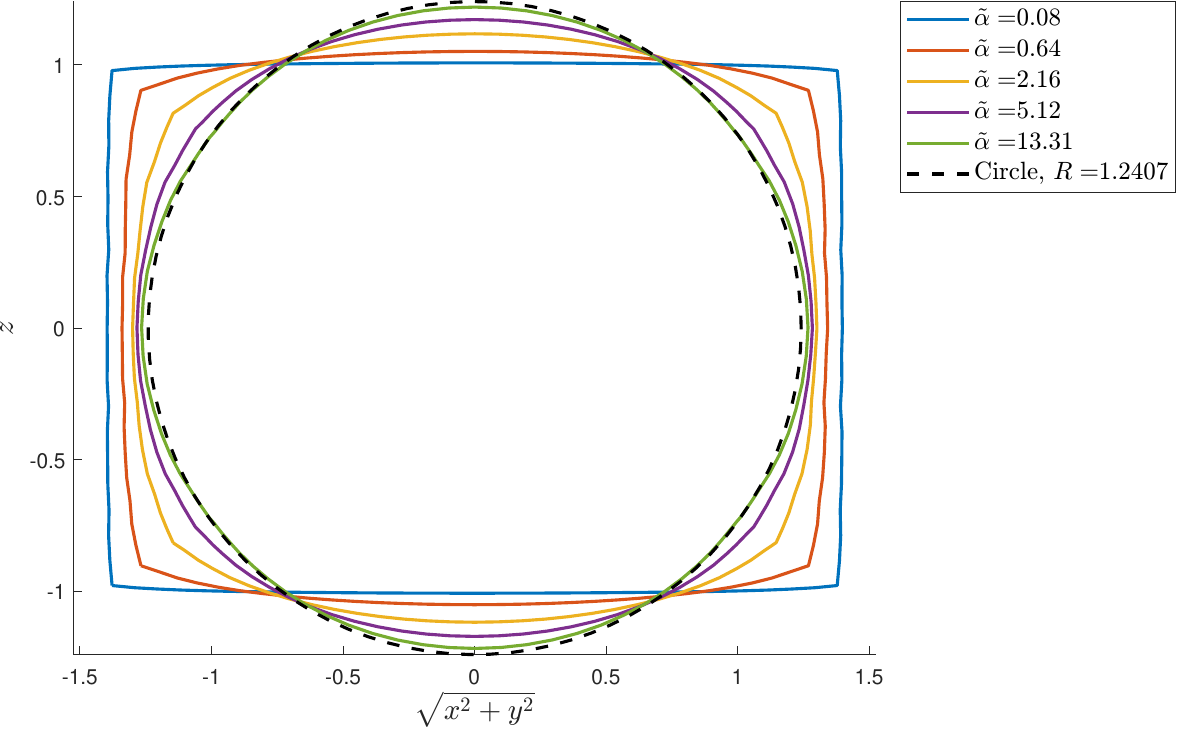}
\subcaption{Outline of the diagonal section of the cube with the strain-dependent model applied on the surface.}
\label{fig:cube_comp_b}
\end{subfigure}
\caption{The surface material parameters are increasingly applied on the surface of an incompressible cube. In Figures \ref{fig:cubeSTa} - \ref{fig:cubeSTd}, the surface tension parameter $\tilde\gamma$ is increased and nonzero, while in Figures \ref{fig:cubeFa} - \ref{fig:cubeFd} increasing strain dependent parameter $\tilde \alpha$ is considered.
Figures \ref{fig:cube_comp_a} and \ref{fig:cube_comp_b} show a comparison of the outlines of the diagonal sections of the cube in the various time steps for both considered models. The outlines are compared to the expected circular profile with a radius corresponding to a sphere of the same volume as the original cube, i.e., with $R=(3/4/\pi a^3)^{1/3}$.  
}
\label{fig_cube_all}
\end{figure}

\subsubsection{Incompressible sphere with anisotropic surface material}
\label{sec:sphere}
In the present section, we test the effect of the anisotropic surface-polyconvex model on the deformation of an incompressible sphere of radius $r$. The unit tangential vector field $\bm{a}$ defining the direction of the anisotropy is given by unit tangents to the parallels of the sphere, see the sketch in Figure \ref{fig:scheme_sphere}. The anisotropic polyconvex model of general form $\tilde \Psi_s = \gamma \hat{J} +\alpha \left|\left|\hat{\bm{F}} \right|\right| +
\eta \left|\left| \hat{\bm{F}} \bm{a} \right|\right| + \beta  \sqrt{\left|\left| \hat{\bm{F}}^T \hat{\bm{F}} \bm{a} \right|\right|}$ is considered in four different variants, always keeping one isotropic and one anisotropic term nonzero, meaning the combinations of nonzero parameters are $(\gamma,\eta)$, $(\gamma,\beta)$, $(\alpha,\eta)$, and $(\alpha,\beta)$. We describe the loading procedure on the combination with nonzero parameters $\gamma$ and $\eta$, but the strategy is analogous for other combinations. 
Firstly, the isotropic parameter $\gamma$ is incrementally increased to reach the value of $\tilde\gamma = \gamma/(\mu r) = 10$. Simultaneously, the ratio of the bulk material parameters $\tilde\kappa = \kappa/\mu$ is increased, reaching the final value of $\tilde\kappa = 1\cdot 10^5$. Subsequently, the anisotropic parameter $\eta$ is increased so that the ratio $\tilde\eta = \eta/\gamma \in \{0, 0.5,1,1.5\}$.

In the simulations, the advantage of rotational symmetry was used so that only one eighth of the whole sphere was discretized into 20394 bulk linear tetrahedron elements and 1710 linear triangular surface elements with a total of 4390 nodes. 

In Figure \ref{fig_sphere_Aniso_Outline}, the outlines of the deformed part of the sphere are displayed for all four "loading" combinations. In Figure \ref{fig_sphere_Aniso_Outline_A}, the combinations including the surface tension model as the isotropic part are visualized. In contrast, in \ref{fig_sphere_Aniso_Outline_B}, the results for the parameter combinations with deformation-dependent isotropic term are plotted. In both plots, the solid lines are associated with the combination for which the anisotropic parameter $\eta=0$, while cross markers are associated with combinations corresponding to vanishing parameter $\beta$.       
\begin{figure}[ht]
\centering%
\begin{subfigure}[b]{0.49\textwidth}
\includegraphics[width=\textwidth, keepaspectratio=true]{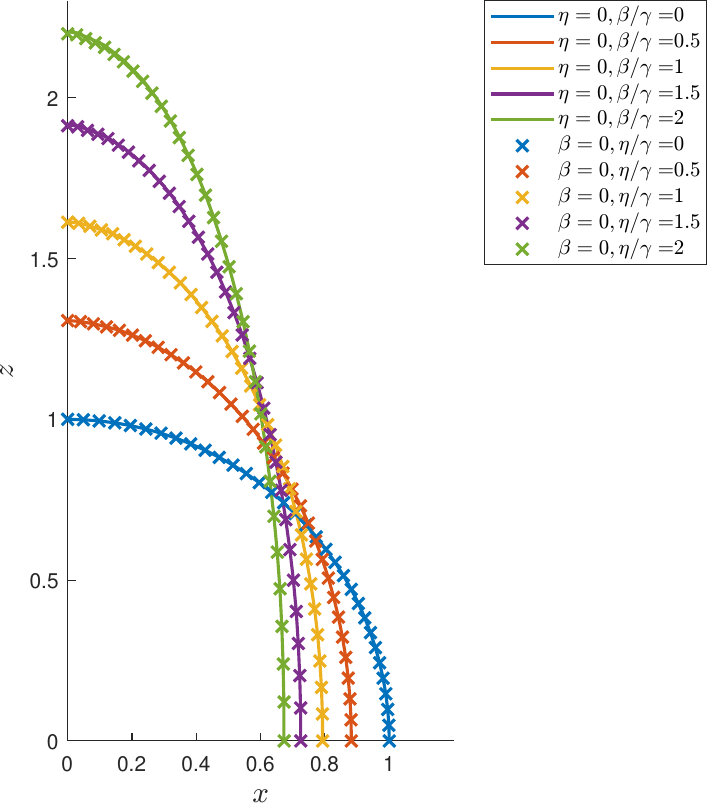}
\subcaption{Surface energy of form $ \tilde \Psi_s = \gamma \hat{J} +
\eta \left|\left| \hat{\bm{F}} \bm{a} \right|\right| + \beta  \sqrt{\left|\left| \hat{\bm{F}}^T \hat{\bm{F}} \bm{a} \right|\right|}$}
\label{fig_sphere_Aniso_Outline_A}
\end{subfigure}
\begin{subfigure}[b]{0.49\textwidth}
\includegraphics[width=\textwidth, keepaspectratio=true]{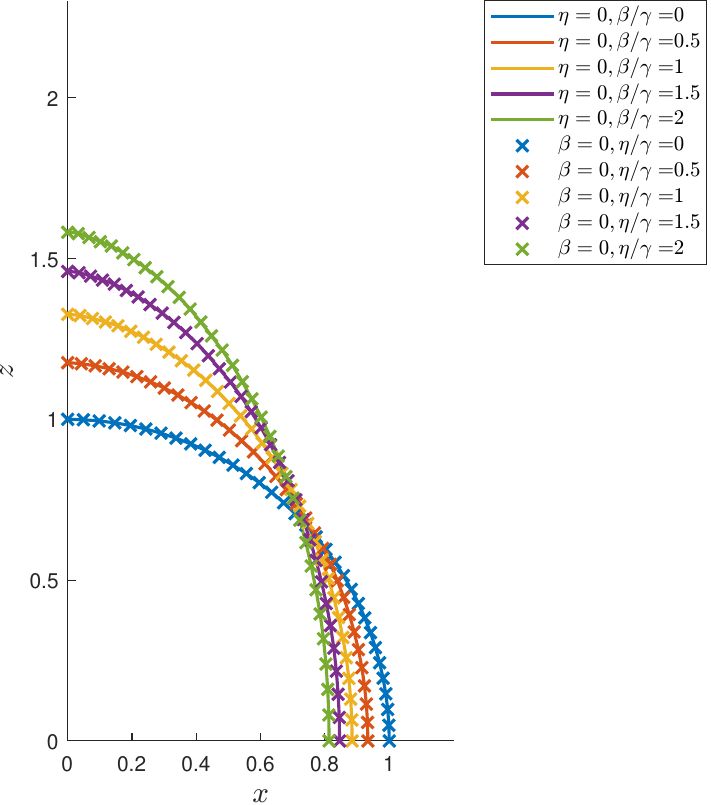}
\subcaption{Surface energy of form $\tilde \Psi_s = \alpha \left|\left|\hat{\bm{F}} \right|\right| +
\eta \left|\left| \hat{\bm{F}} \bm{a} \right|\right| + \beta  \sqrt{\left|\left| \hat{\bm{F}}^T \hat{\bm{F}} \bm{a} \right|\right|}$}
\label{fig_sphere_Aniso_Outline_B}
\end{subfigure}
\caption{A sphere made of incompressible material with anisotropic surface energy is deformed by increasing the ratio of the surface to bulk energy material parameters. 
The ratio of the bulk modulus to the shear parameter is $\tilde\kappa=\kappa/\mu=1\cdot 10^5$. In each of the considered options, the isotropic parameter ($\gamma$ or $\alpha$) is incrementally increased to reach the value $\tilde\gamma = 10$ (or $\tilde\alpha = 10$). Subsequently, the anisotropic parameter ($\eta$ or $\beta$) is increased so that the ratio $\eta/\gamma \in \{0, 0.5,1,1.5\}$ (or $\beta/\gamma$).  
}   
\label{fig_sphere_Aniso_Outline}
\end{figure} 

The results indicate that with increasing anisotropic parameters, the originally spherical shape tends to deform into an ellipsoid-like object with the longer half-axis aligned with the z-axis, i.e., orthogonal to the direction of anisotropy. Moreover, both anisotropic models seem to produce very similar results (even though the results seem to be identical, there is a discrepancy). This, however, does not hold in general. This can be justified by expressing the surface deformation gradient and surface right Cauchy-Green tensor in principal directions
\bea
\hat{\bm{F}} = \lambda_1 \bm{u}_1 \otimes \bm{U}_1 + \lambda_2 \bm{u}_2 \otimes \bm{U}_2 \\
\hat{\bm{C}} = \lambda_1^2 \bm{U}_1 \otimes \bm{U}_1 + \lambda_2^2 \bm{U}_2 \otimes \bm{U}_2 
\eea
with $\lambda_i$, $\bm{U}_i$, and $\bm{u}_i$ being the principal stretches and referential and deformed unit principal directions. Realizing that for the chosen geometry and direction of anisotropy, the principal direction $\bm{U}_1(\bm{X})$ coincides with the vector characterizing the anisotropy $\bm{a}(\bm{X})$ and therefore $\bm{U}_2 \bm{a} = 0$. Consequently, the two generally different terms $\left|\left| \hat{\bm{F}} \bm{a} \right|\right|$ and $\sqrt{\left|\left| \hat{\bm{C}}  \bm{a} \right|\right|}$ are in this case equal since
\bea
\left|\left| \hat{\bm{F}} \bm{a} \right|\right| = \left|\left| \lambda_1 \bm{u}_1 \right|\right| = \lambda_1 = \sqrt{\left|\left| \lambda_1^2 \bm{U}_1 \right|\right|}  =\left|\left| \hat{\bm{C}} \bm{a} \right|\right| 
\eea
The small discrepancy is caused by the discretization error due to which the term $\bm{U}_2 \bm{a}$ does not vanish. 

Based on the previous discussion, it is also clear why the deformed body is of ellipsoid-like shape. 
The extra anisotropic term $\left|\left| \hat{\bm{F}} \bm{a} \right|\right| = \lambda_1$ in the energy causes the deformation to prioritize shapes with smaller principal stretch in the circumferential direction of the anisotropy.

The 3D deformed shapes composed of the eight identical parts are visualized in Figure \ref{fig:sphere_3D} for the case with an anisotropic part governed only by the term $\beta  \sqrt{\left|\left| \hat{\bm{F}}^T \hat{\bm{F}} \bm{a} \right|\right|}$. In Figures \ref{fig:sphere_3D_ST_A}-\ref{fig:sphere_3D_ST_E}, the results obtained with the surface-tension-like isotropic part are displayed, and in Figures \ref{fig:sphere_3D_F_A}-\ref{fig:sphere_3D_F_E}, the results correspond to the deformation-dependent isotropic part. Even though the whole shapes are displayed, the simulations were only performed on one eighth of the sphere.  
\begin{figure}[ht]
\centering%
\begin{subfigure}[b]{0.195\textwidth}
\includegraphics[width=\textwidth, keepaspectratio=true]{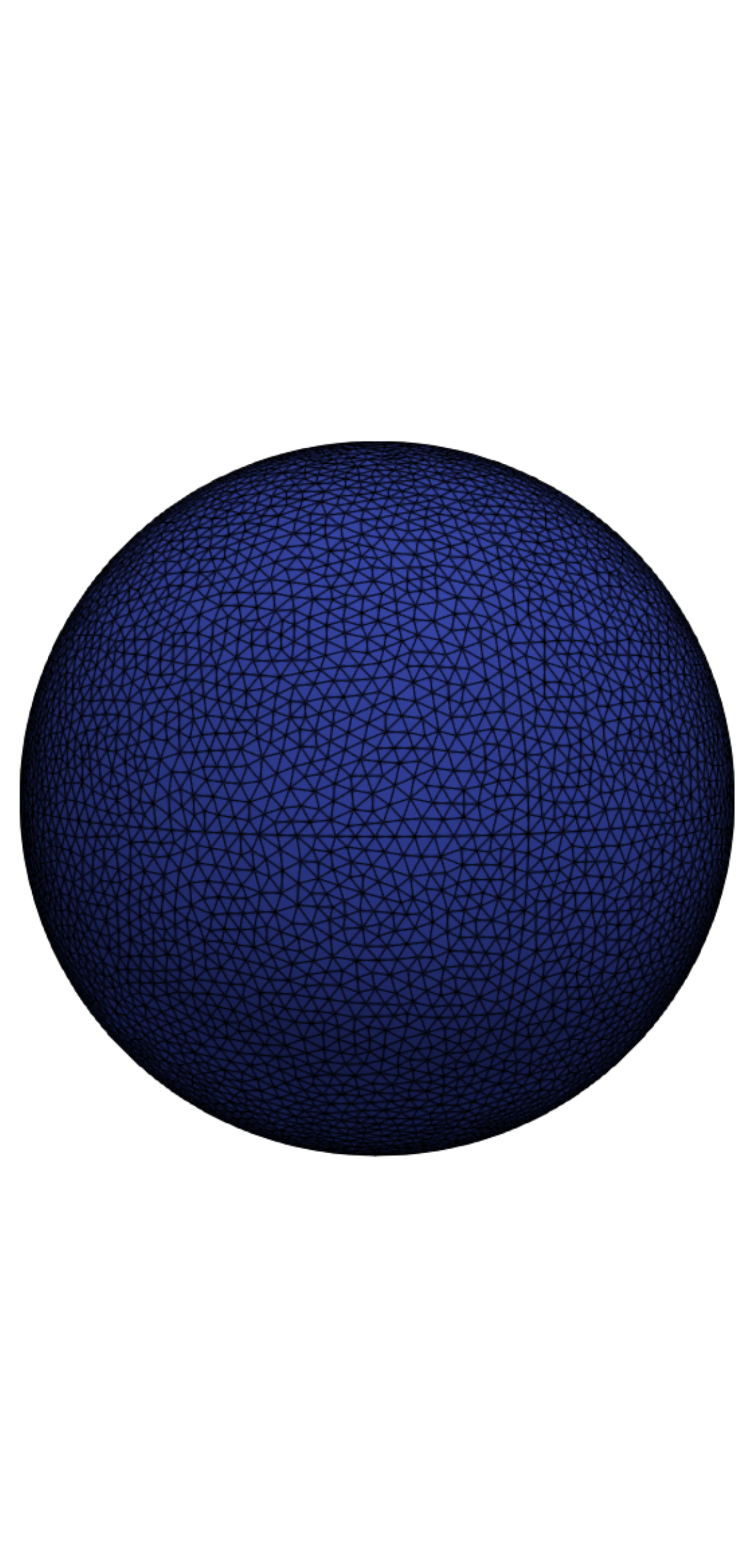}
\subcaption{$\gamma \neq 0$, $\alpha = 0$, $\beta/\gamma=0$}
\label{fig:sphere_3D_ST_A}
\end{subfigure}
\begin{subfigure}[b]{0.195\textwidth}
\includegraphics[width=\textwidth, keepaspectratio=true]{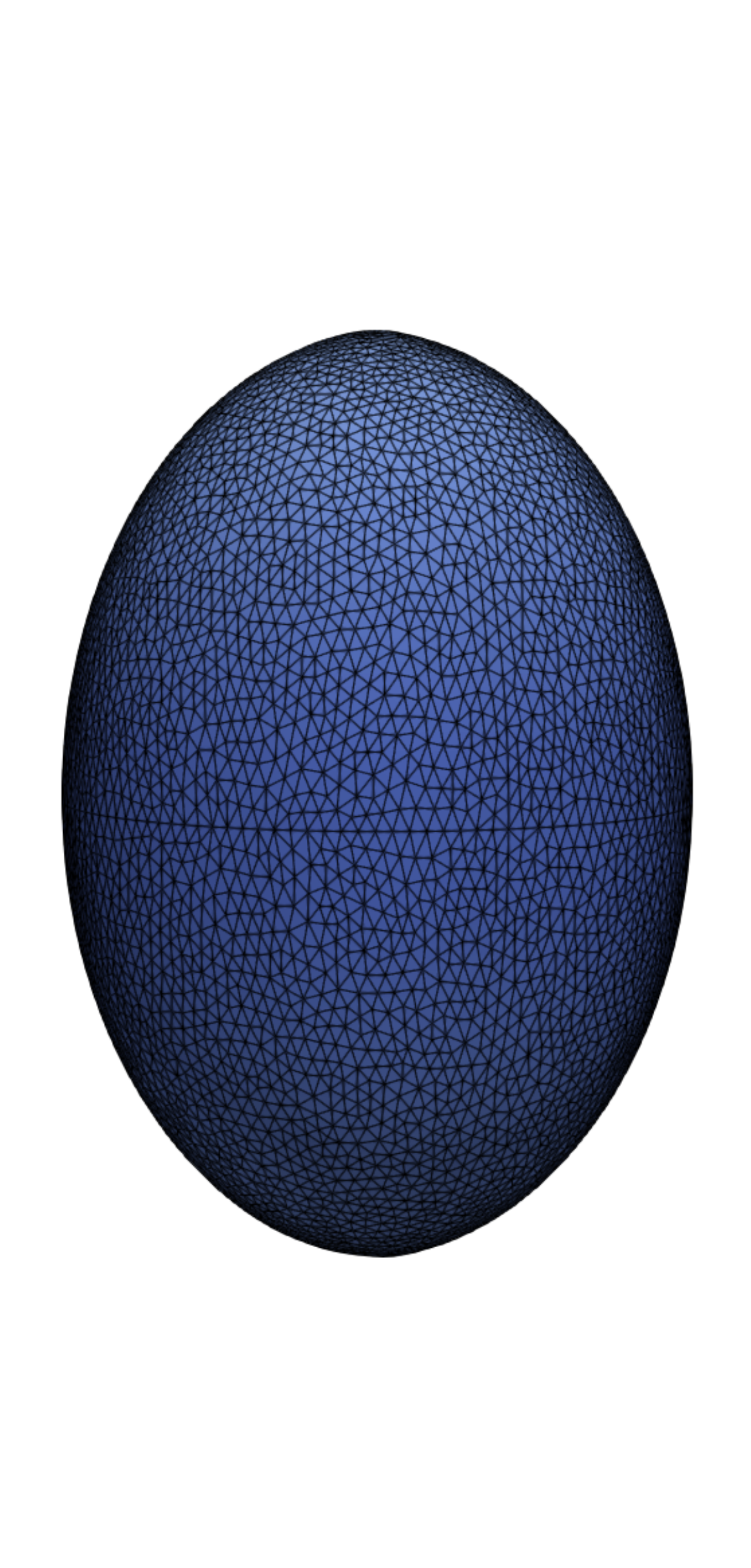}
\subcaption{$\gamma \neq 0$, $\alpha = 0$, $\beta/\gamma=0.5$}
\end{subfigure}
\begin{subfigure}[b]{0.195\textwidth}
\includegraphics[width=\textwidth, keepaspectratio=true]{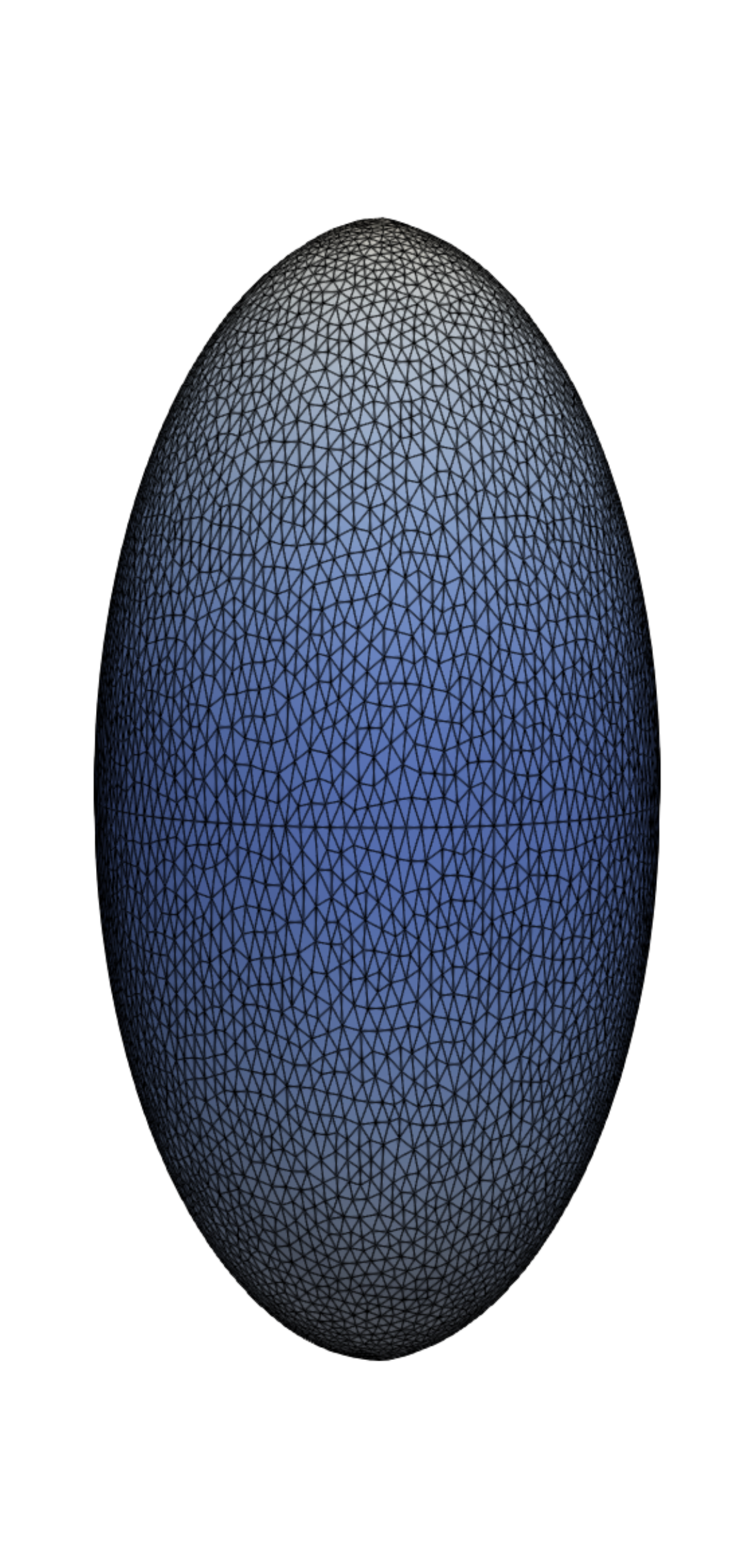}
\subcaption{$\gamma \neq 0$, $\alpha = 0$, $\beta/\gamma=1$}
\end{subfigure}
\begin{subfigure}[b]{0.195\textwidth}
\includegraphics[width=\textwidth, keepaspectratio=true]{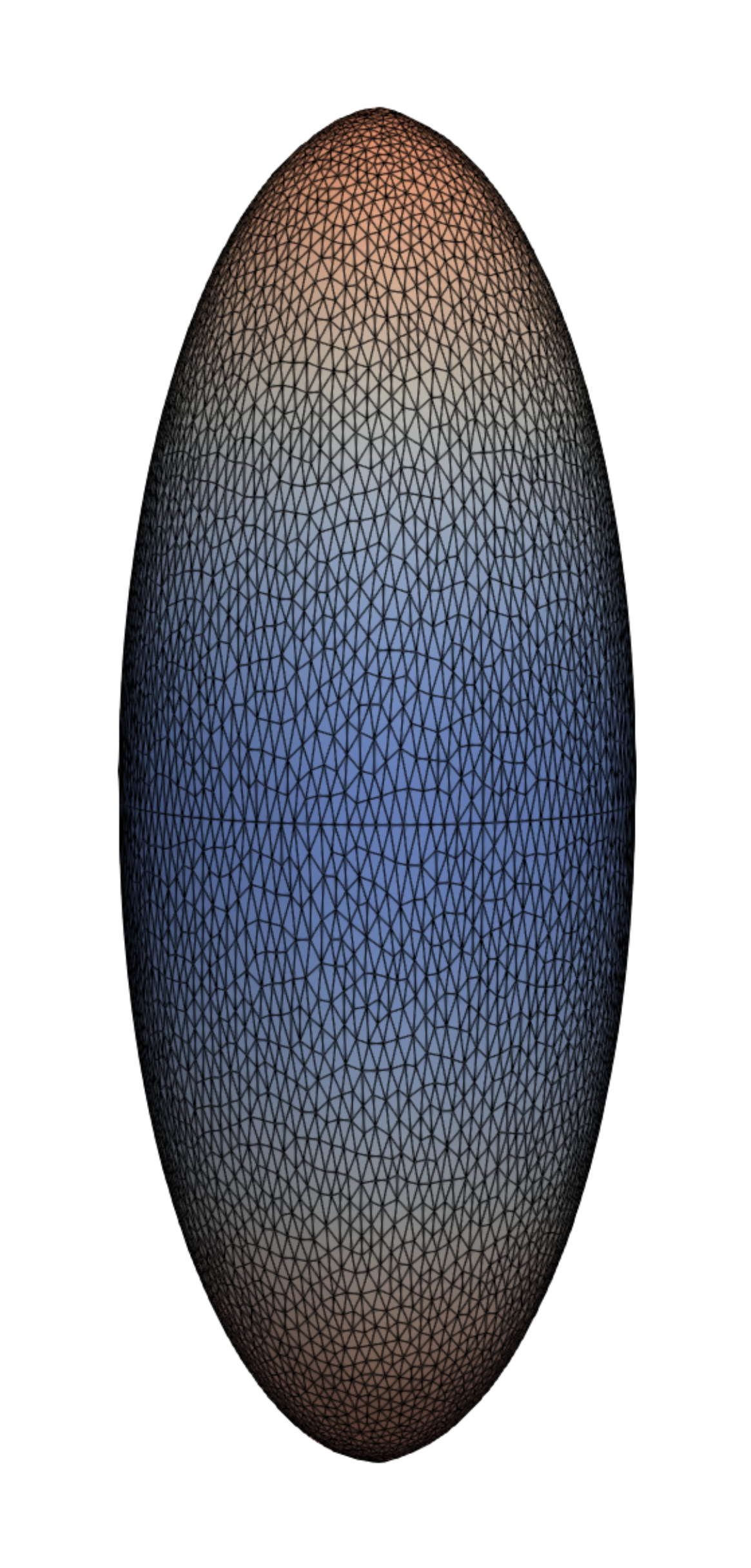}
\subcaption{$\gamma \neq 0$, $\alpha = 0$, $\beta/\gamma=1.5$}
\end{subfigure} 
\begin{subfigure}[b]{0.195\textwidth}
\includegraphics[width=\textwidth, keepaspectratio=true]{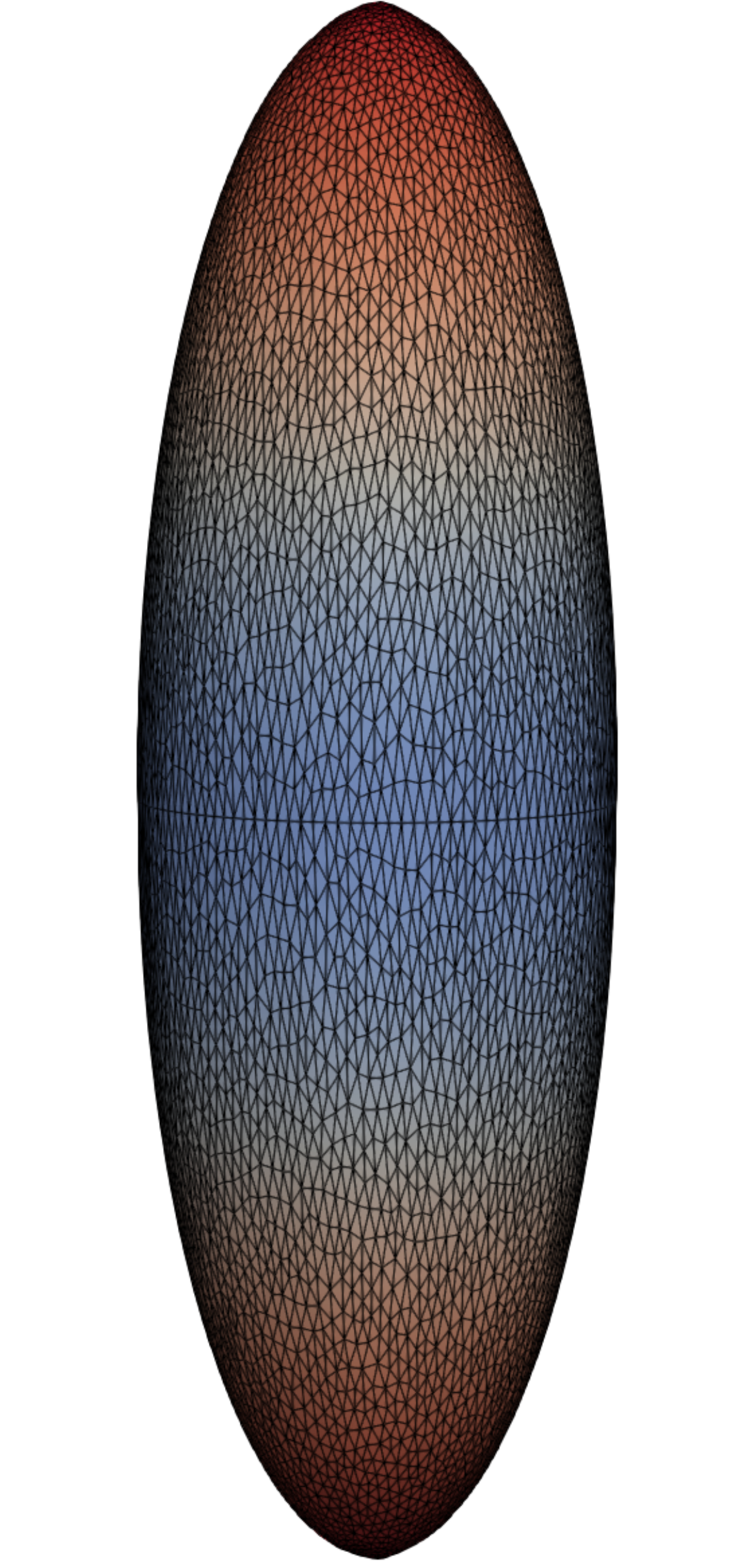}
\subcaption{$\gamma \neq 0$, $\alpha = 0$, $\beta/\gamma=2$}
\label{fig:sphere_3D_ST_E}
\end{subfigure} 
\\ 
\begin{subfigure}[b]{0.195\textwidth}
\includegraphics[width=\textwidth, keepaspectratio=true]{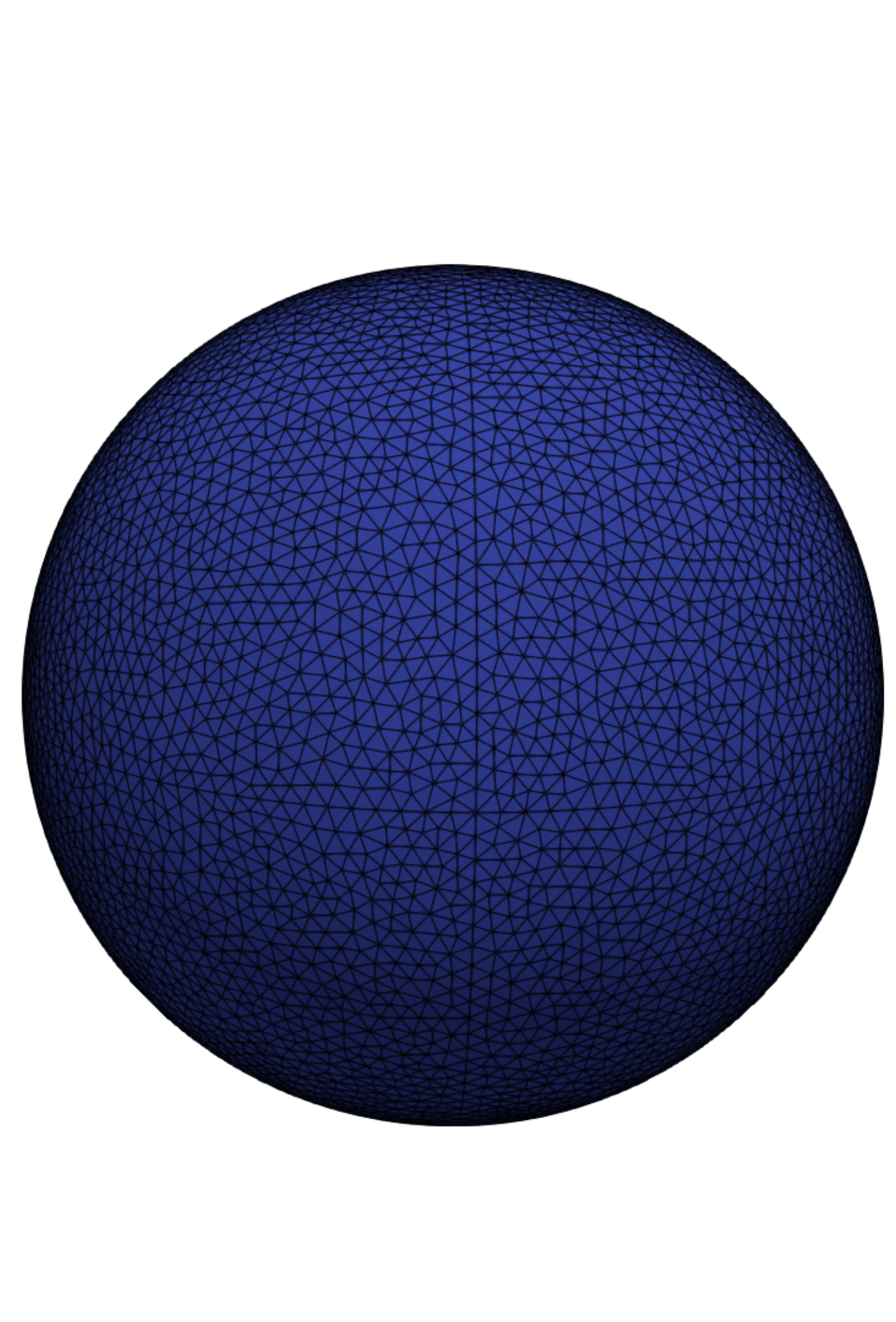}
\subcaption{$\gamma = 0$, $\alpha \neq 0$, $\beta/\alpha=0$}
\label{fig:sphere_3D_F_A}
\end{subfigure}
\begin{subfigure}[b]{0.195\textwidth}
\includegraphics[width=\textwidth, keepaspectratio=true]{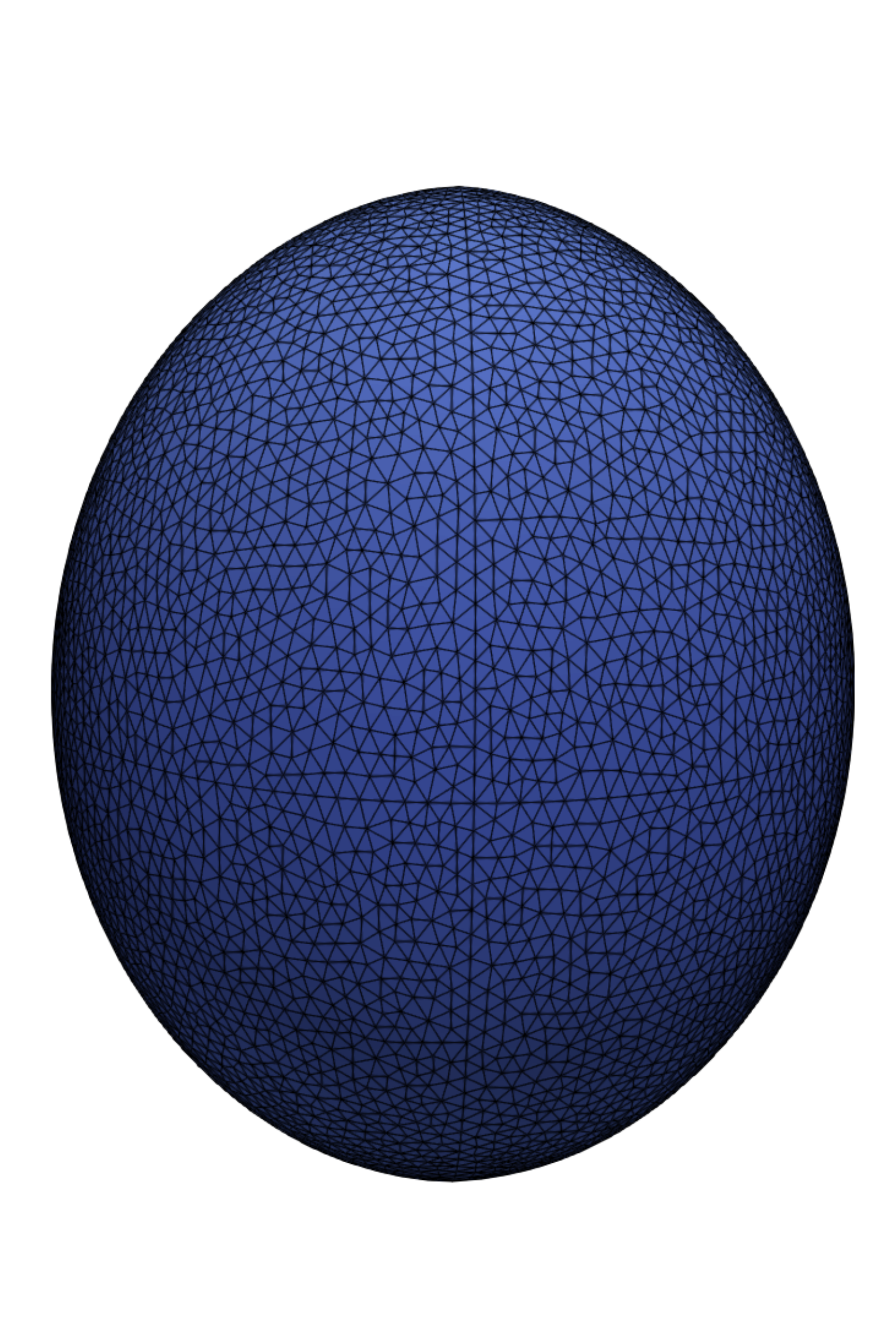}
\subcaption{$\gamma = 0$, $\alpha \neq 0$, $\beta/\alpha=0.5$}
\end{subfigure}
\begin{subfigure}[b]{0.195\textwidth}
\includegraphics[width=\textwidth, keepaspectratio=true]{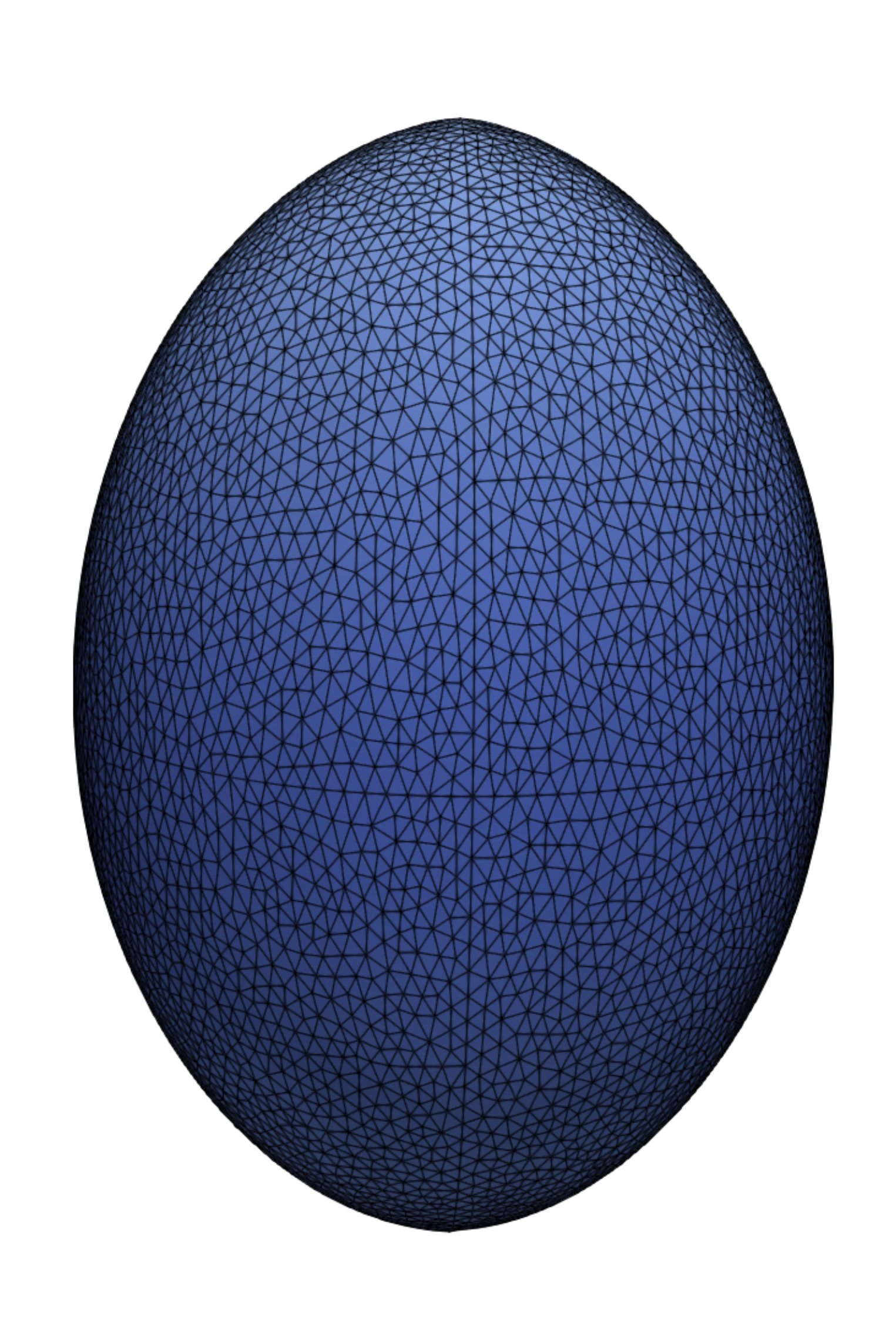}
\subcaption{$\gamma = 0$, $\alpha \neq 0$, $\beta/\alpha=1$}
\end{subfigure}
\begin{subfigure}[b]{0.195\textwidth}
\includegraphics[width=\textwidth, keepaspectratio=true]{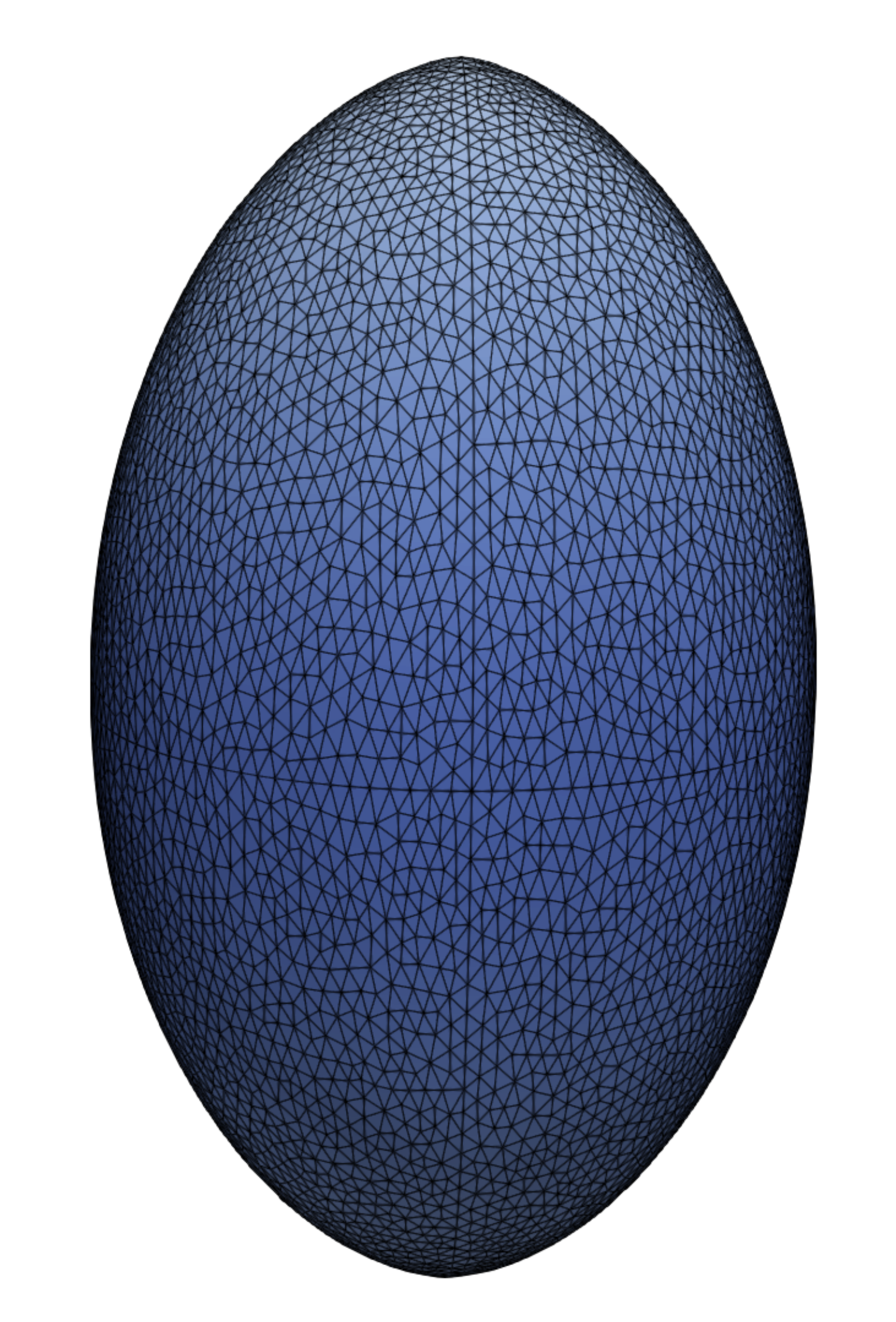}
\subcaption{$\gamma = 0$, $\alpha \neq 0$, $\beta/\alpha=1.5$}
\end{subfigure} 
\begin{subfigure}[b]{0.195\textwidth}
\includegraphics[width=\textwidth, keepaspectratio=true]{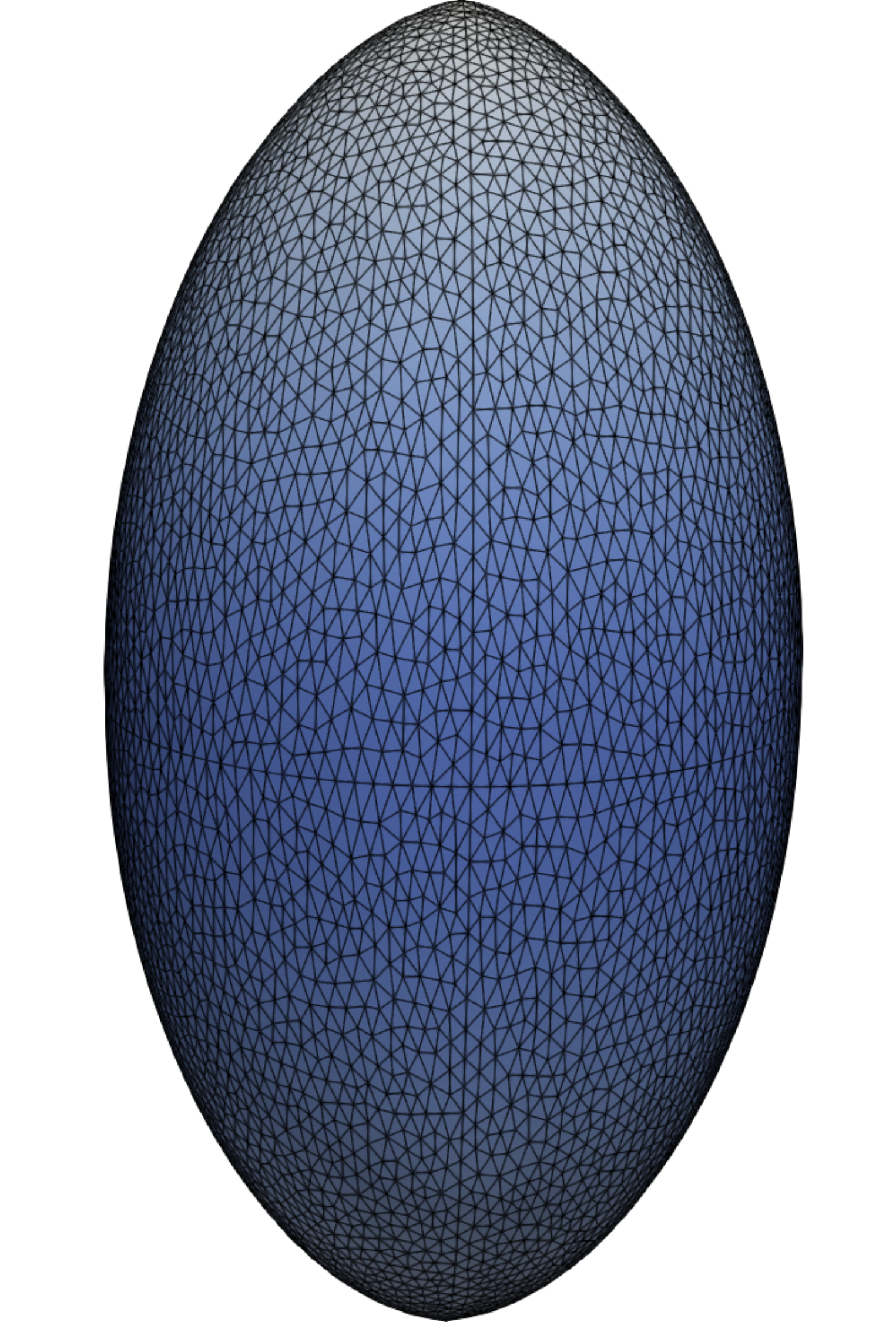}
\subcaption{$\gamma = 0$, $\alpha \neq 0$, $\beta/\alpha=2$}
\label{fig:sphere_3D_F_E}
\end{subfigure} 
\caption{A sphere made of incompressible material with anisotropic surface energy is deformed by increasing the ratio of the surface and the bulk material energy parameters.
}   
\label{fig:sphere_3D}
\end{figure} 
\subsubsection{Stability analysis and bifurcation} \label{Sec:Stability analysis and bifurcation}
In this section, we study the effect of surface energy on the formation of instabilities in a long cylinder. Contrary to the first two examples, the cylinders' sides are only fixed in the horizontal direction, except for the points on the axis of rotation, which are also fixed in the vertical direction to prevent rigid body motions. The cylinder again is of radius $R$ and length $L$; its scheme is depicted in Figure \ref{fig10}.
\begin{figure}[ht]
\centering%
\begin{subfigure}[b]{0.49\textwidth}
\includegraphics[width=\textwidth, keepaspectratio=true]{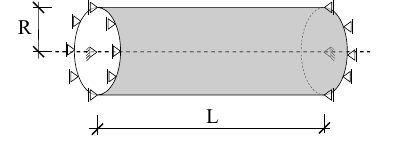}
\subcaption{Full model}
\label{fig10}
\end{subfigure}
\begin{subfigure}[b]{0.49\textwidth}
\includegraphics[width=\textwidth, keepaspectratio=true]{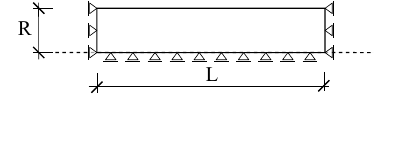}
\subcaption{Axisymmetric view}
\label{fig10b}
\end{subfigure}
\caption{Scheme of a long cylinder.}     
\end{figure} 
In this test case, the cylinder has bulk energy $\tilde\Psi_b$ and surface energy is applied on the whole boundary except the supported sides. The cylinder is also subjected to axial stretching of magnitude $\lambda$.

For better clarity, we normalize all the parameters and results with respect to the shear modulus $\mu$ and radius of the cylinder $R$. The normalized quantities read $\tilde\gamma = \gamma/(\mu R)$, $\tilde\alpha = \alpha/(\mu R)$, $\tilde \kappa = \kappa/\mu$, $\tilde L = L/R$. 

The normalized first Lamé parameter was chosen as $\tilde\kappa=4$, corresponding to Poisson's ratio $\nu = 0.4$. 

For the case of the surface tension model and infinite length, such a problem was already investigated in the literature, and the behavior is well known, see, e.g., \cite{dortdivanlioglu2022plateau}. For low values of the surface tension $\gamma$ the cylinder deforms homogeneously, but when the combination of the surface tension and axial stretch reaches a certain threshold, the homogeneous solution becomes unstable and a different nonhomogeneous stable solution appears. Also, in the case of deformation-dependent models, the onset of bifurcation
can be determined semi-analytically by solving incremental equations. This was demonstrated in \cite{yu2024incremental} for certain choices of surface material model, which, however, are not surface-polyconvex. First, we apply the same procedure to the surface material model obtained by combining the surface tension model with the newly proposed model. The outline of the derivation of the incremental equations and their solution is described in \ref{stabSec}. The analysis leads to a system of two nonlinear algebraic Equations \eqref{eqA16} and \eqref{eqA17}, depending on four variables (not considering the bulk material parameters, which are considered fixed). The parameters are the critical surface material parameters $\tilde\alpha$ and $\tilde\gamma$, critical axial stretch $\lambda$, and ratio $\tilde r=r/R$ of the deformed and undeformed radii of the cylinder. 
The solutions of such a nonlinear system represent the limiting combinations of the parameters for which the homogeneous configuration loses stability and a new, nontrivial stable solution emerges. To determine some of these combinations, we fix two of the four unknown variables, e.g., surface parameter $\tilde \alpha$ and axial stretch $\lambda$, and solve the resulting system of two nonlinear equations for two unknowns, $\tilde r$ and $\tilde \gamma$, by the standard Newton-Raphson method. For $\tilde \alpha$ ranging between $0$ and $0.3$, and $\lambda$ chosen in the interval between $0.5$ and $1.6$, the values of the critical parameter $\tilde \gamma$ are displayed in Figure \ref{fig9}. 
\begin{figure}[ht]
\centering%
\includegraphics[width=0.55\textwidth, keepaspectratio=true]{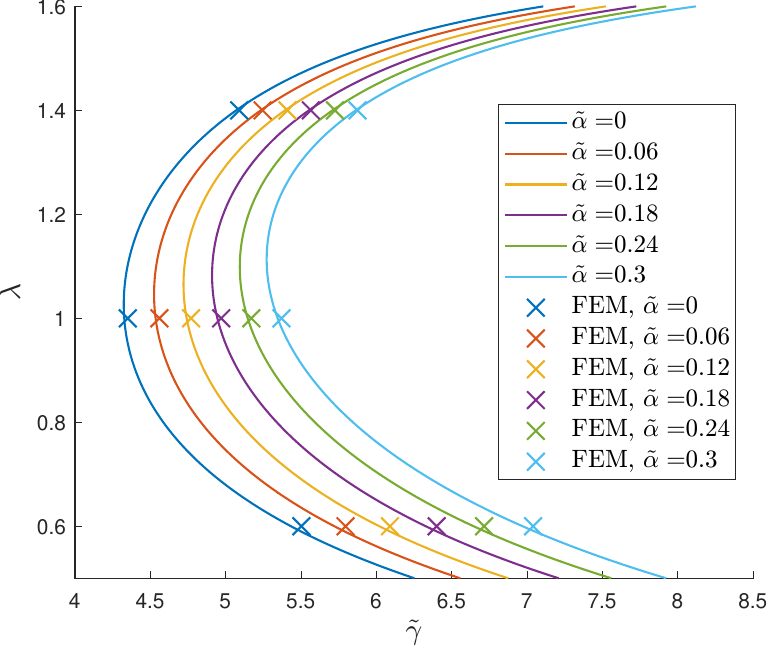}
\caption{Onset of bifurcation for combinations of axial stretch $\lambda$ and parameters $\tilde\alpha$ and $\tilde\gamma$ of the surface energy $\tilde\Psi_s = \gamma\hat{J} + \alpha|\hat{\bm{F}}|$ obtained by solving incremental equations and compared to the FE analysis. The bulk material parameter is $\tilde\kappa=4$.}  
\label{fig9}
\end{figure} 
Each line is associated with a fixed value of parameter $\tilde \alpha$ and visualizes a combination of $\lambda$ and $\tilde \gamma$ for which the homogeneous solution becomes unstable. Each point $(\tilde \gamma, \ \lambda)$ in the $\tilde \gamma-\lambda$ space, which is located on the left side of the associated limit curve, corresponds to a stable homogeneous solution. On the other hand, the homogeneous solution for which the point $(\tilde \gamma, \ \lambda)$ is located on the right side of the limit curve is unstable. Clearly, the proposed strain-dependent surface model has a stabilizing effect. As parameter $\alpha$, which governs the contribution of the deformation-dependent surface energy, increases, a higher surface tension $\gamma$ is required for the homogeneous solution to become unstable.

Next, we present the results of the FE simulations in which we applied the path following and instability tracking methods, which are described in Section \ref{secBif}. In the analysis presented in \ref{stabSec}, it is assumed that the cylinder is of infinite length and also that the wavenumber characterizing the perturbation of the homogeneous solution is $0$, thus the wavelength of the shape at the onset of bifurcation tends to infinity. Clearly, in the FE simulation, we cannot model infinite domains, and it is necessary to consider a cylinder of finite length, which introduces additional error when the results are compared to those presented in Figure \ref{fig9}. To make the simulations more efficient, we considered only the axisymmetric 2D elements since the path following methods become computationally expensive. For the simulation, the dimensionless length was set to $\tilde L = L/R = 30$, and 20 elements were used per one unit of dimensionless length, meaning 20 elements in the radial direction and 600 elements in the longitudinal, overall summing up to 12000 bulk elements and 600 surface elements. In the simulations, parameter $\tilde\alpha$ was fixed to a specific value; stretching $\lambda$ is applied in one step by deforming the cylinder to length $\lambda \tilde L$, and the simulation is then controlled by increasing parameter $\tilde\gamma$ in multiple nonuniform steps. For comparison with the reference solution, the onset of bifurcation is determined for stretches $\lambda=\{0.6,1.0,1.4\}$ and values $\tilde\alpha = \{ 0.0,0.06,0.12,0.18,0.24,0.3 \}$. The considered combinations of $(\tilde\gamma,\ \lambda)$ are depicted in Figure \ref{fig9} by cross markers. By visual comparison, one can conclude that the limit points computed using FE simulation match the semi-analytical results very well. The quality of the match improves with increasing stretch $\lambda$. This comes most likely from the fact that the semi-analytical results are obtained for infinite cylinders, while the numerical simulations are computed for finite structures. 
{The increasing axial stretch elongates the cylinder, which causes decrease in the error caused by analyzing a cylinder of finite length.}
Nevertheless, even the results for the axial stretch $\lambda=0.6$ are in a very good agreement with the reference values. For example, for $\tilde\alpha=0$ the reference value of $\tilde\gamma$ is $5.433$ while the one computed by FE simulation equals 5.502, which makes the relative error of $1.27\%$.

To support the claims discussed in the previous paragraph, we conducted the simulations with parameters $\lambda=0.6$ and $\tilde\alpha=0$ for increasing lengths $\tilde L$ while the mesh of the 20 elements per unit of dimensionless length was fixed. The evaluated limit values of $\tilde\gamma$ and also the relative error are shown in Table \ref{tab:converg}. The relative error indeed decreases for an increasing reference length $\tilde L$; the mismatch for $\tilde L=100$ is only $0.13\%$. 
\begin{table}[ht]
    \centering
    \begin{tabular}{|c|ccccc|c|}
    \hline
        $\tilde L$ & 30   & 40 & 50 & 60 & 100 & $\infty$ (semi-analytical)  \\
        \hline
        $\tilde\gamma$ & 5.502   & 5.472 & 5.458 & 5.451 & 5.440 & 5.433 \\
        \hline
        error [$\%$] & 1.2700  &  0.7178  &  0.4602  &  0.3313 &   0.1288 &   \\
        \hline
      \end{tabular}
    \caption{Relative error in $\tilde\gamma$ between the FE simulation and semi-analytical results evaluated for $\lambda=0.6$ and $\alpha=0$, 20 elements per $\tilde R$.}
    \label{tab:converg}
\end{table}

Using the methods described in Section \ref{secBif}, we can also track the stable solution after the onset of bifurcation. In Figure \ref{fig11}, the deformed shape of the stable solution evaluated for $\lambda=1$ and $\alpha=0$ is visualized for various values of surface tension parameter $\tilde\gamma$ (note that axisymmetry is used in accordance with the scheme shown in Figure \ref{fig10b}). One can see that in stages (a) and (b), the value of surface tension is smaller than the onset of bifurcation, which occurs for $\tilde\gamma=4.35$ (see the blue cross with $\lambda=1$ in Figure \ref{fig9}). Therefore, the stable solution still corresponds to the homogeneously deformed one. On the other hand, in stage (c), $\tilde\gamma$ is larger than the bifurcation limit, hence the stable solution differs from the homogeneously deformed one; one can see that the left part of the cylinder inflates while the right part deflates. In the remaining stages (d) to (g), the nonuniformity of the deformed shape gets more pronounced; simultaneously, the left bulged part tends to localize in smaller regions, while in the right part, sufficiently far from the bulged section, the solution resembles homogeneous deformation.
\begin{figure}[ht]
\centering%
\begin{subfigure}[b]{0.9\textwidth}
\includegraphics[width=\textwidth, keepaspectratio=true]{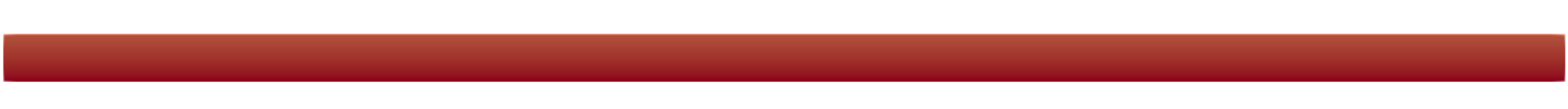}
\subcaption{$\tilde\gamma=1$}
\label{fig11a}
\end{subfigure}\\
\begin{subfigure}[b]{0.9\textwidth}
\includegraphics[width=\textwidth, keepaspectratio=true]{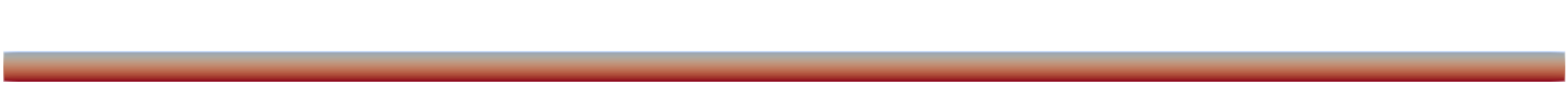}
\subcaption{$\tilde\gamma=4.32$}
\label{fig11b}
\end{subfigure}\\
\begin{subfigure}[b]{0.9\textwidth}
\includegraphics[width=\textwidth, keepaspectratio=true]{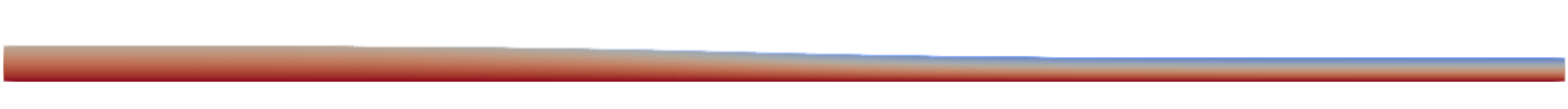}
\subcaption{$\tilde\gamma=4.47$}
\label{fig11c}
\end{subfigure}\\
\begin{subfigure}[b]{0.9\textwidth}
\includegraphics[width=\textwidth, keepaspectratio=true]{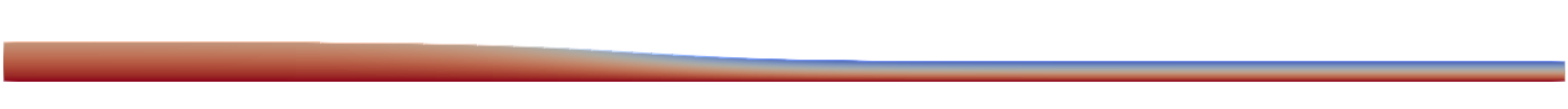}
\subcaption{$\tilde\gamma=4.7$}
\label{fig11d}
\end{subfigure}\\
\begin{subfigure}[b]{0.9\textwidth}
\includegraphics[width=\textwidth, keepaspectratio=true]{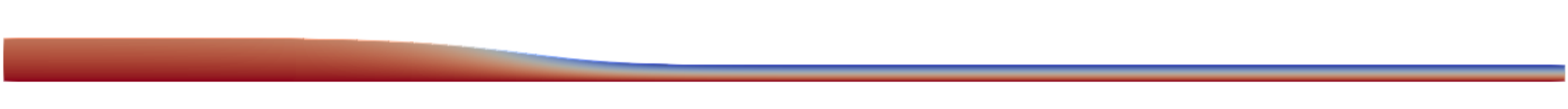}
\subcaption{$\tilde\gamma=5.15$}
\label{fig11e}
\end{subfigure}\\
\begin{subfigure}[b]{0.9\textwidth}
\includegraphics[width=\textwidth, keepaspectratio=true]{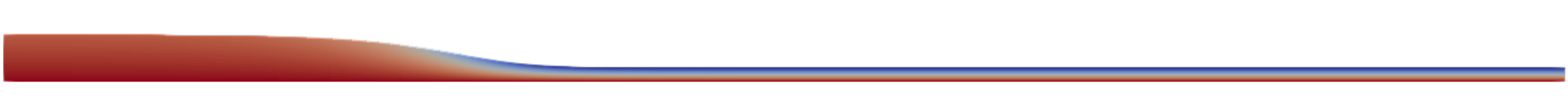}
\subcaption{$\tilde\gamma=5.6$}
\label{fig11f}
\end{subfigure}\\
\begin{subfigure}[b]{0.9\textwidth}
\includegraphics[width=\textwidth, keepaspectratio=true]{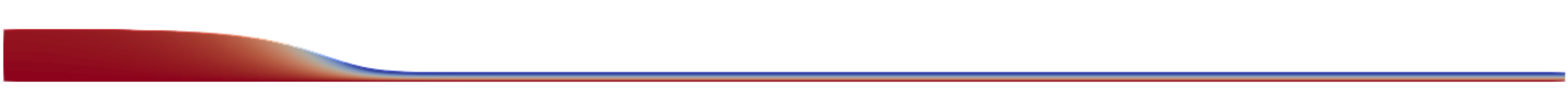}
\subcaption{$\tilde\gamma=7.2$}
\label{fig11g}
\end{subfigure}
\caption{Deformed shapes of the cylinder for increasing parameter $\tilde\gamma$ and fixed parameters $\tilde\alpha=0$, $\lambda=1$, $\tilde\kappa=4$, $\tilde L=30$. }   
\label{fig11}
\end{figure} 

For comparison, the deformed shapes evaluated for the same geometry and axial stretching but different $\tilde\alpha=0.3$ are visualized in Figure \ref{fig12}. As was already pointed out, increasing parameter $\tilde\alpha$ has a stabilizing effect, hence the homogeneous solution remains stable even for larger values of $\tilde\gamma$. For example, in stage (b) of Figure \ref{fig12}, associated with $\tilde\gamma=5.15$, the deformed shape remains homogeneous. In contrast, the stable solution evaluated for the same stretch and $\tilde\gamma$, but with $\tilde\alpha=0$ (see stage (e) of Figure \ref{fig11}) is already bifurcated. Otherwise, the trend in the evolution of the deformed shape with increasing $\tilde\gamma$ remains the same as for the previous case.  
\begin{figure}[ht]
\centering%
\begin{subfigure}[b]{0.9\textwidth}
\includegraphics[width=\textwidth, keepaspectratio=true]{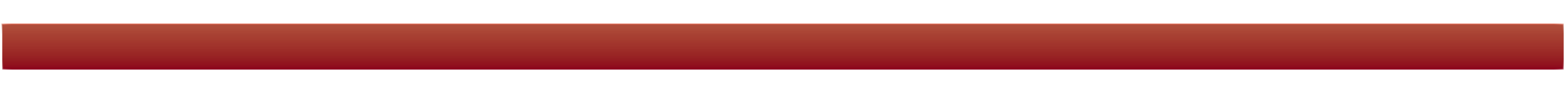}
\subcaption{$\tilde\gamma=1$}
\end{subfigure}\\
\begin{subfigure}[b]{0.9\textwidth}
\includegraphics[width=\textwidth, keepaspectratio=true]{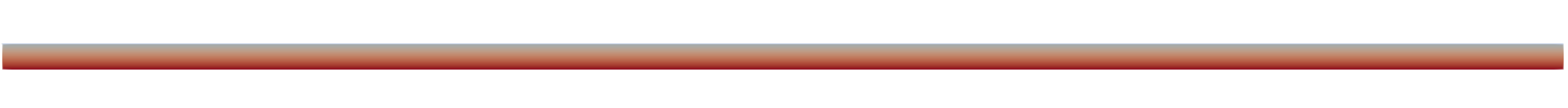}
\subcaption{$\tilde\gamma=5.15$}
\end{subfigure}\\
\begin{subfigure}[b]{0.9\textwidth}
\includegraphics[width=\textwidth, keepaspectratio=true]{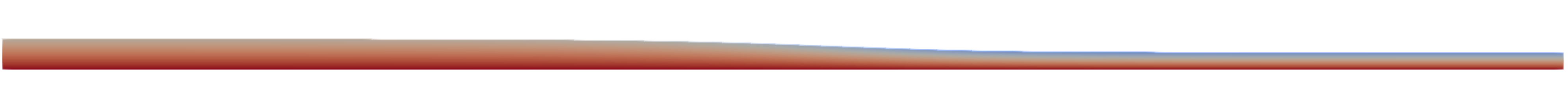}
\subcaption{$\tilde\gamma=5.6$}
\end{subfigure}\\
\begin{subfigure}[b]{0.9\textwidth}
\includegraphics[width=\textwidth, keepaspectratio=true]{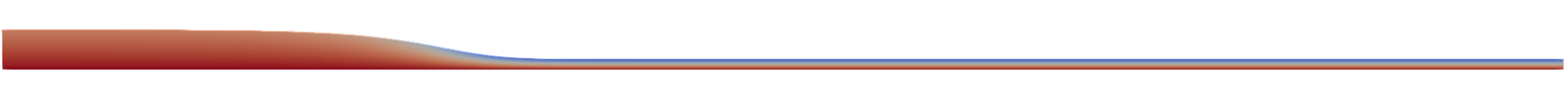}
\subcaption{$\tilde\gamma=7.2$}
\end{subfigure}
\begin{subfigure}[b]{0.9\textwidth}
\includegraphics[width=\textwidth, keepaspectratio=true]{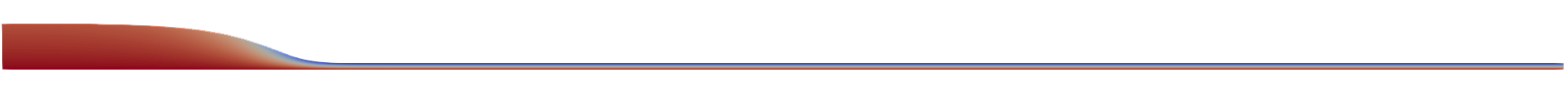}
\subcaption{$\tilde\gamma=10$}
\end{subfigure}
\caption{Deformed shapes of the cylinder for increasing parameter $\tilde\gamma$ and fixed parameters $\tilde\alpha=0.3$, $\lambda=1$, $\tilde\kappa=4$, $\tilde L=30$. }   
\label{fig12}
\end{figure}

\section{Conclusions}
In this work, we proposed a new class of surface energy models for soft materials based on the principle of surface polyconvexity. When combined with polyconvex and coercive bulk energy densities, the condition of surface polyconvexity guarantees the existence of minimizers and ensures a consistent variational formulation. Unlike classic models based solely on surface tension, our framework incorporates deformation-dependent surface energy contributions, enabling a more nuanced and physically meaningful description of surface mechanics.

We formulated the surface energy for isotropic surfaces as a combination of the surface tension and deformation-dependent terms expressed through isotropic norms of the surface deformation gradient. We focused on a representative case involving the Frobenius norm and further extended the formulation to capture anisotropic effects. All variants were designed to satisfy the surface polyconvexity condition, which in turn implies the positive semi-definiteness of the surface Cauchy stress tensor and precludes the existence of a stress-free reference configuration.

To evaluate the proposed models, we implemented them into the OOFEM open-source finite element code and conducted a series of numerical studies. These simulations validated the implementation and illustrated the distinct mechanical behavior of surface-polyconvex models. For a cylindrical surface, we demonstrated excellent agreement between numerical and semi-analytical solutions, highlighting the stiffening response of the deformation-dependent model in contrast to the stretch-invariant nature of classical surface tension. In a cubic body with incompressible bulk and surface energy, both surface energy types led to near-spherical deformations, but with quantitatively distinct outcomes, revealing that even in simple settings, model choice can significantly affect the results. Anisotropic surface energies were explored on a spherical domain, where alignment of the anisotropic direction with the sphere’s parallels led to ellipsoidal equilibrium shapes. Despite the apparent equivalence of the two anisotropic formulations in this symmetric setting, such similarity is not expected in general, underscoring the importance of model selection in anisotropic applications. Finally, we investigated surface-driven instabilities in a compressible hyperelastic cylinder. Numerical and semi-analytical approaches yielded consistent bifurcation thresholds, and post-bifurcation analysis—enabled by eigenvector perturbation, line search, and deflation techniques—further validated the robustness of the proposed framework in capturing complex nonlinear behaviors.

In summary, the surface-polyconvex models introduced here offer a theoretically sound and computationally versatile framework for modeling surface phenomena in soft materials. They bridge the gap between classical surface tension models and the need for deformation-dependent formulations, opening new possibilities for studying surface elasticity, anisotropy, and instability phenomena in soft structures. 

In future work, we plan to investigate the mutual interaction between bulk and surface energies and its implications for the existence of minimizers. 
Our goal is to relax the currently imposed conditions, as surface polyconvexity imposes restrictive constraints on the form of surface energy, primarily due to the requirement of homogeneity of degree one.
We also intend to extend our analysis to incorporate curvature-dependent surface energies.

\clearpage
\section*{Acknowledgements}
Financial support received from the Czech Ministry of Education, Youth and Sports (ERC CZ project No. LL2310) is gratefully acknowledged.  Moreover, M. Kružík’s work was co-funded by the European Union under the ROBOPROX project (reg.~no.~CZ.02.01.01/00/22 008/0004590). The authors would also like to thank Prof.~Yibin Fu for his valuable discussions on the incremental equations in surface elasticity.

\appendix
\section{Surface Jacobian}\label{App:SurfaceJacobian}
The surface Jacobian $\hat{J}$, i.e., the ratio $\mathrm{d}a/\mathrm{d}A$ can be expressed from Equation \eqref{eq25} as
\bea\label{eq35}
\hat{J} = \left|\left|\mathrm{cof} \bm{F}\bm{N}\right|\right|
\eea
It is easy to see that $\left|\left|\operatorname{cof} \mathbf{F} \mathbf{N}\right|\right| = \left|\left|\operatorname{cof} \hat{\mathbf{F}} \mathbf{N}\right|\right|$. However, additional expressions for the surface Jacobian can be derived:
\bea
\hat{J} &=& \left|\left| \text{cof} \bm{F}\bm{N} \right|\right| = \left|\left| \text{cof} \hat{\bm{F}} \bm{N} \right|\right|= \sqrt{\bm{N}\otimes \bm{N} : \text{cof} \hat{\bm{F}}^T \text{cof} \hat{\bm{F}} } = \sqrt{\bm{N}\otimes \bm{N} : \text{cof} \hat{\bm{C}} } = \sqrt{\left(\bm{I}-\hat{\bm{I}}\right) : \text{cof} \hat{\bm{C}} } = \left|\left| \text{cof} \hat{\bm{F}}  \right|\right| 
\nonumber
\\ \label{Eq:surfaceJall}
&=& \sqrt{\text{tr} \left(\text{cof}\hat{\bm{C}}\right)  } = \sqrt{\frac{1}{2}\text{tr}\left(\hat{\bm{C}}\right)^2 -\frac{1}{2}\text{tr}\left(\hat{\bm{C}}^2\right)}
\eea
where we used that $\text{cof} \hat{\bm{C}} = \text{cof} \hat{\bm{F}}^T \text{cof} \hat{\bm{F}}$ and $\hat{\bm{I}} : \text{cof} \hat{\bm{C}} = 0$.
The last equality in \eqref{Eq:surfaceJall} can be easily obtained if we express the trace of the cofactor using the tensor cross product as defined in \eqref{eq:cofTensorCrossProduct} as
\bea
\text{tr} \left(\text{cof}\hat{\bm{C}}\right)&=& \delta_{ij}\frac{1}{2}\epsilon_{ikm}\epsilon_{jln}\hat{C}_{kl}\hat{C}_{mn} =\frac{1}{2}\epsilon_{ikm}\epsilon_{iln}\hat{C}_{kl}\hat{C}_{mn} =\frac{1}{2}\left(\delta_{kl}\delta_{mn}-\delta_{kn}\delta_{lm}\right)\hat{C}_{kl}\hat{C}_{mn}  
\\
&=&\frac{1}{2}\left(\hat{C}_{kk}\hat{C}_{mm}-\hat{C}_{nl}\hat{C}_{ln}\right) = \frac{1}{2}\text{tr}\left(\hat{\bm{C}}\right)^2 -\frac{1}{2}\text{tr}\left(\hat{\bm{C}}^2\right)
\eea
Note also  that the last expression corresponds to the second principal invariant $I_2(\hat{\bm{C}})$ of the surface right Cauchy-Green tensor and consequently 
\bea
\hat{J} = \sqrt{I_2(\hat{\bm{C}})} \label{eq44}
\eea

Alternatively, the surface Jacobian can be defined directly by evaluation of the ratio of deformed and reference areas induced by arbitrary tangential vectors $\bm{t}_1$ and $\bm{t}_2$
\bea
\hat{J} = \dfrac{\left|\left| \hat{\bm{F}} \bm{t}_1 \times \hat{\bm{F}} \bm{t}_2   \right|\right| }{\left|\left| \bm{t}_1 \times \bm{t}_2   \right|\right|}
\eea

\section{Variations and Derivatives of Kinematic Quantities}\label{Appendix:derivatives}
We now introduce a general definition of the derivative $\mathrm{D}$ according to \cite{silhavy2011equilibrium}, which is useful when a scalar or tensorial function, such as energy or surface stress tensor response functions, is differentiated. The definition is analogous to the definition of surface derivative from Section \ref{surfDef}, only assuming the arguments are not referential positions from the surface manifold $\Gamma$, but rather tensorial objects from a manifold $\mathcal{M}$. To be more specific, the objects might be, e.g.,\ pairs $(\hat{\bm{F}}, \bm{N})$ from a manifold specified as
\bea
\mathcal{M} =  \left\{   \left(\hat{\bm{F}}, \bm{N}\right) \in \text{Lin}(\mathbb{R}^3,\mathbb{R}^3)\times \mathbb{R}^3:   \hat{\bm{F}} \bm{N}=\bm{0}  \right\}\label{eq34c}
\eea
Assuming a scalar or tensorial response function $\bm{f}:U \to W$ which maps points from a relatively open subset $U$ of a manifold $\mathcal{M}$ in a finite-dimensional space $\bar{V}$ to a finite-dimensional vector space $W$ (for our purposes $W=\mathbb{R}$, $W=\mathbb{R}^3$, or $W=\mathbb{R}^3 \times \mathbb{R}^3$). The derivative $\mathrm{D}\bm{f}(x) \in \text{Lin}(\bar{V},W) $ evaluated in point $x\in U$ can be interpreted as a linear mapping between spaces $\bar{V}$ and $W$ satisfying
\bea
\lim_{\substack{y\to x \\ y\in \mathcal{M}, y \neq x}}\dfrac{ \left|\left|\bm{f}(y) -\bm{f}(x) - \text{D} \bm{f}(x)\left( x- y \right)\right|\right|}{\left|\left|x - y\right|\right|} = 0
\eea
Moreover, $\text{D} \bm{f}(x)$ is required to be an element of the tangent space $T_x\mathcal{M}$ of manifold $\mathcal{M}$ at point $x$, meaning 
\bea
\text{D} \bm{f}(x) P = \text{D}  \bm{f}(x) \label{eq26b}
\eea
must hold with $P$ being the orthogonal projection to the tangent space $T_x\mathcal{M}$.
For example, if $\bm{f}: G \to \mathbb{R}$ is a scalar function of a pair $(\hat{\bm{F}}, \bm{N}) \in G$, $U$ is a relatively open subset of manifold $\mathcal{M}$, $W=\mathbb{R}$, and the derivative $D f(\hat{\bm{F}}_t,\bm{N}_t )$ at $(\hat{\bm{F}}_t,\bm{N}_t ) \in G$ must lie in the tangent space $T_{(\hat{\bm{F}}_t,\bm{N}_t ) }G$, which is a subspace of $\text{Lin}(\mathbb{R}^3,\mathbb{R}^3)\times\mathbb{R}^3$. Consequently, the derivative $D f$ has two components $( \partial_{\hat{\bm{F}}}  f,\partial_{{\bm{N}}}  f )$, the first associated to the part $\text{Lin}(\mathbb{R}^3,\mathbb{R}^3)$ and the other with the part $\mathbb{R}^3$. Clearly, due to the constraint $\hat{\bm{F}}\bm{N}=\bm{0}$ the components $( \partial_{\hat{\bm{F}}} f,\partial_{{\bm{N}}} f )$ are not independent. 

Next, we express variations of some of the kinematical quantities. By variation $\delta E(\bm{\varphi},\delta\bm{\varphi})$ of some quantity $E(\bm{\varphi})$ we understand
\bea
\delta E(\bm{\varphi},\delta\bm{\varphi}) = \dfrac{\mathrm{d}\left( E(\bm{\varphi} + t \delta\bm{\varphi})  \right)   }{\mathrm{d}t} \bigg \rvert_{t=0}, \ \ \ t \in \mathbb{R}
\eea
Sometimes the arguments of the spatial variation are dropped and we use the abbreviation $\delta E:=\delta E(\bm{\varphi},\delta\bm{\varphi})$. 

Let us now express the variation of the surface Jacobian $\delta \hat{J}$. Starting from Equation \eqref{eq35} we have
\bea
\delta \hat{J} = \delta J \left|\left| \bm{N} \cdot \bm{F}^{-1} \right|\right|+ J \delta\left(\left|\left| \bm{N} \cdot \bm{F}^{-1} \right|\right|\right) \label{eq38}
\eea
The variation of the bulk Jacobian $\delta J$ is obtained using the chain rule and expressing the derivative of the determinant of a tensor with respect to the tensor itself, leading to
\bea
\delta J = J \text{tr} (\bm{F}^{-1}  \nabla \delta \bm{\varphi} ) = J \bm{I}:\left(\nabla \delta \bm{\varphi}\bm{F}^{-1}\right) \label{eq36}
\eea
Moreover, the variation $\delta\left(\left|\left| \bm{N} \cdot \bm{F}^{-1} \right|\right|\right)$ is reformulated to
\bea
\delta\left(\left|\left| \bm{N} \cdot \bm{F}^{-1} \right|\right|\right) &=& \dfrac{1}{\left|\left| \bm{N} \cdot \bm{F}^{-1} \right|\right|} \left(  \bm{N} \cdot \bm{F}^{-1}\right) \cdot \left(\bm{N}\cdot \delta \bm{F}^{-1} \right)=  - \dfrac{1}{\left|\left| \bm{N} \cdot \bm{F}^{-1} \right|\right|} \left(  \bm{N} \cdot \bm{F}^{-1}\right) \cdot \left(\bm{N}\cdot \bm{F}^{-1}\delta \bm{F}\bm{F}^{-1}\right) = \nonumber \\
&=&  - \left|\left| \bm{N} \cdot \bm{F}^{-1} \right|\right| \bm{n} \cdot \left(\bm{n}  \nabla \delta \bm{\varphi}\bm{F}^{-1}\right) = -\left|\left| \bm{N} \cdot \bm{F}^{-1} \right|\right| \bm{n}\otimes\bm{n} :\left(\nabla \delta \bm{\varphi}\bm{F}^{-1}\right) \label{eq37}
\eea
where the formula $\delta \bm{F}^{-1} =- \bm{F}^{-1} \delta \bm{F}\bm{F}^{-1}$ was applied after the second equality sign. Substituting the Equations \eqref{eq36} and \eqref{eq37} back to \eqref{eq38}, we arrive at
\bea
\delta \hat{J}= J  \left|\left| \bm{N} \cdot \bm{F}^{-1} \right|\right|\bm{I}: \left(\nabla \delta \bm{\varphi}\bm{F}^{-1}\right) -J\left|\left| \bm{N} \cdot \bm{F}^{-1} \right|\right|\bm{n}\otimes\bm{n}:\left( \nabla \delta \bm{\varphi}\bm{F}^{-1}\right) = \hat{J}  \hat{\bm{i}} : \left( \nabla \delta \bm{\varphi}\bm{F}^{-1}\right)
\eea
With the use of relations \eqref{eq40} and \eqref{eq41}, the double contraction from the last equation can be recast to
\bea
 \hat{\bm{i}} : \left( \nabla \delta \bm{\varphi}\bm{F}^{-1}\right) =   \nabla \delta \bm{\varphi}^T :\left(\bm{F}^{-1} \hat{\bm{i}}\right) =  \nabla \delta \bm{\varphi}^T :\hat{\bm{F}}^{-1} =  \nabla \delta \bm{\varphi}^T :\left(\hat{\bm{I}}\hat{\bm{F}}^{-1} \right) =  \hat{\bm{F}}^{-T}:\hat{\nabla} \delta \bm{\varphi}
\eea
and the variation $\delta \hat{J}$ reads
\bea
\delta \hat{J} = \hat{J} \hat{\bm{F}}^{-T}:\hat{\nabla} \delta \bm{\varphi}
\eea
Comparing the previous expression with general chain rule expansion of the surface Jacobian 
\bea
\delta \hat{J}\left( \hat{\bm{F}},\hat{\bm{N}}  \right) = \partial_{\hat{\bm{F}}}  \hat{J} :\hat{\nabla} \delta \bm{\varphi}
\eea
we can associate the derivative of the surface Jacobian with respect to the surface deformation gradient
\bea
\partial_{\hat{\bm{F}}}  \hat{J} = \hat{J} \hat{\bm{F}}^{-T} \label{eq53}
\eea

It is also possible to obtain an alternative expression for the derivative $\partial_{\hat{\bm{F}}}  \hat{J}$ by considering the formula in Equation \eqref{eq44} as the starting point, i.e., \ $\hat{J}=(I_2)^{\frac{1}{2}} = ({\frac{1}{2}\text{tr}(\hat{\bm{C}})^2 -\frac{1}{2}\text{tr}(\hat{\bm{C}}^T\hat{\bm{C}})})^{\frac{1}{2}}$. The expression for the derivative $\partial_{\hat{\bm{F}}}  \hat{J}$ is then analogical to the derivative of the bulk second principal invariant and reads
\bea
\partial_{\hat{\bm{F}}}  \hat{J} = \frac{1}{\hat{J}}\left( \left(\text{tr}\hat{\bm{C}}\right) \hat{\bm{F}} - \hat{\bm{F}}\hat{\bm{C}} \right) = \frac{1}{\hat{J}} \hat{\bm{F}} \left( \text{tr}\hat{\bm{C}}\hat{\bm{I}} - \hat{\bm{C}} \right) \label{eq46}
\eea
The term inside the last parentheses can also be reformulated to
\bea
\text{tr}\hat{\bm{C}}\hat{\bm{I}} - \hat{\bm{C}} =\hat{\bm{I}} \left(\text{tr}\hat{\bm{C}}{\bm{I}} - \hat{\bm{C}} \right) = \hat{\bm{I}}\left( \hat{\bm{C}} \bm{\times} \bm{I} \right) = \hat{\bm{C}} \bm{\times} \left( \bm{N} \otimes \bm{N} \right) = \hat{\bm{\epsilon}} \hat{\bm{C}}  \hat{\bm{\epsilon}}^T
\eea
where $\hat{\bm{\epsilon}} = \bm{\epsilon} \cdot \bm{N}$ denotes the tangential second order antisymmetric tensor. The third equality in the previous equation can be shown by decomposing the Levi-Civita tensor, arising from the definition of the tensor cross product in Equation \eqref{eq45}, as $\bm{\epsilon} = \hat{\bm{\epsilon}} \wedge \bm{N}$ and realizing that $\hat{\bm{C}} \bm{N} =  \bm{N}\hat{\bm{C}} = \bm{0}$, where $\wedge$ denotes the standard wedge product. Therefore, the derivative of the surface Jacobian can alternatively be expressed by
\bea
\partial_{\hat{\bm{F}}}  \hat{J} = \frac{1}{\hat{J}} \hat{\bm{F}}  \hat{\bm{\varepsilon}} \hat{\bm{C}}  \hat{\bm{\varepsilon}}^T\label{eq47}
\eea

The second derivative of the surface Jacobian with respect to the surface deformation gradient is obtained by differentiating one of the expressions for the first derivative, here we chose to start with the formula in \eqref{eq46} to obtain 
\bea
\left(\partial_{\hat{\bm{F}}} \otimes \partial_{\hat{\bm{F}}} \right)\hat{J} = \frac{1 }{\hat{J}} \left( -\partial_{\hat{\bm{F}}}  \hat{J} \otimes \partial_{\hat{\bm{F}}}  \hat{J} + \bm{I}~ \overline{\otimes} 
 \left(  \text{tr}\hat{\bm{C}}\hat{\bm{I}} - \hat{\bm{C}}\right) + 2 \hat{\bm{F}} \otimes \hat{\bm{F}} - \hat{\bm{F}} \underline{\otimes}    \hat{\bm{F}}^T  - \left(\hat{\bm{F}}\hat{\bm{F}}^T \right)\overline{\otimes} \hat{\bm{I}} \right)
\eea
Alternatively, when starting from expression \eqref{eq47} the second derivative reads
\bea
\left(\partial_{\hat{\bm{F}}} \otimes \partial_{\hat{\bm{F}}} \right)\hat{J} 
 = \frac{1}{\hat{J}}\left(-\partial_{\hat{\bm{F}}}  \hat{J} \otimes \partial_{\hat{\bm{F}}}  \hat{J} +\bm{I}~\overline{\otimes}\left(\hat{\bm{\varepsilon}} \hat{\bm{C}}  \hat{\bm{\varepsilon}}^T\right) - \left(\hat{\bm{F}}\hat{\bm{\varepsilon}}\right)\underline{\otimes}\left(\hat{\bm{F}}\hat{\bm{\varepsilon}}\right)^T + \left(\hat{\bm{F}}\hat{\bm{\varepsilon}}\hat{\bm{F}}^T\right)\overline{\otimes}~\hat{\bm{\varepsilon}} \right)
\eea
The nonstandard tensor products are defined by coordinate expressions
\bea
[\bm{A} \overline{\otimes} \bm{B}]_{ijkl} = [\bm{A}]_{ik}[\bm{A}]_{jl}, \ \ \ [\bm{A} \underline{\otimes} \bm{B}]_{ijkl} = [\bm{A}]_{il}[\bm{A}]_{jk}
\eea


Moreover, besides the derivatives of the surface Jacobian also the first and second derivatives of quantities $||\hat{\bm{F}} ||$, $||\hat{\bm{F}} \hat{\bm{a}} ||$, and $\sqrt{||\hat{\bm{C}} \hat{\bm{a}} ||}$, with $\hat{\bm{a}}\in \mathbb{R}^3$ satisfying $\hat{\bm{I}}\hat{\bm{a}}=\hat{\bm{a}}$, are relevant for subsequent sections and read
\bea
\partial_{\hat{\bm{F}}} \left|\left|\hat{\bm{F}} \right|\right| &=& \dfrac{\hat{\bm{F}} }{\left|\left|\hat{\bm{F}} \right|\right| } \label{eq52} \\ 
\partial_{\hat{\bm{F}}} \left|\left|\hat{\bm{F}} \hat{\bm{a}} \right|\right| &=& \dfrac{\hat{\bm{F}} \hat{\bm{a}} \otimes \hat{\bm{a}} }{\left|\left|\hat{\bm{F}}\hat{\bm{a}}  \right|\right| } \label{eq63}\\ 
\partial_{\hat{\bm{F}}} \sqrt{\left|\left|\hat{\bm{C}} \hat{\bm{a}} \right|\right|} &=& \frac{1}{2\left|\left|\hat{\bm{C}}\hat{\bm{a}}  \right|\right| \sqrt{\left|\left|\hat{\bm{C}}\hat{\bm{a}}  \right|\right| }} \hat{\bm{F}}\left( \hat{\bm{C}}\hat{\bm{a}} \otimes  \hat{\bm{a}}+ \hat{\bm{a}}\otimes\hat{\bm{C}}\hat{\bm{a}}  \right)\label{eq64} \\
(\partial_{\hat{\bm{F}}} \otimes \partial_{\hat{\bm{F}}} ) ||\hat{\bm{F}} || &=& \dfrac{1}{\left|\left|\hat{\bm{F}} \right|\right| } {\bm{I}} ~\overline{\otimes} ~\hat{\bm{I}} - \dfrac{1}{\left|\left|\hat{\bm{F}} \right|\right|^3 } \hat{\bm{F}} {\otimes} \hat{\bm{F}} \\ 
(\partial_{\hat{\bm{F}}} \otimes \partial_{\hat{\bm{F}}} ) ||\hat{\bm{F}}\hat{\bm{a}} || &=&
 \dfrac{1}{\left|\left|\hat{\bm{F}}\hat{\bm{a}} \right|\right| } {\bm{I}} ~\overline{\otimes} ~\left(\hat{\bm{a}}\otimes \hat{\bm{a}}\right) - \dfrac{1}{\left|\left|\hat{\bm{F}}\hat{\bm{a}} \right|\right|^3 } \hat{\bm{F}}\hat{\bm{a}} {\otimes}\hat{\bm{a}}\otimes\hat{\bm{F}}\hat{\bm{a}} \otimes\hat{\bm{a}}\label{eq65}
  \\ 
(\partial_{\hat{\bm{F}}} \otimes \partial_{\hat{\bm{F}}} ) \sqrt{\left|\left|\hat{\bm{C}} \hat{\bm{a}} \right|\right|} &=&
-\frac{3}{4\left|\left|\hat{\bm{C}}\hat{\bm{a}}  \right|\right|^3 \sqrt{\left|\left|\hat{\bm{C}}\hat{\bm{a}}  \right|\right| }} \hat{\bm{F}}\left( \hat{\bm{C}}\hat{\bm{a}} \otimes  \hat{\bm{a}}+ \hat{\bm{a}}\otimes\hat{\bm{C}}\hat{\bm{a}}  \right) \otimes\hat{\bm{F}}\left( \hat{\bm{C}}\hat{\bm{a}} \otimes  \hat{\bm{a}}+ \hat{\bm{a}}\otimes\hat{\bm{C}}\hat{\bm{a}}  \right) \label{eq66}  \\
 &+& 
\frac{1}{2\left|\left|\hat{\bm{C}}\hat{\bm{a}}  \right|\right| \sqrt{\left|\left|\hat{\bm{C}}\hat{\bm{a}}  \right|\right| }} \Bigg(
\bm{I} ~\overline{\otimes} \left( \hat{\bm{C}}\hat{\bm{a}} \otimes  \hat{\bm{a}}+ \hat{\bm{a}}\otimes\hat{\bm{C}}\hat{\bm{a}}\right) +\nonumber\\
 &+&
 \hat{\bm{F}} ~\underline{\otimes} \left(\hat{\bm{a}}\otimes\hat{\bm{F}}\hat{\bm{a}}\right)
 +
 \hat{\bm{F}}\hat{\bm{F}}^T ~\overline{\otimes} \left(\hat{\bm{a}}\otimes\hat{\bm{a}}\right)
+
 \left(\hat{\bm{F}}\hat{\bm{a}}\otimes \hat{\bm{F}}\hat{\bm{a}}\right) ~\overline{\otimes} \bm{I}
 + 
 \hat{\bm{F}}\hat{\bm{a}} \otimes \hat{\bm{F}}^T \otimes \hat{\bm{a}}
\Bigg)\nonumber 
\eea


\subsection{Surface Divergence Theorem}
For tensor fields $\hat{\bm{W}}: \Gamma \to \mathbb{R}^3 \times W$ and $\forall \bm{a} \in W$ the surface divergence $\widehat{\text{Div}} \hat{\bm{W}}: \Gamma \to W$ is defined by
\bea
\bm{a}\cdot \widehat{\text{Div}} \hat{\bm{W}} = \text{tr}\left( \hat{\nabla}\left(\hat{\bm{W}}^T \bm{a} \right)  \right) \label{eq54}
\eea
where the transposition is given by
\bea
\bm{b}\cdot \hat{\bm{W}}^T \bm{a} = \hat{\bm{W}} \bm{b} \cdot \bm{a}, \ \ \ \forall \bm{b} \in \mathbb{R}^3
\eea
Symbol $\cdot$ has either the meaning of an inner product in $V$ or a contraction in $\mathbb{R}^3$.
To transform integrals on the part of the boundary surface $\mathcal{R} \subset \Gamma$, we apply the surface divergence theorem, stating that if the tensor field $\hat{\bm{W}}$ is superficial, meaning $\hat{\bm{W}} = \hat{\bm{W}} \hat{\bm{I}}$, the identity 
\bea
\int_\mathcal{R}  \widehat{\text{Div}} \hat{\bm{W}} ~\mathrm{d}\mathcal{A} = \int_{\partial \mathcal{R}} \hat{\bm{W}} \hat{\bm{N}} ~\mathrm{d}{\mathcal{L}}
\eea
holds. Here $\hat{\bm{N}}$ denotes the unit normal to the boundary curve $\partial \mathcal{R}$ belonging to the cotangent space of $\mathcal{R}$. Using the identity
\bea
 \widehat{\text{Div}} \left( \hat{\bm{W}}^T \bm{a} \right) = \bm{a}\cdot \widehat{\text{Div}}  \hat{\bm{W}} + \hat{\bm{W}} : \hat{\nabla} \bm{a}
\eea
the divergence theorem is reformulated to 
\bea
\int_\mathcal{R} {\hat{\bm{W}} :\widehat{\nabla}\bm{a}} ~\mathrm{d}\mathcal{A} = -\int_\mathcal{R} {\bm{a} \cdot \widehat{\text{Div}}\hat{\bm{W}}} ~\mathrm{d}\mathcal{A} +
\int_{\partial \mathcal{R}} {\bm{a} \cdot \hat{\bm{W}}  \hat{\bm{N}}} ~\mathrm{d}\mathcal{L}\label{eq29}
\eea

Furthermore, in order to transform superficial tensor fields $\hat{\bm{W}}$ from the reference to deformed configuration, we use the surface Piola transformation, e.g.,
\bea \label{eq:piola}
\hat{\bm{W}}^{\bm{\varphi}} = \frac{1}{\hat{J}} \hat{\bm{W}} \hat{\bm{F}}^T \label{eq55}
\eea
Consequently, the transformed field $\hat{\bm{W}}^{\bm{\varphi}}$ is superficial in the deformed configuration, i.e.,\ $\hat{\bm{W}}^{\bm{\varphi}} \hat{\bm{i}}=\hat{\bm{W}}^{\bm{\varphi}}$, and 
\bea
\widehat{\text{Div}}^{\bm{\varphi}}\hat{\bm{W}}^{\bm{\varphi}} = \frac{1}{\hat{J}}\widehat{\text{Div}}\hat{\bm{W}} 
\eea
holds. The spatial divergence $\widehat{\text{Div}}^{\bm{\varphi}}$ of a spatial tensor field $\hat{\bm{W}}^{\bm{\varphi}}$  is defined analogously to the material divergence as in \eqref{eq54} by replacing the material surface gradient with the spatial one.

\section{Analytical solutions of cylindrical surface}
The cylindrical surface examined in Section \ref{BridgeExam} is solved directly by a variational approach considering two forms of the surface energy response function $\tilde\Psi_s$. In both cases, the complete solution cannot be obtained analytically, and at some point, numerical techniques need to be applied to either find a root of an algebraic equation or to solve system of ordinary differential equations (ODEs). Both of these procedures can be conducted using standard tools. The cylinder's length and radius are denoted by $L$ and $R$; the geometry and boundary conditions are sketched in Figure \ref{Scheme1}. 
\subsection{Surface tension
}
\label{apA1}
The energy of the standard fluid-like surface tension model $\tilde\Psi_s = \gamma \hat{J}$ is proportional to the surface Jacobian $\hat{J}$ defined as the ratio of the areas of the deformed and undeformed surface elements. In cylindrical coordinates with $(\theta,x,y)$ being the angular, longitudinal, and radial directions and $\bm{R}$ and $\bm{r}$ the undeformed and deformed position vectors, the Jacobian reads 
\bea
\hat{J} = \dfrac{\left|\left|\frac{\partial \bm{r}}{\partial x} \mathrm{d}x \times\frac{\partial \bm{r}} {\partial \theta} \mathrm{d}\theta\right|\right|}{\left|\left|\frac{\partial \bm{R}}{\partial x} \mathrm{d}x \times\frac{\partial \bm{R}} {\partial \theta} \mathrm{d}\theta\right|\right|} =
\dfrac{\mathrm{d}x~ y\mathrm{d}\theta \left|\left|\frac{\partial \bm{r}}{\partial x}  \times \bm{e}_\theta\right|\right|}{\mathrm{d}x~ R\mathrm{d}\theta \left|\left| \bm{e}_x \times \bm{e}_\theta \right|\right|} = 
\dfrac{y \frac{\partial s}{\partial x}  \left|\left|\frac{\partial \bm{r}}{\partial s}  \times \bm{e}_\theta\right|\right|}{R\left|\left| \bm{e}_x \times \bm{e}_\theta \right|\right|} = \dfrac{1}{R}y\sqrt{1+y'^2}
\eea
where $\bm{e}_x$ and $\bm{e}_\theta$ are the unit vectors in $x$ and $\theta$ direcitons. Since $s$ denotes the longitudinal arc length coordinate of the deformed cylinder, the tangent vector $\frac{\partial \bm{r}}{\partial s}$ has unit length. 
The total energy functional to be minimized is then expressed by
\beq
E=\int_S \tilde\Psi_s \,\mbox{d}\mathcal{A} = \int_0^{2\pi}\int_{-\frac{L}{2}}^{\frac{L}{2}}\dfrac{\gamma}{R}y\sqrt{1+y'^2} \,\mbox{d}x\mbox{d}\theta= 
 \dfrac{2\pi\gamma}{R}\int_{-\frac{L}{2}}^{\frac{L}{2}} y \sqrt{1+y'^2}\,\mbox{d}x = 
\dfrac{2\pi\gamma}{R} \int_{-\frac{L}{2}}^{\frac{L}{2}} g(y,y')\,\mbox{d}x
\eeq
with boundary conditions $y(-L/2)=R$ and $y(L/2)=R$.
One can express the Euler-Lagrange equations and further simplify them using the Beltrami identity since $g(y,y')$ does not directly depend on $x$. Therefore the resulting differential equation reads
\bea
g - y'\frac{\partial g}{\partial y'} &=& C \nonumber \\ 
y \sqrt{1+y'^2} - yy'^2\frac{1}{\sqrt{1+y'^2}} &=& C\\ 
y' &=& \frac{1}{C}\sqrt{y^2 - C^2}\nonumber
\eea
This equation can be solved by the separation of variables approach. Separating the variables, formally integrating, and using the substitution $y = C\cosh \xi$, we get
\bea
\int \frac{C \,\mbox{d}y}{\sqrt{y^2 - C^2}} &=& \int \mbox{d}x \nonumber\\
\int \frac{C\sinh \xi \, \mbox{d}\xi}{\sqrt{\cosh^2 \xi - 1}} &=& x + D \\
C \xi &=& x + D \nonumber \\ 
y &=& C\cosh \frac{x + D}{C}\nonumber
\eea
Finally, using the two boundary conditions
\bea
C\cosh \frac{-L/2 + D}{C}  = R, \ \ \ C\cosh \frac{L/2 + D}{C} = R 
\eea
one can conclude that $D = 0$, since $\cosh x$ is an even function. To summarize, the solution reads
\beq
y = C\cosh \frac{x}{C}
\eeq
where $C$ is obtained by solving the equation
\beq
C\cosh \frac{L/2 }{C} = R 
\eeq
which needs to be done numerically. For $L/R = 3/2.5$ the value of $C/R$ is 0.7451.


\subsection{Deformation dependent model 
}
\label{apA2}
Let us now analyze the surface-polyconvex deformation-dependent model with energy $\tilde\Psi_s = \alpha||\hat{\bm{F}}||$. The term $||\hat{\bm{F}}||$ can be expressed using principal surface stretches $\hat\lambda_1$ and $\hat\lambda_2$, i.e., singular values of $\hat{\bm{F}}$, as
\beq
||\hat{\bm{F}}|| = \sqrt{\hat\lambda_1^2+\hat\lambda_2^2}
\eeq
Therefore, without the presence of bulk material, the sample tends to deform to minimize the principal surface stretches. 

Using the cylindrical coordinates again and denoting $u(x)$ the axial displacement and $w(x)$ the displacement in the radial direction, the principal surface stretches and $||\hat{\bm{F}}||$ can be expressed as
\bea
\hat\lambda_1^2 = \left(1+u'(x)\right)^2+w'(x)^2, \;\;\hat\lambda_2^2= \left(1-\frac{w(x)}{R}\right)^2, \;\; ||\hat{\bm{F}}|| = \sqrt{\left(1+u'\right)^2+w'^2 + \left(1-\frac{w}{R}\right)^2}= f(w(x),u'(x),w'(x))
\eea
Subsequently, the energy functional to be minimized reads
\beq
E=\int_S \tilde\Psi_s \,\mbox{d}\mathcal{A} = \int_{0}^{2\pi} \int_{-\frac{L}{2}}^{\frac{L}{2}}\alpha f(w,u',w') \,\mbox{d}x\,\mbox{d}\theta = 4\pi \alpha \int_{0}^{\frac{L}{2}} \sqrt{\left(1+u'\right)^2+w'^2 + \left(1-\frac{w}{R}\right)^2} \,\mbox{d}x
\eeq
where the symmetries were used. The boundary conditions read
\bea
u(0)=0 ,\;\; w'(0)=0 ,\;\; u(L/2)=0,\;\; w(L/2)=0 \nonumber
\eea
The Euler-Lagrange equations then have the generic form
\bea
\frac{\partial}{\partial x}\left(\frac{\partial f}{\partial u'}\right) &=& 0 \nonumber \\
\frac{\partial f}{\partial w}-\frac{\partial}{\partial x}\left(\frac{\partial f}{\partial w'}\right) &=& 0 
\eea
Substituting and evaluating the individual terms, the system is reformulated to


\beq
\begin{bmatrix}
 f^2-(1+u')^2 & -(1+u')w' \\
 (1+u')w' & w'^2-f^2 
\end{bmatrix}
\begin{bmatrix}
u'' \\ w''
\end{bmatrix}=
\begin{bmatrix}
-\frac {(1+u')(1-\frac{w}{R})w'}{R }\\ 
\frac {(1-\frac{w}{R})(w'^2+f^2)}{R }
\end{bmatrix}
\eeq
Such a system of second-order ODEs can be inverted to an explicit form
\bea
u'' &=& -\frac{2w'(1+u')}{R-w} \\ 
w'' &=& -\frac{2R^2w'^2 + R^2 - 2Rw+w^2}{R^2(R-w)} 
\eea
which can be solved numerically using, e.g.,\ inbuilt MATLAB solver. 


\section{Incremental equations}\label{stabSec}
In this section, we revise the theory of incremental equations to analytically study the instabilities of materials with surface energy. The derivation follows closely \cite{yu2024incremental}. Initially, the general form of the incremental equations is introduced, followed by examining the specific case of a stretched solid cylinder. In \cite{yu2024incremental}, the authors solved a similar test case, however,  with surface energy which is not surface-polyconvex.

\subsection{Incremental equilibrium equations}
The equilibrium equations, together with the boundary conditions associated with a deformation $\bm{\varphi}$ mapping from the reference configuration $\Omega$ to the deformed configuration $\Omega^{\bm{\varphi}}$ are given in equations \eqref{eq:equil1}-\eqref{eq:equil2}.
To derive the corresponding incremental equations, it is assumed that the deformed body undergoes an additional deformation ${\bar{\bm{\varphi}}}:\Omega^{\bm{\varphi}} \to \bar\Omega^{\bm{\varphi}}$, which maps the deformed configuration $\Omega^{\bm{\varphi}}$ to the final deformed configuration $\bar\Omega^{\bm{\varphi}}$. The composed mapping from the reference configuration to the final deformed configuration is denoted by ${\bar{\bm{\varphi}}}_t = \bar{\bm{\varphi}}({\bm{\varphi}}) :\Omega \to \bar\Omega^{\bm{\varphi}}$.
The position of the material points in the final deformed configuration is thus given by $\bar{\bm{x}} = {\bar{\bm{\varphi}}}(\bm x) = {\bar{\bm{\varphi}}}_t(\bm{X})$. The difference between the material points in the configurations $\Omega^{\bm{\varphi}}$ and $\bar\Omega^{\bm{\varphi}}$ is described by the displacement $\bm{u}(\bm{x})$, therefore  
\beq
\bar{\bm{x}} = \bm{x}(\bm{X}) + \bm{u}(\bm{x})
\eeq
 The deformation gradient of the mapping ${\bar{\bm{\varphi}}}_t$ reads
 \beq\label{eq1}
 \bar{\bm{F}} = \nabla {\bar{\bm{\varphi}}}_t = \left({\bm{I}} + \bm{\eta}\right){\bm{F}}
 \eeq
where ${\bm{F}} = \nabla {{\bm{\varphi}}}$ denotes the deformation gradient of the mapping between configurations $\Omega$ and $\Omega^{\bm{\varphi}}$ and ${\bm{\eta}} = \text{grad}~\bm{u}$ stands for the incremental displacement gradient with respect to the intermediate deformed configuration $\Omega^{\bm{\varphi}}$. The tensor ${\bm{I}} + \bm{\eta}$ thus has the meaning of the gradient of the mapping ${\bar{\bm{\varphi}}}$.

The relation between the surface deformation gradients $\hat{\bm{F}} = {\bm{F}}\hat{\bm{I}}$ and ${\hat{\bar{\bm{F}}}} = \bar{\bm{F}}\hat{\bm{I}}$ is obtained by projecting Equation \eqref{eq1} to the surface by the referential surface identity tensor $\hat{\bm{I}}$ and using the relation $\hat{\bm{F}} = \hat{\bm{i}}\hat{\bm{F}}$ with $\hat{\bm{i}}$ being the surface identity tensor in configuration $\Omega^{\bm{\varphi}}$ 
 \beq
 {\hat{\bar{\bm{F}}}}  = \left({\bm{I}} + \bm{\eta}\right)\hat{\bm{F}} = \left(\hat{\bm{i}}+ \bm{\eta}\hat{\bm{i}}\right)\hat{\bm{F}} = \left(\hat{\bm{i}}+ \hat{\bm{\eta}}\right)\hat{\bm{F}}
 \eeq

The first Piola-Kirchhof stresses $\bar{\bm{P}}$ and ${\bm{P}}$ associated with the deformations ${\bar{\bm{\varphi}}}_t$ and ${{\bm{\varphi}}}$ are defined as 
\beq
\bar{\bm{P}} = \partial_{{\bm{F}}} \tilde\Psi_b\bigg\rvert_{\bar{\bm{F}} }, \quad {\bm{P}} = \partial_{{\bm{F}}}\tilde \Psi_b \bigg\rvert_{{\bm{F}} } \label{eqA3}
\eeq
The surface first Piola-Kirchhof stresses $\hat{\bar{\bm{P}}}$ and $\hat{\bm{P}}$ are defined analogously by
\beq
\hat{\bar{\bm{P}}} = \partial_{\hat{\bm{F}}} \bar\Psi_s\bigg\rvert_{\hat{\bar{\bm{F}}} }\hat{\bm{I}}, \quad \hat{\bm{P}} = \partial_{\hat{\bm{F}}} \tilde\Psi_s \bigg\rvert_{\hat{\bm{F}} } \hat{\bm{I}} \label{eqA4}
\eeq
Note that in this section we do not distinguish between the first Piola-Kirchhoff stress tensor fields and the corresponding response functions. 

The incremental counterpart of Equations \eqref{eq:equil1}-\eqref{eq:equil2} expressed in the intermediate deformed configuration $\Omega^{\bm{\varphi}}$ in terms of the incremental stress 
\beq
\bm{\chi} = \dfrac{1}{{J}} \left( \bar{\bm{P}}- {\bm{P}}\right){\bm{F}}^{T}
\eeq
and surface incremental stress 
\beq
\hat{\bm{\chi}} = \dfrac{1}{\hat{J}} \left( \hat{\bar{\bm{P}}}- \hat{\bm{P}}\right)\hat{\bm{F}}^{T}
\eeq
are derived in \cite{yu2024incremental} and \cite{ogden1997nonlinear} and read
\bea
\text{div} \bm{\chi} &=& \bm{0}  \ \text{in} \ \Omega^{\bm{\varphi}}\label{eq4} \\
\bm{\chi} \bm{n} &=& {\bm{0}} \ \text{on} \ \bm{\varphi}\left(\Gamma\setminus S \right) \\
\widehat{\text{div} }\hat{\bm{\chi}}-\bm{\chi} \bm{n} &=&  {\bm{0}}  \ \text{on} \ \bm{\varphi}(S) \label{eqA1} \\
\hat{\bm{\chi}} \hat{\bm{n}} &=& {\bm{0}} \ \text{on} \ \bm{\varphi}(\partial S) \label{eq5} 
\eea
The operators $\text{div}$ and $\widehat{\text{div}}$ denote the divergence and surface divergence with respect to the intermediate deformed configuration $\Omega^{\bm{\varphi}}$.

\subsection{Linearized constitutive law}
The incremental equilibrium Equations \eqref{eq4} - \eqref{eq5} are valid for arbitrary deformation. However, to analyze only the onset of bifurcation and conduct as many derivations analytically as possible, in this and upcoming section, it is assumed that the displacement $\bm{u}$ is small, which allows to linearize the general constitutive laws $\bm{\chi} = \bm{\chi}(\bm{\eta})$ and  $\hat{\bm{\chi}} =\hat{\bm{\chi}}(\hat{\bm{\eta}})$ around the configuration $\Omega^{\bm{\varphi}}$, leading to relations
\bea
\bm{\chi} &=& \bm{C}:\bm{\eta}\label{eqA5} \\ 
\hat{\bm{\chi}} &=& \hat{\bm{C}}:\hat{\bm{\eta}}\label{eqA6}
\eea
where $\bm{C}$ and $\hat{\bm{C}}$ are the first-order instantaneous moduli related to bulk and surface, respectively. In the Cartesian reference frame 
their components read
\bea
C_{ijkl} &=&\dfrac{1}{{J}}{F}_{jm} {F}_{ln} \dfrac{\partial \tilde\Psi_b}{\partial F_{im}\partial F_{kn}}\bigg\rvert_{{\bm{F}} } \\ 
\hat{C}_{ij kl} &=&\dfrac{1}{{\hat{J}}}{\hat{{F}}} _{j m} {\hat{{F}}}_{l n} \dfrac{\partial \tilde\Psi_s}{\partial \hat{F}_{im}\partial \hat{F}_{kn}}\bigg\rvert_{{\hat{\bm{F}}} } 
\eea

\subsection{Hyperelastic cylinder with surface energy}
Let us now consider a particular example of an infinite hyperelastic cylinder with reference radius $R$. 
The bulk is made of material with bulk energy of neo-Hookean type
\bea
\tilde\Psi_b(\bm{F}) = \frac{1}{2} \mu \left(\bm{F}:\bm{F}-3 -2 \ln J  \right)+\frac{1}{2} \kappa \left( \frac{1}{2}\left(J^2-1 \right)-\ln J \right) \label{eqA7}
\eea
where $\kappa = 2\mu \nu/(1-2\nu)$ and $\mu$ are the Lamé parameters and $\nu$ denotes the Poisson's ratio. 
The material of the surface of the cylinder possesses the energy of form
\bea
\tilde\Psi_s = \gamma {\hat{J}}+\alpha||{\hat{\bm{F}}}|| \label{eqA8}
\eea
where $\gamma$ is the fluid-like surface tension and $\alpha$ is the surface parameter quantifying the deformation-dependent part. It is convenient to state the final results in terms of normalized parameters 
$\tilde\gamma = \gamma/\mu/R$, $\tilde\alpha = \alpha/\mu/R$, $\tilde \kappa = \kappa/\mu$. 

The derivations can be significantly simplified by adopting the deformed cylindrical coordinate system with $(\theta,z,r)$ being the angular, longitudinal, and radial coordinates and $(\bm{e}_\theta, \bm{e}_z, \bm{e}_r)$ the associated orthonormal cylindrical basis. Instead of the coordinate symbols $(\theta,z,r)$, sometimes the directions are denoted by numbers $(1,2,3)$.

The deformation of the cylinder is governed by the surface energy and by application of longitudinal stretch $\lambda$. Until the onset of instability, the cylinder deforms homogeneously with deformation gradient
\bea
\bm{F}=a \bm{e}_\theta \otimes \bm{e}_\theta + \lambda \bm{e}_z \otimes \bm{e}_z +a \bm{e}_r \otimes \bm{e}_r
\eea
Its surface counterpart reads
\bea
\hat{\bm{F}}=a \bm{e}_\theta \otimes \bm{e}_\theta + \lambda \bm{e}_z \otimes \bm{e}_z 
\eea
where $a$ denotes the stretch in the radial direction. The relation between the stretches $a$ and $\lambda$ and the surface material parameters $\gamma$ and $\alpha$ is obtained by expressing the boundary condition \eqref{eq:equil3}, which for the particular case simplifies to
\bea
P_{33} = -\dfrac{\hat{P}_{11}}{R}
\eea
The components $P_{33}$ and $\hat{P}_{11}$ of the stress and surface stress tensors $\bm{P}$ and $\hat{\bm{P}}$ are obtained by evaluating \eqref{eqA3} and \eqref{eqA4}. For the particular choice of bulk and surface energies \eqref{eqA7} and \eqref{eqA8}, the last equation reads
\bea
\tilde\gamma \lambda -\dfrac{\frac{\tilde\kappa }{2}+1}{a}+a\left(\frac{\tilde\kappa a^2\lambda ^2}{2}+1\right)+\dfrac{a\tilde\alpha }{\sqrt{a^2+\lambda ^2}} = 0 \label{eqA16}
\eea

The displacement vector field $\bm{u}$ between the two solutions has, due to symmetry, the form
\bea
\bm{u} = u(r,z) \bm{e}_r+v(r,z) \bm{e}_z
\eea
where the functional values $u(r,z)$ and $v(r,z)$ are assumed to be small in order to apply the linearized constitutive laws \eqref{eqA5} and \eqref{eqA6} for the incremental stresses. The components of the incremental displacement gradient and its surface counterpart then read
\bea
\left[ \bm{\eta} \right]= \begin{bmatrix}
 \frac{u}{r} & 0 & 0\\
 0 & \partial v /  \partial z & \partial v /  \partial r \\ 
 0 & \partial u /  \partial z & \partial u /  \partial r
\end{bmatrix}, \quad 
\left[ \hat{\bm{\eta}} \right]= \begin{bmatrix}
 \frac{u}{r} & 0 & 0\\
 0 & \partial v /  \partial z &0\\ 
 0 & \partial u /  \partial z &0
\end{bmatrix} \label{eqA9}
\eea

The incremental equilibrium Equations \ref{eq4} expressed in the cylindrical coordinates have the form
\bea
\dfrac{\partial \chi_{22}}{\partial z} + \dfrac{\partial \chi_{23}}{\partial r} +\dfrac{ \chi_{23}}{r} &=&0\nonumber \\
\dfrac{\partial \chi_{33}}{\partial r} + \dfrac{\partial \chi_{32}}{\partial z} +\dfrac{ \chi_{33}-\chi_{11}}{r} &=&0 \label{eqA11}
\eea
and the incremental surface boundary condition \ref{eqA1}  on the surface of the cylinder where $r=a R$ read as
\bea
\chi_{23}=\dfrac{\partial \hat\chi_{22}}{\partial z},\ \chi_{33}=\dfrac{\partial \hat\chi_{32}}{\partial z}-\dfrac{ \hat\chi_{11}}{r} \label{eqA10}
\eea

To find the onset of bifurcation, we assume the functions $u(r,z)$ and $v(r,z)$ describing the infinitesimal perturbation of harmonic form, i.e.,
\bea
u(r,z) = f(r) \exp{(\mathrm{i}kz)}, \ v(r,z) = g(r) \exp{(\mathrm{i}kz)}
\eea
with $\mathrm{i}$ being the imaginary unit. Subsequently, we evaluate the incremental displacement gradient \eqref{eqA9}, then express the components of the incremental stresses from \eqref{eqA5} and \eqref{eqA6}, and substitute these relations into incremental equilibrium Equations \eqref{eqA11} and incremental boundary conditions \eqref{eqA10}. The system of two second-order ODEs \eqref{eqA11} can be reformulated to one fourth-order ODE
\bea\label{eqA12}
\left( \frac{\mathrm{d}^2}{\mathrm{d}\tilde r^2} + \frac{1}{\tilde r}\frac{\mathrm{d}}{\mathrm{d}\tilde r}-\left(\frac{1}{\tilde r^2}+ \tilde k^2 q_1^2 \right) \right)
\left( \frac{\mathrm{d}^2}{\mathrm{d}\tilde r^2} + \frac{1}{\tilde r}\frac{\mathrm{d}}{\mathrm{d}\tilde r}-\left(\frac{1}{\tilde r^2}+ \tilde k^2 q_2^2 \right) \right) \tilde f(\tilde r) = 0
\eea
with boundary conditions on $\tilde r=a$
\bea
\tilde f'''+c_1 \tilde f'' +c_2 \tilde f' +c_3 \tilde f &=& 0 \label{eqA13}\\
\tilde f'''+c_4 \tilde f'' +c_5 \tilde f' +c_6 \tilde f &=& 0\label{eqA14}
\eea
where
$\tilde f (\tilde r) = f(\tilde r R)/ R$, $\tilde r = r/R$, $\tilde k = k R$ and 
\bea
q_1 = \frac{\lambda}{a}, \ q_2 = \sqrt{\dfrac{2+\tilde\kappa + 2\lambda^2 + a^4\lambda^2 \tilde\kappa}{2+\tilde\kappa + 2a^2 + a^4\lambda^2 \tilde\kappa}}
\eea
The coefficients $c_1$-$c_6$ are not given due to the length of the expressions. Note that Equation \eqref{eqA12} is identical to Equation (93) in \cite{yu2024incremental} since the bulk energy is considered of the same form, while the coefficients $c_1$-$c_6$ are different because they include the effect of surface energy \eqref{eqA8}, which differs from the one considered in \cite{yu2024incremental}.

The solution of the system is sought as a combination 
\bea
\tilde f(\tilde r) = d_1 I_1(\tilde k q_1 \tilde r)+d_2 I_1(\tilde k q_2 \tilde r) 
\eea
of modified Bessel functions of the first kind $I_1(x)$,
which implicitly satisfies Equation \eqref{eqA12}.  Substituting such ansatz into boundary conditions \eqref{eqA13} and \eqref{eqA14}, a system of two algebraic equations is obtained, which admits a nontrivial solution for coefficients $d_1$ and $d_2$ only if the determinant of the coefficient matrix vanishes, resulting in equation 
\bea
\Theta(\tilde k, a, \lambda,\tilde \gamma, \tilde \alpha) = 0 \label{eqA15}
\eea
where the exact form of function $\Theta(\tilde k, a, \lambda,\tilde \gamma, \tilde \alpha) $ is again not shown due to conciseness reasons. Following the arguments of \cite{yu2024incremental}, that the first onset of bifurcation occurs for zero wave number, Equation \eqref{eqA15} is reformulated to 
\bea
\Theta_0(a, \lambda,\tilde \gamma, \tilde \alpha) = 0
\eea
with
\bea
\Theta_0 &=& \lambda \left(a^2-\lambda ^2\right) \bigg[4\tilde\kappa {\left(a^2+\lambda ^2\right)}^{3/2}+4{\left(a^2+\lambda ^2\right)}^{3/2}+4a^2{\left(a^2+\lambda ^2\right)}^{3/2}+\tilde\kappa ^2{\left(a^2+\lambda ^2\right)}^{3/2}+4\lambda ^2{\left(a^2+\lambda ^2\right)}^{3/2}+\nonumber\\
&&4a^2\lambda ^2{\left(a^2+\lambda ^2\right)}^{3/2}+12a^2\tilde\alpha \lambda ^2+4a^2\tilde\alpha \lambda ^4+8a^4\tilde\alpha \lambda ^2+2a^2\tilde\kappa {\left(a^2+\lambda ^2\right)}^{3/2}+2\tilde\kappa \lambda ^2{\left(a^2+\lambda ^2\right)}^{3/2}+\nonumber\\ &&4a^4\tilde\kappa ^2\lambda ^2{\left(a^2+\lambda ^2\right)}^{3/2}  -5a^8\tilde\kappa ^2\lambda ^4{\left(a^2+\lambda ^2\right)}^{3/2}+8a^4\tilde\kappa \lambda ^2{\left(a^2+\lambda ^2\right)}^{3/2}+6a^4\tilde\kappa \lambda ^4{\left(a^2+\lambda ^2\right)}^{3/2}+ \label{eqA17}\\&&2a^6\tilde\kappa \lambda ^2{\left(a^2+\lambda ^2\right)}^{3/2}+ 16a^3\tilde\alpha \tilde\gamma \lambda ^3+6a^2\tilde\alpha \tilde\kappa \lambda ^2+30a^6\tilde\alpha \tilde\kappa \lambda ^4-8a^2\tilde\gamma ^2\lambda ^2{\left(a^2+\lambda ^2\right)}^{3/2}- \nonumber\\&& 16a^5\tilde\gamma \tilde\kappa \lambda ^3{\left(a^2+\lambda ^2\right)}^{3/2}\bigg] \bigg/\left({8a^4{\left(a^2+\lambda ^2\right)}^{3/2}\sqrt{\tilde\kappa a^4\lambda ^2+2a^2+\tilde\kappa +2}\sqrt{\tilde\kappa a^4\lambda ^2+2\lambda ^2+\tilde\kappa +2}}\right) \nonumber
\eea

To summarize, we are now left with the system of two algebraic Equations \eqref{eqA16} and \eqref{eqA17} depending on two surface material parameters $\tilde\alpha$, $\tilde\gamma$ and two stretches $a$, $\lambda$. In order to find a solution, two of the four unknown parameters are chosen, while the second two are determined by numerically solving the two algebraic equations.

 \bibliographystyle{elsarticle-num-names.bst} 
 \bibliography{Reference.bib,all.bib}









\end{document}